\documentclass[letterpaper,12pt]{article}
\usepackage{fancybox,amsmath,enumerate,amssymb,cite,epsfig}
\usepackage{latexsym} 
\newcommand{\roughly}[1]%
        {\mathrel{\raise.4ex\hbox{$#1$\kern-.75em\lower1ex\hbox{$\sim$}}}}

\def\dnk{{\rm d}^d k\ }

\def\pds{{\it PDS}\ }
\def\ms{MS}\def\){\right)}
\def\({\left(}
\def\]{\right]}
\def\[{\left[}
\def\si{{}^1\kern-.14em S_0}
\def\siii{{}^3\kern-.14em S_1}
\def\diii{{}^3\kern-.14em D_1}
\newcommand\bfpi{{\boldsymbol \pi}}
\newcommand\bfl{{\boldsymbol \ell}}

\newcommand\bfk{{\bf k}}
\newcommand\bfK{{\bf K}}
\newcommand\bfp{{\bf p}}
\newcommand\bfP{{\bf P}}

\newcommand\bfr{{\bf r}}
\newcommand\bfv{{\bf v}}

\newcommand\Tr{{\rm Tr}}

\newcommand\CD{{\cal D}} 
\newcommand\CE{{\cal E}} 
 
\newcommand\CH{{\cal H}} 
 
\newcommand\CL{{\cal L}} 
 
\newcommand\CA{{\cal A}} 
\newcommand\CO{{\cal O}} 
 
\newcommand\CZ{{\cal Z}} 
\newcommand\half{\frac{1}{2}} 
 
\newcommand\beq{\begin{eqnarray}} 
\newcommand\eeq{\end{eqnarray}} 
\newcommand\Bbeq{\begin{Beqnarray}} 
\newcommand\Beeq{\end{Beqnarray}} 
\newcommand\eqn[1]{\label{eq:#1}} 
 
\newcommand\eq[1]{eq. (\ref{eq:#1})} 
 
\newcommand\bra[1]{\langle #1 |}
\newcommand\ket[1]{| #1\rangle}

\newcommand\bfq{{\bf q}}

\newcommand{\mybar}[1]%
        {\kern 0.8pt\overline{\kern -0.8pt#1\kern -0.8pt}\kern 0.8pt}
\newcommand{\sla}[1]%
        {\raise.15ex\hbox{\kern+.15em$/$}\kern-.56em #1}
\newcommand{\dsl}{\sla{\partial}}
\newcommand{\Dsl}{\raise.15ex\hbox{\kern+.15em$/$}\kern-.72em D}
\newcommand{\fm}{{\rm ~fm }}
\newcommand{\eV}{{\rm ~eV }}

\newcommand{\GeV}{{\rm ~GeV }}
\newcommand{\TeV}{{\rm ~TeV }}
\newcommand{\MeV}{{\rm ~MeV }}

\newcommand\vev[1]{\langle #1 \rangle}

\newcommand\expect[3]{\langle #1|#2|#3\rangle}
\newcommand{\eqsii}[2]{eqs.~(\ref{eq:#1}, \ref{eq:#2})}

\begin{document}

\title{Five lectures on effective field theory}
\author{David B. Kaplan}
\date{}
\maketitle
\abstract{Lectures delivered at the $17^{th}$ National Nuclear Physics
Summer School 2005, Berkeley, June 6-17, 2005.}

\tableofcontents

\vfill\eject

\section{ Introduction}
\label{sec:1}

\bigskip
\hrule
\bigskip

\subsection{What is an effective field theory?}
\label{sec:1a}

The content of a quantum theory is  encoded in its correlation
functions, which  in general depend in a complicated way 
on the momenta of incoming and outgoing particles.  In
particular they
exhibit cuts and poles and various other nonanalytic
behavior, which arise when the kinematics allow for  physical
intermediate states.  When that happens, the $1/(p^2-m^2)$ propagators
for the intermediate states become singular which leads to the
nonanalytic behavior of the correlation function \footnote{This
  discussion is for a relativistic 
  theory, but analogous statements may be made for nonrelativistic
  theories.}.  Once you realize that this is the source of most of the
complicated behavior in a correlation function, it is apparent that
when the kinematics are far from being able to produce a propagating
heavy state, the contribution of that heavy state to the correlation
function of interest will be relatively simple, well approximated by the
first few terms in a Taylor expansion in the incoming momenta of the
scattering problem.  Such will be the case, for example, when
considering neutron decay at tree level in the Standard Model, where a $d$ quark decays into a
$u$ quark and a virtual $W$, which then turns into an electron and
anti-neutrino.  The
energy released is about $10^{-6}$ times the $W$ mass, and so the $W$
propagator may be approximated by $1/(p^2-M_W^2) \simeq -1/M_W^2 -
p^2/M_W^4 + \ldots$. 

While such a Taylor expansion makes the process
slightly simpler to analyze, the benefits of expanding each amplitude seem minimal, and it does
not seem obvious how to generalize the procedure to 
nonperturbative physics. 

Instead of Taylor expanding each amplitude it turns out
to be much more profitable to expand the Lagrangian in local
operators that only involve the light degrees of freedom, where the
expansion is in powers of the external momenta of the light fields
(appearing as derivatives in the Lagrangian) divided by the scale of
heavy physics.  Such a Lagrangian is called an effective field
theory. In principle it consists of
 an
infinite sum of local operators  which are typically of diminishing importance 
with higher dimension.  In practice this series is truncated according
to the accuracy desired for the process in question. With this
Lagrangian one can then compute any low energy amplitude one wishes.
 
There are many situations in which effective field theories are of
utility:
\begin{itemize}
\item They allow one better understand complicated problems involving
  a lot of different length scales, and to understand the results
  qualitatively from dimensional analysis;
\item They allow one to  compute
   low energy scattering amplitudes without having a detailed
   understanding of short distance physics, or to avoid wasting time
   calculating 
   tiny effects from known short distance physics;
\item In nonperturbative theories (such as low energy QCD) one can
  construct a predictive effective field theory  for low energy
  phenomena by combining a power counting of operators with the
  symmetry constraints of the underlying theory (such as the chiral
  Lagrangian for pion physics). 
\item By regarding theories of known physics as effective field
  theory descriptions of more fundamental underlying physics,  one can
  work from bottom up, extrapolating from 
  observed rare processes to a more complete theory of short distance physics.
\end{itemize}

In these lectures I will try to explain in more detail what is an
effective theory, how to construct it, and how to use it. Effective
field theory is a tool, and writing a comprehensive review 
of its development and applications would be like writing a treatise
on the hammer. So this is not a review.  Instead I have tried to
give you a general idea how effective field theory works and  can be used, citing papers I find
particularly interesting or pedagogical without trying to be
historical.  I  often
focus on ways that I have used EFT in my own research. There are
available a
number of other equally idiosyncratic reviews, and I recommend that you read  them to gain
perspective and to encounter a wider range of applications.  Favorites
mine are found in refs.
\cite{Georgi:1994qn,Cohen:1993uk,Manohar:1996cq,Polchinski:1992ed,Schafer:2003yh,Lepage:1997cs}.

\subsection{Local operators and scaling dimension in a relativistic theory}
\label{sec:1b}

If an effective field theory is given by a Lagrangian with an infinite
number of operators in it, to be useful there must be a criterion for
ignoring most of them.  So as a prototypical example of such a theory,
consider the Lagrangian (in four dimensional Euclidean spacetime, after a Wick rotation
to imaginary time) for  relativistic  scalar field with a
$\phi\to-\phi$ symmetry:
\beq
\CL_E = 
\half (\partial\phi)^2 + \half m^2 \phi^2 +\frac{\lambda}{4! }\phi^4 +
\sum_n \left(\frac{c_n}{\Lambda^{2n}} \phi^{4+2n} +
  \frac{d_n}{\Lambda^{2n}} (\partial\phi)^2 \phi^{2+2n} +
  \ldots\right) 
\eqn{scalar}\eeq
 I have introduced a scale $\Lambda$,
the momentum cutoff of the theory so that 
the couplings $l$, $c_n$ and $d_n$ are all dimensionless, since the mass
dimension of $\phi$ is 1, while the mass dimension of $\CL$ is 4.
\footnote{To determine the mass dimension, start with the fact that
  relativity requires $p$ and $E$ to have mass dimension one, so that
  the uncertainty relations with $\hbar=1$ require $x$ and $t$ to
  have mass dimension -1.  Then  since the action $S=\int\,d^dx\,\CL$
  in $d$-dimensions, must be
  dimensionless, it follows that $(\partial\phi)^2$ has dimension $d$,  and so a scalar field has
  mass dimension $(d/2-1)$.}.  I will assume for now that the theory
is perturbative, namely that
 $\lambda\ll 1$, $c_n\ll 1$ and $d_n\ll 1$.  

\begin{figure}[t]
\centerline{\epsfxsize=2.25in \epsfbox{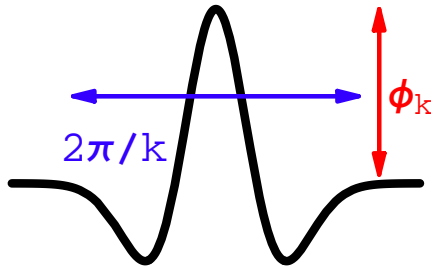}}
\caption{{\it A sample configuration contributing to the path integral for
  the scalar field theory in \eq{scalar}.  Its amplitude is $\phi_k$
  and 
  has wave number $\sim k$ and spatial extent $\sim 2\pi/k$.}}
\label{fig:1.1}
\vskip .2in
\end{figure}

How do we figure out which of the interactions in $\CL$ are the
most important?  To compute correlation functions in this theory one
performs the path integral:
\beq
\int \, D\phi\, e^{-S_E}\ ,\qquad S_E = \int d^4x\, \CL_E\ .
\eeq
Now consider a particular field configuration $\tilde\phi$ that
contributes to this path integral, where $\tilde\phi$ is localized to a
spacetime volume of size $L^4$, where $L\simeq 2\pi/k$ with wavenumber $|k_\mu|\sim k$, and has amplitude
$\phi_k$. See figure \ref{fig:1.1}, where I sketched such a ``wavelet''.
With this configuration, the Euclidean action is given by
\beq
S_E\simeq (2\pi)^4 \left[ \frac{\widehat \phi_k^{\,2}}{2} +
  \frac{m^2}{k^2}\widehat\phi_k^{\,2} + \frac{\lambda}{4!}{\widehat\phi_k}^{\,4} + \sum_n \left(c_n
  \left(\frac{k^2}{\Lambda^2}\right)^n {\widehat\phi_k}^{\,4+2n} +  d_n
  \left(\frac{k^2}{\Lambda^2}\right)^n
  {\widehat\phi_k}^{\,4+2n}+\ldots\right)\right]\ ,\cr
\eqn{act1}\eeq
where
\beq
\widehat\phi_k \equiv \phi_k/k\ .
\eeq
Now for the path integral, consider the ordinary integration
over the amplitude $\widehat \phi_k$ for a particular $k$:
\beq
\int\,d\widehat\phi_k\, e^{-S_E}\ .
\eqn{pint1}
\eeq  
The integral is dominated by those values of $\widehat\phi_k$ for
which $S_E\lesssim 1$. Which are the important terms in $S_E$ in this
region?   First, assume that the particle is relativistic, $k\gg
m$. Then evidently, as the amplitude $\widehat\phi_k$ gets large, the
first term in $S_E$ to become become large is the kinetic term,
$(2\pi)^4 \widehat \phi_k^{\,2}$. It determines that the integral gets
its maximum contribution at $ \phi_k\sim k/(2\pi)^2$. It is because the kinetic term 
controls the fluctuations of the scalar field that we ``canonically
normalize'' the field such that the kinetic term is $\half
(\partial\phi)^2$, and perturb in the coefficients of the other
operators in the theory.

Since we
assume that $k\le \Lambda$, $\Lambda$ being the
momentum cutoff, and that the $c_n$ and $d_n$ couplings are $\ll 1$, we
see that the integrand will become small when the terms in $S_E$ which
are quadratic in $\widehat\phi_k$ become $O(1)$, at which point the
terms with higher powers of $\widehat\phi_k$ are still small.

What happens as we vary $k$? We see from \eq{act1} that as $k$ is
reduced, the $c_n$ and $d_n$ terms, proportional to
$(k^2/\Lambda^2)^n$, get smaller.  Such operators are termed
``irrelevant'' in Wilson's language, because they become unimportant
in the infrared (low $k$, long wavelength). In contrast, the mass term  becomes more
important; it is called a ``relevant'' operator.  The kinetic term and the
$\lambda\phi^4$ interaction do not change; such operators are called
``marginal''.

Classification
of operators as irrelevant, marginal, or relevant may be done in a
more straightforward way by a simple scaling exercise.  Consider an arbitrary
 field configuration $\phi(x)$ contributing to the path
integral. The action for this field configuration is
$S_E( \phi(x); m, \lambda, c_n, d_n, \ldots)$, where the functional $S_E$
is given in \eq{act1}.  I have made a point of listing as its
arguments not only the field $\phi$, but all of the couplings that
characterize the theory. Now consider the family of
field configurations 
\beq
\phi_\xi(x) =  \phi(\xi x)\ .
\eqn{phixi}
\eeq
 For example, if $\phi(x)=e^{i k\cdot x}$, 
then $\phi_\xi(x) = 
e^{i\xi {k\cdot x}}$, so that $\xi\to 0$ corresponds
to looking at configurations with longer wavelength $k'=\xi k$ (the
``IR'').  The action for this family of configurations is
%
\begin{equation}
\begin{aligned}
S_E(\phi_\xi(x); \Lambda,m^2,&\lambda,c_n,d_n,\ldots) \cr & =
\int d^4 x\,
\frac{1}{2}(\partial_{x}\phi(\xi x))^2 + \half m^2 
\phi(\xi x)^2 + \frac{\lambda }{4!}\phi(\xi x)^4 \cr&\qquad+\sum_n c_n
 \frac{\phi^{4+2n}(\xi x)}{\Lambda^{2n}} +
d_n 
\frac{(\partial_{x}\phi(\xi x))^2\phi^{2n}(\xi x)}{\Lambda^{2n}} \cr 
&=
\int d^4 x'\,
\half(\partial_{x'}\phi'(x'))^2 + \half m^2 \xi^{-2}
{\phi'(x')}^2 + \frac{\lambda}{4!}{\phi'(x')}^4
\cr&
\qquad +\sum_n c_n \xi^{2n} \frac{{\phi'}^{4+2n}(x')}{\Lambda^{2n}} +
d_n \xi^{2n}
\frac{(\partial_{x'}{\phi'(x')})^2{\phi'}^{2n}(x')}{\Lambda^{2n}}\ ,
\end{aligned}
\end{equation}
where $\phi'(x)\equiv \xi^{-1}\phi(x)$, and I made the change of
integration variable $x'=\xi x$.  Since $x'$ is a dummy variable, we
can drop the prime and recognize that  the above action equals the
original action with rescaled fields and couplings:
\beq
S_E(\phi(\xi x); \Lambda,m^2,&\lambda,c_n,d_n,\ldots) =
S_E\left(\xi^{-1}\phi(x); \xi^{-2} m^2,\lambda, c_n\xi^{2n},d_n \xi^{2n},\ldots\right)\ .
\eeq
so that
\beq
\phi\to \xi^{-1} \phi\ ,\quad m^2\to \xi^{-2} m^2\ ,\quad \lambda\to\lambda\ ,\quad  c_n \to \xi^{2n}
c_n\ ,\quad d_n \to \xi^{2n} d_n\ 
.
\eqn{scaling}\eeq 
Now as  we scale to the infrared (long wavelength, low energy
processes) by taking $\xi \to 0$, we see again that the mass term grows
in importance (relevant), the $c_n$ and $d_n$ couplings fall like
$\xi ^{2n}$ (irrelevant, with higher $n$ being more irrelevant), and the
kinetic terms and $\lambda\phi^4$ interaction not changing (marginal).  Using
this analysis one can also easily see that if there was a constant $\CE_0$
added to our Lagrangian (vacuum energy, or a cosmological constant
term), it would scale as $\CE_0\to \xi^{-4}\CE_0$, and hence would be
very relevant.

It is convenient to define a scaling dimension, which is the negative
of the power of $\xi$ with which a quantity scales,
and I will denote the this scaling dimension with square brackets
$[\cdots]$.  Thus $[x]=-1$ and \eq{scaling} tells us
\beq
[\phi]=1\ ,\quad [m^2]=2\ ,\quad [\lambda]=0\ ,\quad [c_n] =
  [d_n]=-2n\ .
\eeq
Marginal operators have coefficients with scaling dimension zero;
coefficients of relevant operators have positive scaling dimensions;
irrelevant operators have negative scaling dimension. Note that this
scaling dimension for a classical, relativistic action 
is just the same as the mass dimension when $\hbar =c=1$.

You might wonder about the above derivation:  why did I choose to
scale $\phi'=\xi^{-1}\phi$ instead of some other power? Then the
kinetic term would have picked up powers of $\xi$, but other operators
might not have, depending on the choice of scaling for $\phi$.  The answer
is that the kinetic term was chosen to be the
scale invariant term because, for a weakly interacting relativistic
system,  it is the kinetic term that dominates the
size of fluctuations in the path integral (see \eq{act1}).  This sort of scaling
argument will clearly fail when $k\lesssim m$, at which point the mass term
dominates. So we will have to provide a different scaling argument for
nonrelativistic theories.
 
\subsubsection{Scaling in a nonrelativistic theory}

Going back to our wavelets, what happens happens when $k$ is taken
below $m$, and the 
particle becomes nonrelativistic?  In Minkowski spacetime we have $\CL
= \half(\dot\phi^2-(\nabla\phi)^2 -m^2 \phi^2 -\ldots)$, and for a
nonrelativistic particle $\phi$ has the expansion $\phi =
(a_k\,e^{-iEt}+ a_k^\dagger e^{i E t})$, with $E\simeq m$. Note that
 since $E\simeq m$, both  $\dot\phi^2$ and $m^2\phi^2$ are large but nearly
cancel in the Minkowski action.  This confuses the power
counting. Therefore it is convenient to change variables to absorb the
uninteresting rapid oscillation associated with the particle's rest
energy.
This leads us to the substitution 
\beq
\phi(x,t) =  \frac{1}{\sqrt{2m}}\left(e^{-imt}\psi(x,t)+c.c\right)\ ,
\eeq
where $\psi$ is complex and $\dot\psi\ll m \psi$. Because of this last
condition, we can drop from our Lagrangian terms with unequal numbers
of $\psi$ and $\psi^*$, since they would be multiplied by nonzero
powers of $e^{-imt}$ and would  vanish when the Lagrangian is
integrated over time.  The Minkowski
spacetime Lagrangian may now be rewritten as
\beq
\CL_M = \psi^*\left(i\partial_t + \frac{\nabla^2}{2m}\right)\psi
-\frac{\lambda}{8m^2}(\psi^\dagger\psi)^2 + \ldots
\eqn{nonrel}
\eeq
and the Euclidean action is given by
\beq
S_E(\psi(x,t);\lambda,\ldots) = \int dt\,d^3x\, \psi^*\left(\partial_t
  - \frac{\nabla^2}{2m}\right)\psi 
+\frac{\lambda}{8m^2}(\psi^\dagger\psi)^2 + \ldots
\eqn{act2}\eeq

The $\partial_t $ and $\frac{\nabla^2}{2m}$ operators must be
considered to be of the same size, one corresponds to the energy $E$
and the other to $p^2/2m$. Thus to analyze how things scale, we must
have time and space scale differently, namely $x\to x'=\xi x$, $t\to t' =
\xi ^2 t$.  As before, we determine the scaling properties of $\psi$ by
requiring that the kinetic term be scale invariant, because we believe
that it dominates the fluctuations in the path integral.  Thus we
consider the family of field configurations 
\beq
\psi_\xi (x,t) =
\psi(\xi x,\xi ^2t)\ .
\eeq
The new action is
\begin{equation}
\begin{aligned}
S_E(\psi_\xi &(x,t);\lambda,\ldots) 
\cr &
=\int dt\,d^3x\ 
\,\psi^*(\xi x,\xi ^2t)\left(\partial_t -
  \frac{\nabla^2_x}{2m}\right)\psi(\xi x,\xi ^2t) 
+\, \frac{\lambda}{8m^2}(\psi^\dagger(\xi x,\xi ^2t)\psi(\xi x,\xi ^2t))^2 + \ldots
\cr &=
 \int dt'\,d^3x'\, \xi^{-3}\psi^*(x',t')\left(\partial_{t'} - \frac{\nabla_{x'}^2}{2m}\right)\psi(x',t')
+\xi^{-5} \frac{\lambda}{8m^2}(\psi^\dagger(x',t')\psi(x',t'))^2 + \ldots
\cr &=
S_E(\psi'(x,t);\xi \lambda,\ldots)\ .
\end{aligned}
\end{equation}
where $\psi'(x) \equiv \xi^{-3/2} \psi(x)$.  Thus in this case we find
\beq
[x]=-1\ ,\quad [t]=-2\ ,\quad [\psi]=\frac{3}{2} ,\quad [\lambda]=-1\ .
\eeq
 We see then that the $\phi^4$
interaction, which was marginal for 
relativistic scalars, becomes the  irrelevant $|\psi|^4$ interaction
for nonrelativistic particles. In 
one of the problems I have provided, you are to show that that the
$\lambda  (\psi^\dagger\psi)^2$ interaction corresponds to a
$\delta$-function potential between two particles, and that the
scaling properties we have derived make sense from the point of view
of the Schr\"odinger equation.

Nonrelativistic EFT's are commonly used to describe nonrelativistic
interactions between atoms or nuclei; for analyzing bound states such
as positronium in QED (the EFT is referred to as ``NRQED'', for
nonrelativistic QED); and for heavy quarkonia---bound states of a heavy
quark-antiquark pair---in NRQCD.  Such theories are more complicated
that what I just described due to the interactions of light particles,
such as the photon in NRQED and the gluon in NRQCD.

\subsection{HQET}
\label{sec:1b2}

A somewhat different EFT is encountered in the analysis of hadrons
containing a $c$ or $b$ quark whose mass is greater than the QCD scale
$\Lambda_{QCD}$, which characterizes the binding energy of hadrons.
While the heavy quark is nonrelativistic in the hadron rest frame,
such hadrons can also include light quarks ($u$,  
$d$, $s$) which are not. Isgur and Wise noticed that there was a
symmetry between mesons with a $b$ quark, and mesons with a $c$ quark
in the limit that both were heavy, but with unequal masses.  That
there should be a symmetry is easy to see:  as far as the light
degrees of freedom are concerned, the heavy quark just looks like a
static color source, pretty much independent of its (heavy)
mass. Therefore the light quarks in a $D$ meson and those in a $B$
meson will have the same wavefunction, up to $1/M$ corrections.  The
system seems ideal for an EFT treatment, with $\Lambda_{QCD}/M$ being
the expansion parameter, where $M$ is the heavy quark mass.

Since we are talking about bound states of both heavy and light degrees
of freedom, a
nonrelativistic EFT is inadequate.  A better EFT formalism was introduced
by Georgi, who wrote the momentum of the heavy quark as \cite{Georgi:1990um}
\beq
p^\mu = m v^\mu + k^\mu\ ,
\eeq
where $m$ is the heavy quark mass, $v_\mu$ is the 4-velocity of the
heavy quark satisfying $v_\mu v^\mu=1$, and $k^\mu=O(\Lambda_{QCD}$ is
the ``residual momentum'' characterizing how off-shell the heavy quark
is. He noted that in the large $m$ limit, the velocity $v^\mu$ is
unchanged in strong interaction processes, which involve momentum
transfer of $O(\Lambda_{QCD})$.  Therefore he introduced a different
field for each velocity $v$:
\beq
h_v(x) = e^{i m \sla{v} v_\mu x^\mu} q(x)\ ,
\eeq
where $q(x)$ is the usual Dirac spinor for the heavy quark.   Note that
for $v=\(1,0,0,0\)$ the exponential prefactor is just $e^{imt
  \gamma_0}$, so the above transformation eliminates the fast time
behavior for both  particles and antiparticles at the same time.
Then $h_v$ can be decomposed into spinors $h_v^+$ which annihilates
heavy quarks with velocity $v$, and $h_v^-$ which creates heavy
antiquarks with velocity $v$, via the projection operators
\beq
h_v^\pm = \left(\frac{1\pm \sla{v}}{2}\right) h_v\ .
\eeq
One can then drop operators with interactions between $h^+_v$ and
$h^-_v$ as pair creation and annihilation is beyond the purview of the
EFT.  The free kinetic term then looks like
\beq
\CL_0 = i\mybar{h^+_v} \dsl h^+_v +  i\mybar{h^-_v} \dsl h^-_v =
i\mybar{h_v} \sla{v} v_\mu\partial^\mu  h_v\ .
\eeq
One great virtue of this EFT is that $b$ and $c$ quarks now look
similar, despite the difference of the quark masses, and one can
easily see the Isgur-Wise symmetry between them.  HQET is used
extensively in a $\Lambda_{QCD}/m$ expansion to analyze heavy meson
phenomenology which is important for determining the CKM angles
of the standard model. If you want to find out about HQET, see ref. \cite{Manohar:2000dt}.

\subsection{Some qualitative  applications}
\label{sec:1c}

This has all been rather formal.  How are the scaling properties we
have derived reflected in physical applications?  I will consider here
several qualitative  examples.

\subsubsection{Fermi's effective theory of the weak interactions}

The term ``weak interactions'' refers in general to any interaction mediated by
the $W^\pm$ or $Z^0$ bosons, whose masses are approximately $80$ GeV and $91$ GeV
respectively. The currents they couple to are called ``charged 
currents'' and {\it neutral currents} respectively.  The charged currents
are given by\footnote{Careful with factors of 2! I give here the
  currents that the $W^\pm$ and $Z$ boson couple to; however, weak
  currents were historically defined to be twice these expressions, long before the standard
  model was written down.}
\beq
J^\mu_{\pm} = \frac{j^\mu_1\mp i\,j^\mu_2}{\sqrt{2}}\ ,
\eeq
where
\beq
j^\mu_a = \frac{e}{\sin\theta_w}\, \sum_{\psi} \mybar\psi \gamma^\mu
\left(\frac{1-\gamma_5}{2}\right) \frac{\tau_a}{2} \psi\ ,a=1,2,3\ ,
\eeq
with $\tau_a$ being the first two Pauli matrices and  $g_2\equiv e/\sin\theta_w$ being the
$SU(2)$ gauge coupling (written in terms of the electromagnetic
coupling $e$ and the weak angle $\theta_w$). The fermions $\psi$
participating in the charged current interaction are the leptons
\beq
\psi = \begin{pmatrix} \nu_e\cr e \end{pmatrix}\ ,\quad
\begin{pmatrix} \nu_\mu\cr \mu \end{pmatrix}\ ,\quad \begin{pmatrix}
  \nu_\tau\cr \tau \end{pmatrix}\ ,
\eeq
and the quarks
\beq
\psi= \begin{pmatrix} u\cr d' \end{pmatrix}\ ,\quad  \begin{pmatrix}
  c\cr s' \end{pmatrix}\ ,\quad  \begin{pmatrix} t\cr b'
\end{pmatrix}\ ,
\eeq
with the ``flavor eigenstates'' $d'$, $s'$ and $b'$ being related to
the mass eigenstates $d$, $s$ and $b$ by the unitary Cabibbo-Kobayashi-Maskawa
(CKM) matrix:
\beq
q'_i = V_{ij} q_j\ .
\eeq
The elements of the CKM matrix are named after which quarks they
couple through the charged current, namely  $V_{11}\equiv V_{ud}$,
$V_{12}\equiv V_{us}$, $V_{21}\equiv V_{cd}$, etc.

The $Z^0$ boson has a mass $M_Z = M_W\cos\theta_w$ and couples to the
current
\beq
J^\mu_Z = \frac{e}{\sin\theta_w\,\cos\theta_w} \left(j_3 -
  \sin^2\theta_w j_{\text{em}}\right)
\eeq
where $j_{\text{em}}$ is the  electromagnetic current, where the
neutrinos, charged leptons, up-type quarks and down-type quarks have
$Q_{\text{em}}=0,-1,\frac{2}{3}$ and $-\frac{1}{3}$ respectively.

For many  processes the dominant weak interaction is given
by the tree level exchange of a $W$ or $Z$ boson. If the process is at
low energy  (where the momentum exchanged in any
channel satisfies $p^2\ll M_W^2$), then the gauge boson propagators
may be approximated by a constant, by Taylor expanding in $p^2/M^2$
\beq
\frac{1}{p^2-M^2} = -\frac{1}{M^2} + \frac{p^2}{M^4} + \ldots
\eeq
and keeping only the leading term.  Since the Fourier transform of a
constant is a $\delta$ function, the weak boson exchange gives rise to
a point-like current-current interaction:
\beq
\CL_{\text{eff}}^\text{weak} &=& 8\frac{G_F}{\sqrt{2}} \left(
 J^\mu_+ J_{-\mu} +\half J^\mu_Z J_{Z\mu}\right)\ ,\quad G_F =
\frac{\sqrt{2} e^2}{8 \sin^2\theta_w M_W^2} = 1.166\times
10^{-5}\,\GeV^2\ .
\eqn{fermi}\eeq
\begin{figure}[t]
\centerline{\epsfysize=1in€\epsfbox{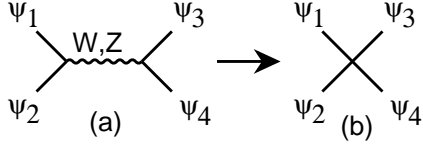}}
\caption{{\it (a) Tree level $W$ and $Z$ exchange between four fermions. (b)
The effective vertex in the low energy effective theory (Fermi interaction).}}
\label{fig:1.2}
\vskip .2in
\end{figure}

The charged current part, written in terms of leptons and nucleons
instead of leptons and quarks, was postulated by Fermi to explain
neutron decay. Neutral currents were proposed in the 60's and
discovered in the 70's. The relation between $G_F$ and $M_W$ is
derived by ``matching'' --- requiring that the two processes in
Fig.~\ref{fig:1.2} give the same $S$-matrix elements.

Since neutrinos carry no electric or color charge, in the standard
model all of their low energy interactions are contained in 
$\CL_{\text{eff}}^\text{weak}$ in \eq{fermi}.  Thus the neutrino
cross-section $\sigma$  must be proportional to $G_F^2$ which has
dimension -4.  But a cross section has dimensions of area, or mass
dimension -2.  If the scattering process of interest is relativistic,
then the only other scale around is the center of mass energy
$\sqrt{s}$.  
Therefore on dimensional grounds, the cross-section must scale with energy as
\beq
\sigma_\nu \simeq G_F^2 s\ .
\eqn{snu}
\eeq
This explains why low energy neutrinos are so hard to detect, and 
the weak interactions are weak;  they won't be at the LHC, though,
where $G_F s >1$ and the Taylor expansion of the $W$ and $Z$
propagators is unjustified.



\subsubsection{The blue sky}

Why the sky is blue? The full explanation is somewhat complicated (see
Jackson's {\it Classical Electrodynamics}, 3rd ed.), but the basic
reason is that blue light from the sun scatters more strongly from
atoms in the atmosphere than red. Consider the
problem of low energy light scattering from neutral atoms in their ground
state (Rayleigh scattering). By ``low energy'' I mean that the photon energy $E_\gamma$  is
much smaller than the excitation energy $\Delta E$ of the atom, which is of
course much smaller than its inverse size or mass:
$$ E_\gamma\ll \Delta E\ll a_0^{-1}\ll  M_{atom}\ .$$
Thus the process is necessarily elastic scattering, and to a good approximation
we can ignore that the atom recoils, treating it as infinitely heavy. Let's
construct an ``effective Lagrangian'' to describe this process.    This means
that we are going to write down a Lagrangian with all interactions describing
elastic photon-atom scattering that are allowed by the symmetries of the world
--- namely Lorentz invariance and gauge invariance.  Photons are described by a
field $A_\mu$ which creates and destroys photons; a gauge invariant object
constructed from $A_\mu$ is the field strength tensor $F_{\mu\nu} =
\partial_\mu A_\nu - \partial_\nu A_\mu$.  The atomic field is defined as
$\phi_v$, where $\phi_v$ destroys an atom with four-velocity $v_\mu$
(satisfying $v_\mu v^\mu=1$, with $v_\mu=(1,0,0,0) $ in the rest-frame of the
atom), while $\phi^{\dagger}_v$ creates an atom with four-velocity
$v_\mu$.  The kinetic terms in the effective Lagrangian are then just
\beq
\CL_0 = \pi_v^\dagger i v^\mu\partial_\mu \phi_v -\frac{1}{4}
F_{\mu\nu}F^{\mu\nu}\ .
\eeq
Note that the energy of the atom at rest is defined to be zero (the
equation of motion is $\partial_t \phi_v=0$ in the atom's rest frame).

 So
what is the most general form for interactions in $\CL^{\text{eff}}$?
Since the atom is electrically neutral, gauge invariance implies that $\phi$
can only be coupled to $F_{\mu\nu}$ and not directly to $A_\mu$.   So
$\CL_{eff}$ is comprised of all local, Hermitian monomials in
$\phi^\dagger_v\phi_v$, $F_{\mu\nu}$, $v_{\mu}$, and
$\partial_\mu$. We can neglect operators which vanish by the equations
of motion, to the order we work.   Thus we
are led to consider the interaction Lagrangian
\beq
\CL_{\text{int}} &=& c_1 \phi^{\dagger}_v\phi_v
F_{\mu\nu}F^{\mu\nu} + c_2  \phi^{\dagger}_v\phi_v v^\alpha F_{\alpha\mu}
v_\beta F^{\beta\mu}\cr
&&+c_3 \phi^{\dagger}_v\phi_v(v^\alpha\partial_\alpha)  F_{\mu\nu}F^{\mu\nu} +
\ldots
\eqn{rayleigh}
\eeq

The above expression involves an infinite number of operators and an infinite
number of unknown coefficients!  Nevertheless, dimensional analysis allows us
to identify the leading contribution to low energy scattering of light by
neutral atoms.  Since the light is relativistic, and the atom is
basically a static scatterer, we use the relativistic power
counting, $[x]=[t]=-1$. Thus the scaling dimensions are given by
$$[\partial_\mu] = 1\ ,\qquad [F_{\mu \nu}]=2\ ,\qquad [\phi]=\frac{3}{2}\ .$$
The first follows from the fact that $\partial_\mu$ scales like
$1/x_\mu$. The second  two follow from $\CL_0$ above, and the fact
that  $[\CL]=4$.

Since the effective Lagrangian has dimension 4, the coefficients $c_1$, $c_2$
etc. also have dimensions.  It is easy to see that they all have negative mass
dimensions:
$$[c_1] =[c_2] = -3\ ,\qquad [c_3] = -4$$
and that operators involving higher powers of $\partial\cdot v$ would have
coefficients of even more negative dimension. It is crucial to note that these
dimensions must be made from dimensionful parameters describing the atomic
system --- namely its size $r_0$ and the energy gap $\delta E$ between the
ground state and the excited states. The other dimensionful quantity,
$E_\gamma$, is explicitly represented by the derivatives $\partial_\mu$ acting
on the photon field.  Thus for $E_\gamma\ll \Delta E, r_0^{-1}$ the dominant
effect is going to be from the operator in $\CL$ which has the {\it
lowest} dimension.  There are in fact two leading operators, the first two in
eq. \eq{rayleigh}, both of dimension 7.   Thus low energy scattering is dominated by
these two operators, and we need only compute $c_1$ and $c_2$.(The
independence of $c_{1,2}$
means that electric and magnetic scattering can have independent strengths).

What are the sizes of the coefficients?  To do a careful analysis one needs to
go back to the full Hamiltonian for the atom in question interacting with
light, and ``match'' the full theory to the effective theory.  We will discuss
this process of matching later, but for now we will just estimate the sizes of
the $c_i$ coefficients.  We first note that   extremely low energy photons
cannot probe the internal structure of the atom, and so the cross-section ought
to be classical, only depending on the size of the scatterer.  Since such low
energy scattering can be described entirely in terms of the coefficients $c_1$
and $c_2$, we conclude that \footnote{In fact, magnetic scattering
  will in general be weaker than electric, since for small atoms the
  electrons within couple to magnetic fields with a
  $v/c\sim \alpha$ suppression.  However, this cannot be seen from
  the low energy point of view, but must be derived from matching
  conditions from the full theory.}
$$c_1 \simeq c_2\simeq r_0^3\ .$$
The effective Lagrangian for low energy scattering of light is therefore
\beq
\CL_{eff} = r_0^3 \left( a_1 \phi^{\dagger}_v\phi_v
F_{\mu\nu}F^{\mu\nu} + a_2  \phi^{\dagger}_v\phi_v v^\alpha F_{\alpha\mu}
v_\beta F^{\beta\mu}\right)
\eqn{rayleighii}
\eeq
where $a_1$ and $a_2$ are dimensionless, and expected to be $\CO(1)$.  The
cross-section (which goes as the amplitude squared) must therefore be
proportional to $r_0^6$.  But a cross section $\sigma$ has dimensions of area,
or $[\sigma]=-2$, while $[r_0^6]= -6$.  Therefore the cross section must be
proportional to
\beq
\sigma\propto E_\gamma^4 r_0^6 \ ,
\eqn{rcross}
\eeq
growing like the fourth power of the photon energy.  Thus blue light is
scattered more strongly than red, and the sky looks blue.

Is the expression \eq{rcross} valid for arbitrarily high energy?  No, because we
left out terms in the effective Lagrangian we used.  To understand the size of
corrections to \eq{rcross} we need to know the size of the $c_3$ operator (and the
rest we ignored).  Since $[c_3]=-4$, we expect the effect of the $c_3$ operator
on the scattering amplitude to be smaller than the leading effects by a factor
of $E_\gamma/\Lambda$, where $\Lambda$ is some energy scale.  But does
$\Lambda$ equal $M_{atom}$, $r_0^{-1}\sim \alpha m_e$ or $\Delta E\sim \alpha^2
m_e$?   The latter is the smallest scale and hence the most important.  We
expect our approximations to break down as $E_\gamma \to \Delta E$ since for
such energies the photon can excite the atom.  Hence we predict
\beq
\sigma\propto E_\gamma^4 r_0^6\left(1 +\CO(E_\gamma/\Delta E)\right) .
\eqn{rcrossii}
\eeq
The Rayleigh scattering formula ought to work pretty well for blue light, but
not very far into the ultraviolet.  Note that eq. \eq{rcrossii} contains a lot of
physics even though we did very little work.  More work is needed to compute
the constant of proportionality.

\subsubsection{The binding energy of charmonium in nuclei}

Closely related to the above example is the calculation of the binding energy
of charmonium (a $\bar c c$ bound state, where $c$ is the charm quark) to
nuclei \cite{Luke:1992tm}.  In the limit that the charm quark mass $m_c$ is very heavy, the
charmonium meson can be thought of as a Coulomb bound state, with size $\sim
\alpha_s(m_c) m_c$, where $\alpha_s(m_c)$ is a small number (more on this
later).   When inserted in a nucleus, it will interact with the nucleons by
exchanging gluons with nearby quarks.  Typical momenta for gluons in a nucleus
is set by the QCD scale $\Lambda_{QCD}\simeq 200$ MeV.  For large $m_c$ then,
the wavelength of gluons will be much larger than the size of the charmonium
meson, and so the relevant interaction is the gluon-charmonium analogue of
photon-atom scattering considered above.  The effective Lagrangian is just
given by \eq{rayleighii}, where $\phi$ now destroys charmonium mesons, and
$F_{\mu\nu}$ is replaced by $G_{\mu\nu}^a$, the field strength for gluons of
type $a=1,\ldots,8$.  The coefficients $a_{1,2}$ may be computed from QCD.  To
compute the binding energy of charmonium we need to compute the matrix element
$$\bra{N,\bar c c} \int d^3 x\,\phi^{\dagger}\phi G^a_{\mu\nu}
G^{a\,\mu\nu}\ket{N,\bar c c} $$
(as well as the matrix element of the other operator in \eq{rayleighii}, which we
do not know how to do precisely since the system is strongly interacting.  We
can estimate its size by dimensional analysis though, getting
$$E_B \sim r_0 ^3 \Lambda_{QCD}^4 \simeq \frac{\Lambda_{QCD}^4  }{ (\alpha_s
m_c)^3}\ .$$

\vfill
\eject
\subsection{Problems}

\bigskip
\hrule
\bigskip
\noindent
{\bf I.1)} Redo the analysis of the relativistic scalar field theory for
arbitrary dimension $d$.  What is striking about the result for $d=2$?
 In what dimension is $\phi^3$ a marginal operator?

\bigskip
\noindent
{\bf I.2)} One defines the ``critical dimension'' $d_c$ for an operator to be the
spacetime dimension for which that operator is marginal.  How will
that operator behave in dimensions $d$ when $d>d_c$ or $d<d_c$? In a
theory of interacting relativistic scalars, Dirac fermions, and gauge 
bosons, determine the critical dimension for  the following operators:
\begin{enumerate}
\item A gauge coupling to either a fermion or a boson through the
  covariant derivative in the kinetic term;
\item A Yukawa interaction, $\phi\mybar\psi\psi$;
\item An anomalous magnetic moment coupling $\mybar \psi
  \sigma_{\mu\nu}F^{\mu\nu}\psi$ for a fermion;
\item A four fermion interaction, $(\mybar\psi\psi)^2$.
\end{enumerate}

\bigskip
\noindent
{\bf I.3)} The Lagrangian \eq{nonrel} is written as a field theory (second
quantized form).  Translating into first quantized form, show that the
interaction is equivalent to a $\delta^d(r_1-r_2)$ potential between
two particles in $d$ spatial dimensions, and relate the strength of the
potential to $\lambda$.  (Match the Born
approximation (tree-level) amplitudes). What does the classification of
the interaction as
relevant, marginal or irrelevant correspond to when solving the
Shr\"odinger equation in $d$ dimensions with a $\delta$-function
potential? What is the critical dimension $d$  for a
$\delta^d(r_1-r_2)$ potential? (Hint: you can think of the
$\delta$-function as being a limit of a 
sequence of square well potential that get deeper and narrower in such
a way that the spatial integral of the potential is kept constant. How
would the depth vs. width of a square well have to scale if you wanted to keep the
physics constant, e.g. the energy of a bound state, or the
scattering length?)  

\bigskip
\noindent
{\bf I.4)} Perform a scaling analysis for a relativistic Dirac fermion with
mass $m$ in $d$ dimensions and a $(\mybar \psi \psi)^2$ interaction.
How does the scaling change in the nonrelativistic limit?  Compare
with $\phi^4$ interaction in the bosonic case; is your answer
consistent with the fact that the Schr\"odinger equation for
nonrelativistic particles is the same whether they be fermions or bosons?

\bigskip
\noindent
{\bf I.5)} Derive \eq{fermi} and the relation between $G_F$ and $M_W$
by requiring that the scattering amplitudes calculated
from the full and effective theories (Fig.~\ref{fig:1.2}) match.
 
\eject

\section{ Loops, symmetries, and matching}
\label{sec:2}

\bigskip
\hrule
\bigskip

Up to now we have ignored quantum corrections in our effective
theory. A Lagrangian such as \eq{scalar} is what used to be termed a
``nonrenormalizable'' theory, and to be shunned. The problem was that
the theory needs an infinite number of counterterms to subtract all
infinities, and was thought to be unpredictive.  In contrast, a ``renormalizable''
theory contained only marginal and relevant operators, and needed only
a finite number of counterterms, one per marginal or relevant operator
allowed by the symmetries. A
``super-renormalizable'' theory contained only relevant operators, and
was finite beyond a certain order in perturbation theory.   However Wilson
changed the view of renormalization.  In a perturbative theory,
irrelevant operators are renormalized, but stay irrelevant.  On the
other hand, the coefficients of relevant operators are renormalized to take on values
proportional to powers of the cutoff, unless forbidden by
symmetry. Thus in Wilson's view the relevant operators are the
problem, since giving them small coefficients requires fine tuning.

\subsection{Quantum corrections to scaling}
\label{sec:2a}

We now turn to quantum corrections in an effective theory, such as
\eq{scalar}.  It is evident that when inserted into loops, the
different operators can renormalize each other.  For example, the
operator $c_1\phi^6/\Lambda^2$ can shift the  $\lambda\phi^4$
interaction  at one-loop
by 
\beq
\Delta\lambda\sim  \frac{c_1}{(4\pi)^2}\ .\eeq
  Here I have estimated a factor of $1/(4
\pi)^2$ from the loop, and have noted that the quadratic divergence
from the scalar loop is
cut off at $p=\Lambda$, contributing a factor of $\Lambda^2$ which
cancels the $1/\Lambda^2$ in front of the $\phi^6$ operator. I began
by assuming that all the dimensionless couplings in the theory were
small, so the above correction is perturbative.  That does not mean
that $\Delta\lambda/\lambda \ll 1$, however, because I may have
specified a very small $\lambda$.  What we see here is that the
``natural size'' for $\lambda$ is something at least as big as $
\frac{c_1}{(4\pi)^2}$.  I can of course choose the bare $\lambda$ to
  nearly cancel against $\Delta\lambda$ in order to have a small
  physical $\phi^4$ coupling, but that will involve fine tuning.

Similarly, at one loop the $\phi^4$ interaction can
multiplicatively renormalize the $\phi^6$ interaction, yielding
\beq
\Delta c_1 \sim \frac{c_1 \lambda}{(4 \pi)^2} \ln \Lambda \ .
\eeq
The form of the renormalizations follow simply from dimensional
analysis. Evidently the radiative contributions
to the dimensional couplings are all $\ll 1$, provided I start with
small tree level couplings. However, there are a 
couple of exceptions.

\subsubsection{Relevant operators and naturalness}

The first has to do with the scalar mass term (or relevant operators
in general).  The mass term receives corrections of the form
\beq
\Delta m^2 \sim \left(\frac{\lambda}{(4 \pi)^2 } + \frac{c_1}{(4\pi)^4}
    + \ldots\right) \Lambda^2\ .
\eeq
This correction is very big compared to $m^2$ even though the coupling constants are
small, because of the factor of $\Lambda^2$.  This is called an {\it
  additive} renormalization.  You see, I have cheated
in \eq{scalar}: whereas all interactions aside from the mass term
involved dimensionless couplings times the appropriate power of
$\Lambda$, I wrote the mass term as $m^2 \phi^2$.  I should have
written it as $c_{-2} \Lambda^2 \phi^2$, where $c_{-2}$ is a
dimensionless coupling.  Then I can rewrite the above equation as 
\beq
\Delta c_{-2} \sim   \left(\frac{\lambda}{(4 \pi)^2 } + \frac{c_1}{(4\pi)^4}
    + \ldots\right).
\eeq
This tells us that we cannot have $m\ll \Lambda$ for this scalar
without fine tuning, unless all of the particles interactions are
extremely weak.  Or: turn it around --- the Higgs boson in the
standard model, for example,
has a $\phi^4$ coupling of size $\lambda\simeq M_H^2/(M_W^2/g^2)$
where $g=e/\sin\theta_w\simeq 1$ is the $SU(2)$ gauge coupling.   The
theory will have to be fine-tuned if $\Delta M_H^2 \gg M_H^2$, or
equivalently if $\Lambda\gg 4 \pi M_W/g\simeq 1\TeV$.  Of course, this
doesn't mean that there can be no momenta above $1\TeV$ --- rather,
the cutoff in the effective theory is simple the scale of short-range
physics that has been omitted from the effective theory.  So this
suggests that the LHC should see new physics above a $\TeV$...if you
believe that nature doesn't like a finely tuned theory! 

In contrast to scalar masses, fermion masses are not fine-tuned.  At
first sight, this is surprising, since a fermion mass term is $m
\mybar\psi \psi$, and $[\mybar \psi\psi]=3$. Thus it also looks like a
relevant operator and one might expect $\Delta m \sim
\frac{g^2}{(4\pi)^2} \Lambda$, where $g$ is a gauge or Yukawa
  coupling. However, the kinetic term for the fermions obeys a chiral
  symmetry, under which $\psi\to e^{i\alpha\gamma_5}\psi$, which is
  broken by the fermion mass.  This means that if $m=0$, and there are
  no other interactions which violate chiral symmetry, then there can
  be no radiative corrections to the fermion mass at all.  It follows
   that if $m\ne 0$ is the only chiral symmetry violating operator
  in the theory, then radiative corrections to $m$ must be
  proportional to $m$:  $\Delta m\propto m$.  On dimensional grounds
  then, the dependence of the radiative corrections to the fermion
  mass on the cutoff $\Lambda$ can be at most logarithmic. Similar
  statements can be made about anomalous magnetic moment operators,
  since $\mybar \psi \sigma^{\mu\nu}\psi$ also violates chiral
  symmetry.

 This
  distinction between log and power law dependence of radiative
  corrections on $\Lambda$ is not a small one.  If we were to take
  $\Lambda=M_{pl} = 10^{19}\GeV$ and $m= 100\GeV$, then power law
  divergences in the scalar mass imply a fine tuning of one part in
  $\frac{\lambda^2}{(4\pi)^2} m^2/\Lambda^2 =
    10^{-34}\frac{\lambda^2}{(4\pi)^2}$ which is tiny; in contrast, the log
      corrections to a fermion mass are of size $\frac{\alpha}{4\pi}
      \log 10^{17} = 40 \frac{\alpha}{4\pi}$, which is typically $O(1)$.

Similar arguments can be applied to the most relevant of all
operators:  the vacuum energy density, a constant term in the
Lagrangian. The natural size of the vacuum energy is $\CE_0 =
\Lambda^4(\lambda/(4\pi)^4 + \ldots$.  The actual vacuum energy
(cosmological constant), as
measured in cosmology, is $\CE_0\simeq (10^{-3}\eV)^4$, {\it much}
lower than the fourth power of any sensible cutoff in the standard
model.  Apparently $\CE_0$ is fine tuned  by at least 60 orders of
magnitude ($(1\TeV/10^{-3}\eV)^4$)!  Maybe Nature doesn't mind
fine-tuning so much? (More on that later).

\subsubsection{Logarithmic corrections and running couplings}

Further attention is warranted for the
logarithmic radiative corrections, such as $\ln \Lambda/\mu$.  They are
especially interesting because they depend on the IR scale $\mu$, which
means that if one subtracts a radiative correction at scale $\mu$ from
one at scale $\mu'$, there is a finite difference proportional to $\ln
\mu'/\mu$. This is real physics, involving only the light degrees of
freedom in the effective theory, and which cannot be absorbed into the
redefinition of some local operator (in contrast, power law
divergences can be, and so their values are scheme dependent...in
minimal subtraction schemes,
for example, power law divergences all vanish!). In cases where
$\alpha/4\pi \ln
\mu'/\mu$ is large,
the logs can be resummed using the renormalization group.  Such
corrections are ubiquitous, and occur for basically every operator in
the theory. You can think of these logarithms as changing the scaling
dimension of an operator.  Consider a four fermion operator, such as a
flavor changing interaction in the effective theory of the weak
interactions.  This is a dimension 6 operator, and so will appear
suppressed by $1/\Lambda^2$, 
\beq
\frac{c}{\Lambda^2} (\mybar \psi \psi)^2\ .
\eeq
Now suppose this operator received a logarithmic renormalization of
the form  $$\Delta c =- c \frac{\alpha}{4\pi} \ln \mu/\Lambda\ ,$$ where
$\alpha$ is some other coupling in the theory.  For
example, a correction like this  could arise from a one-loop graph where the
four fermion interaction is dressed with a gluon running  between two
 fermion propagators.  In the effective action then one would have the
 corrected coefficient
\beq
\frac{c(1- \frac{\alpha}{4\pi}\ln \mu/\Lambda)}{\Lambda^2} (\mybar \psi \psi)^2\ .
\eeq
Since I am assuming that  perturbation theory in $\alpha$ is valid, to
$O(\alpha)$ we can rewrite
\beq
\frac{c\,(1- \frac{\alpha}{4\pi}\ln \mu/\Lambda)}{\Lambda^2} =
\frac{c\,e^{ -\frac{\alpha}{4\pi}\ln \mu/\Lambda}}{\Lambda^2}=
c\,\frac{1}{\Lambda^2}\left(\frac{\mu}{\Lambda}\right)^{-\alpha/4\pi}\ .
\eeq
To one-loop order then, we see (by counting powers of $\Lambda$) that
this four fermion operator scales 
not with scaling dimension 6, but rather
$(6-\frac{\alpha}{4\pi})$, making it more important in the IR than
naively expected.  Note that this analysis assumes the
validity of perturbation theory in $\alpha$.

Quantum corrections to the scaling dimension of irrelevant
operators, such as sketched above,   can be important to include when
predicting the rate of rare processes. An example would be in the
study of  flavor changing
hadronic interactions such as $b\to s\gamma$, where one would like to
uncover new weak-scale physics (such as supersymmetry)  hiding behind the standard model.  

It is when considering  marginal operators that quantum corrections to
the scaling dimension are particularly dramatic, as they make the operator either become relevant
(strong in the IR, as in the case of the QCD coupling, or that of any
nonabelian gauge theory with not too many matter fields) or irrelevant
(as is the case with QED, $\phi^4$ and Yukawa interactions). Hopefully
you have encountered the renormalization group and $\beta$ functions
in a quantum field theory class;  I will not review them here.
However, I would like to demystify asymptotic freedom a bit by showing
how it occurs in a mundane quantum mechanics problem, without the
added complications inherent in a relativistic field theory.

Consider a particle in $d$ spatial dimensions in a $\delta^d(r)$ potential
with no angular momentum (for $d>1$).  In a second quantized language,
a $\delta$-function potential appears as a $(\lambda/8m^2)\left(\psi^\dagger
  \psi\right)^2$ interaction as in \eq{nonrel}. Generalizing yesterday's scaling
arguments for arbitrary $d$ one finds  the scaling dimension of
the interaction $[\left(\psi^\dagger
  \psi\right)^2]= (d-2)$.  This means that the critical dimension is
$d=2$, and that the interaction is relevant in $d=1$ and irrelevant in
$d=3$.

Let's see that directly from the Schrodinger equation:
\beq
-\nabla^2 \Psi -g \delta^d(\bfr) \Psi =2M E \Psi\ .
\eqn{sch}
\eeq
On rescaling $\bfr$, we know that $\nabla^2$ scales like $1/r^2$,
while $\delta^d(\bfr)$ scales like $1/r^d$, since $\int
d^d{\bfr}\,\delta^d(\bfr)=1$. To better understand the meaning of
scaling a $\delta$-function, we  replace it by  a
square barrier of height $V_0$ and radius $r_0$, satisfying $V_0
S_d(r_0)=1$, where $S_d(r_0)$ is the volume of a sphere of radius
$r_0$ in $d$ dimensions.  Since $S_d(r_0)\propto r_0^d$, so
$V_0\propto 1/r_0^d$.  For an attractive $\delta$ function potential,
one has a square well, instead of a square barrier.  The finite size
of $r_0$ regulates the singular
$\delta$-function, and $1/r_0$ plays the role that our momentum
cutoff $\Lambda$ played in prior discussion.   Now take $r_0\to 0$,
varying $V_0$ appropriately, and compare the relative
importance of the kinetic versus the potential potential energy in the
Schr\"odinger equation. Clearly $d=2$ is special, for only in $d=2$ do
the kinetic and potential terms in the Schr\"odinger equation scale the
same way.

In $d=1$, the kinetic term dominates for short wavelength, and so high
energy scattering will resemble a free theory (no potential); on the
other hand, long distance physics is dominated by the potential term,
and for an attractive interaction we find a bound state $\Psi = exp (-g|x|/2)$ with binding energy
$g^2/8M$ in the limit $r_0\to 0$. 

In $d=3$, with the scaling law we chose for $V_0$ we get nonsense if
the interaction is attractive!
The
potential well has depth $V_0\propto 1/r_0^3$ and so potential energy  dominates at
short distance, and we get an infinite number of bound
states. Quite appropriate for an irrelevant operator: it is sensitive
to {\underline{short distance}} physics. In the second quantized form, we can see
that the energy is not bounded from below. This problem is not
encountered with a repulsive $\delta^3(\bfr)$ potential.

Since we will later be talking about an effective theory for
nucleon-nucleon scattering in $d=3$, it is useful to point out that the $d=3$ theory
with an attractive interaction can be forced to make sense by taking 
liberty with what we mean by a $\delta$ function interaction, and by  changing
our scaling law for $V_0$.  In particular, if we require that be a
bound state at some fixed energy as we take $r_0\to 0$, we find that
we must take $V_0$ to scale as $1/r_0^2$, and not $1/r_0^3$.  This
means we are replacing the $g \delta^3(\bfr)$ potential by a regulated
potential $g' r_0 \delta^3_{r_0}(\bfr)$.  This is equivalent to saying
that our ``bare'' coupling $g' r_0$ goes to zero as we remove the
regulator (that is, take $r_0\to 0$) for fixed renormalized coupling
$g'$, whose value is determined by the renormalization condition that
we find a bound state with a specified energy.

The $d=2$ case is particularly interesting we have seen that $d=2$ is the critical
dimension for a $\delta$-function potential.  If we replace the
$g\delta^{(2)}(\bfr)$ in \eq{sch} by a 2d square well of depth $V_0= g/(\pi r_0^2)$
and radius $r_0$ and solve the Schr\"odinger equation for a bound state with fixed
binding energy  $E$ independent of $r_0$ (our renormalization condition) we find a
solution in terms of Bessel functions which is nonsingular at the origin,  bounded at $r=\infty$
and continuous at $r=r_0$ of the form 
\beq
\Psi_< = J_0(pr)\ ,\quad \Psi_> =\frac{J_0(pr_0)K_0(qr)}{K_0(q r_0)}\
,\qquad p\equiv\sqrt{-2ME}\ ,\qquad q\equiv\sqrt{g/(\pi r_0^2)-p^2}\ ,
\eeq
 subject to the condition that the first derivative is continuous at $r=r_0$:
\beq
\Psi'_<(r_0) = \Psi'_>(r_0)\ .
\eqn{cond}
\eeq
Assuming that the coupling $g$ is small, we can expand and solve the
above continuity 
equation to linear order in $g$, and then expand that solution for
small $r_0$.  Doing this, I find
\beq
g = \frac{2\pi}{\half - \gamma - \ln(p r_0/2)} + O(r_0) = \frac{2\pi}{\ln(1/r_0\Lambda)}+ O(r_0)\ ,\qquad \Lambda\equiv
  \frac{p}{ 2e^{1/2-\gamma}} \simeq 0.54\, p
\eqn{geq}
\eeq
where $\gamma=0.5772\ldots$ is the Euler $\gamma$-function.  Evidently,
our bare $g$ has to vanish like an inverse logarithm of $r_0$
as $r_0\to 0$. It also appears to blow up in the IR at
$r_0=1/\Lambda$, but this behavior is not to be trusted since I
employed perturbation theory to derive the result. This  behavior closely 
resembles that of the QCD coupling constant, when
one fixes the mass of some hadron, such as a glueball. The running QCD
coupling vanishes like an inverse logarithm at short distance
(asymptotic freedom), with $\Lambda_{QCD}$ playing the role of
$\Lambda$ in the above equation. 

 Note that if I replace $r_0$ by
$1/\mu$ in \eq{geq}, then $g(\mu)$ obeys the equation
\beq
\mu \frac{d g}{d\mu} = -\frac{g^2}{2\pi}
\eqn{rg2d}\eeq
which is the renormalization group equation for the running coupling
constant in this theory.  The fact that the right hand side of the
above equation is negative indicates asymptotic freedom.  Apparently
if we had analyzed a repulsive interaction ($g\to -g$), fixing
something physical, such as the scattering length,  then the RG
equation would have given an asymptotically {\it unfree} theory...the
coupling would get stronger in the UV, and one would not be able to
take the $r_0\to 0$ limit, as one would encounter a
``Landau pole'' as in QED ---  the coupling $g$ would become
infinite at finite $r_0$.

Hopefully these examples will convince you that the regularization,
renormalization and running couplings encountered in quantum field
theory have to do with the singular nature of local interactions, and
have nothing to do with relativity, or the fact that relativistic
quantum theories are many-body theories.

\subsection{Integrating out massive fields and matching}
\label{sec:2b}

\begin{figure}[t]
\centerline{\epsfysize=1.5in\epsfbox{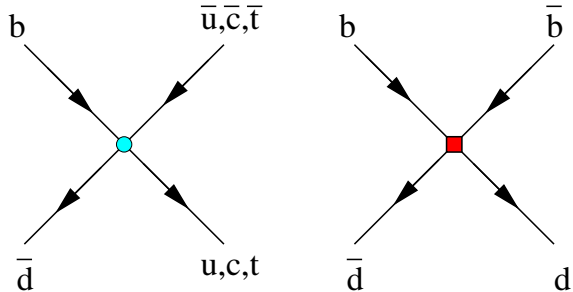}}
\caption{{\it $\Delta B=1$ and $\Delta B=2$ vertices in the effective theory.}}
\label{fig:bgraphs}
\vskip .2in
\end{figure}

One important application for  effective field theories  is to use
them to  sum
up the logarithmic corrections to irrelevant operators of
phenomenological interest.  As an
example consider $\Delta B=2$ processes in the standard model (where
$B$ is $b$-quark number),
which occur at one loop and which
contribute to $B-\mybar B$ meson mixing. The effective theory contains
contact interactions for both $\Delta B=1$ and $\Delta B=2$ processes (Fig.~\ref{fig:bgraphs});
both are dimension 6 operators, and will be multiplied by a
dimensionless number to be determined, divided by the cutoff of the
effective theory, which is $M_W$.  To determine the coefficients is
called ``matching'', and one does it in perturbation theory (a loop
expansion).   

\begin{figure}[t]
\centerline{\epsfysize=2.5in\epsfbox{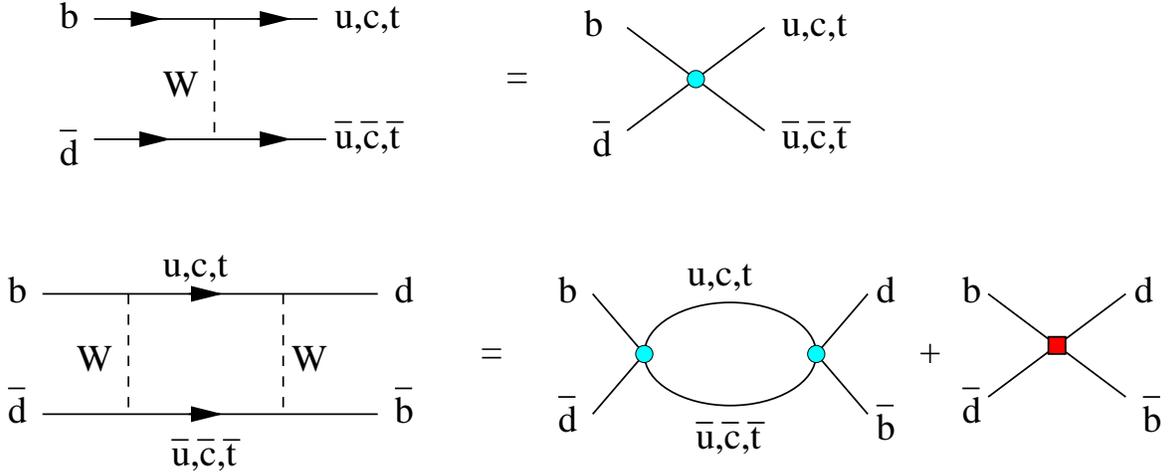}}
\caption{{\it At tree level one matches the $\Delta B=1$ diagram from the
  standard model onto the $\Delta B=1$ contact interaction of the
  effective theory.  At one loop level one determines the $\Delta B=2$
  contact term of the effective theory by matching the $\Delta B=2$
  box diagram from the standard model onto the one-loop graph
  involving the two $\Delta B=1$ vertices in the effective theory,
  plus the $\Delta B=2$ contact term. The calculation requires
  specification of renormalization scale, conveniently taken to be $\mu=M_W$.}}
\label{fig:bbox}
\vskip .2in
\end{figure}

First one calculates the tree diagram for $\Delta B=1$
processes in the standard model, and matches onto the $\Delta B=1$
contact interaction in the effective theory (Fig.~\ref{fig:bbox}).  In
doing so one only keeps the leading part of the tree diagram in a
$p^2/M_W^2$ expansion. This is identical to how
one would determine the relation between $M_W$ and $G_F$ in the Fermi
theory. 

Next one does 1-loop matching.  It turns out we do not have to
consider 1-loop contributions to the $\Delta B=1$ operator for a
leading order calculation of $\Delta B=2$ processes. However, we do
need to compute the coefficient of the $\Delta B=2$ contact
interaction at this order.  In the standard model, we have the box
diagram in Fig.~\ref{fig:bbox}, which must be equated to the sum of
the loop diagram in the effective theory with two tree-level matched
$\Delta B=1$ vertices, plus the $\Delta B=2$ contact interaction,
thereby fixing the coefficient of the latter. One can do the matching
with zero momentum flowing through the external propagators to leading
order in the $p^2/M_W^2$ expansion. The diagram in the effective
theory is divergent, and a renormalization scale $\mu$ must be chosen;
it is convenient to choose $\mu=M_W$, the scale at which one is
performing the matching so that $\ln (\mu/M_W)$ factors vanish..

Note that in the 1-loop  matching, the loop diagram in the effective theory
has the same light internal degrees of freedom as the full box
diagram, and so will exactly reproduce the infrared physics.  on the
other hand, we have butchered the UV physics by replacing the $W$
propagator with a point-like vertex, and the $\Delta B=2$ contact
interaction is adjusted to fix that up.

Now that the $\Delta B=2$ vertex is determined to one-loop accuracy,
one can consider a gluon line dressing it.  This is formally at the
2-loop level.  However, if one wishes to take a matrix element of the
$\Delta B=2$ operator in the $B$ meson state, gluon loops renormalized
at $\mu=M_W$ will
contribute factors of $\alpha_s(M_W) \ln M_W/M_B$, where the log is pretty
big.  Therefore the appropriate thing to do is to compute the
anomalous dimension of the $\Delta B=2$ operator due to one gluon
exchange between quark legs, and run the coupling down to $\mu=M_B$.
This procedure sums up powers of  $\alpha_s(M_W) \ln M_W/M_B$ and
leaves one with an expansion in $\alpha_s(M_B)$.

This whole procedure is possible because one has separated cleanly the
short distance physics ($W$ exchange) from the long distance physics
(gluon exchange) by means of the effective theory.  This procedure was
first worked out by Gilman and Wise for the $\Delta S=1$ Hamiltonian
\cite{Gilman:1979bc}. A detailed analysis of the similar
$\Delta S=2$ effective theory is found in ref. \cite{Manohar:1996cq}.

\subsection{Power counting in a nonperturbative theory}
\label{sec:2c}

We have seen that the logarithmic renormalizations are interesting in
a perturbative theory, as they determine whether naively marginal
operators become relevant or irrelevant when quantum corrections are
included.  However, in a strongly coupled, nonperturbative theory,
there may in fact be no relation between the scaling dimensions of
operators which were derived in the classical action, and the true
scaling in the quantum theory.  In general, when the couplings of the
effective theory are $O(1)$ times the appropriate power of $\Lambda$,
there will be a first order phase transition in the theory, and 
modes will will become heavy with masses $O(\Lambda)$, and the theory
is of no use for describing physics below the cutoff.  However, in
some cases the theory will exhibit a second order phase transition,
with the concomitant diverging correlation lengths, which can be
interpreted in terms of particles whose masses become vanishingly
small compared to the cutoff $\Lambda$ as the couplings of the theory
are tuned to the critical values.  In this case the physical light degrees of
freedom might be a composite of the original degrees of freedom, and
the scaling dimension will be radically different than the naive
prediction.  For example, if a weakly coupled scalar emerges which is a bound state of
fermion $\psi$ and antifermion $\mybar \psi$, then $[\mybar
\psi\psi]\simeq 1$, even the the perturbative scaling dimension is
$[\mybar\psi\psi]=3$ (in $d=4$ spacetime dimensions).  In general, an
EFT in terms of the constituent variables in a strongly coupled theory
is not useful.  Usually the best one can do is express the EFT in
terms of the composite degrees of freedom, and write down the most
general set of operators consistent with the symmetries of the
underlying theory. Such an EFT can still be predictive, as I will
discuss in lecture 3.

\subsection{Symmetry}
\label{sec:2d}

Symmetry is an invaluable tool in effective field theory.  We saw, for
example, that chiral symmetry makes a fermion mass term in four
dimensions behave like a
marginal operator, rather than a relevant one.  Symmetries also have
implications for what operators are allowed in the effective theory
and what the light degrees of freedom are.  There are several basic
ways symmetries play a role, and so I will offer here a brief survey.

\subsubsection{Gauge symmetries}

Gauge symmetries are not really symmetries, they are constraints that
allow one to describe forces with a redundancy of variables, allowing
one to maintain  manifest relativistic covariance. For example, gauge
symmetry allows us to describe the two polarization states of a photon
in terms of the four gauge potential functions $A_\mu$. In general, gauge
symmetries cannot be broken, and when
constructing an effective theory to provide a low energy description
of a more general theory, any gauge symmetry of the more complete
theory will be inherited by the low energy theory.  However, this can
be done deviously.  For example, compare QCD  and the effective theory for pion physics (the
chiral Lagrangian). Both manifestly respect electromagnetic gauge
invariance.  Both also respect $SU(3)_c$ gauge invariance, but in a
more trivial way --- the theory of pions only consists of color
neutral objects, so saying that the theory is gauge invariant is
rather trivial.

 Less obvious might be what happens to gauge invariance
when passing from the $SU(2)\times U(1)$ gauge theory of electroweak
interactions, to Fermi's effective theory.  The kinetic terms for
quarks  in the former involve gauge covariant derivatives, but not in
Fermi's theory.  However, to derive Fermi's theory from  $SU(2)\times
U(1)$ one actually has to ``gauge fix'' first...this doesn't change
the physics, but makes you lose manifest gauge invariance.  If you
didn't gauge fix, there would be all sorts of spurious degrees of
freedom in the effective theory that would unnecessarily complicate
the description of low energy physics.  We put up with them in the
$SU(2)\times U(1)$ theory in order to maintain relativistic
covariance, but we do not need them in the low energy theory since
there are no long range weak forces, and  we do without.


\subsubsection{Exact and approximate global symmetries}

The existence of global symmetries (as opposed to gauged) means that
there are charges
 which commute with the Hamiltonian.  If the short distance theory
 possesses an exact symmetry, then the low energy EFT must as well,
 although it might be trivially realized in that all the light degrees
 of freedom are neutral under that symmetry (for example, at energy
 scales below the pion mass, an EFT for the standard model would
 consist of the electron, muon three flavors of neutrino, and the
 photon; baryon number is an exact symmetry in this theory, but none
 of the degrees of freedom carry baryon number.)

It often happens that a symmetry is not exact at short distance.  If
the breaking is in some sense small, then one can perturb in the
symmetry breaking parameter in the low energy EFT, and still exploit
the symmetry to constrain the allowed operators at any order in the expansion. 
There are several general  cases to consider:

\begin{enumerate}[i.]
\item
The symmetry could be  softly broken: that is to say, it is broken by a
relevant operator. 
Chiral symmetry in QCD is such an example, which is a symmetry of the fermion
kinetic terms but not of the fermion masses.  Such breaking
becomes more important as one scales to low energies.  In chiral
perturbation theory, for example, which is formulated at a scale
$\Lambda\sim 1\GeV$, there is no sense in which chiral symmetry is
approximate for the $t$, $b$ or $c$ quarks, as their masses are
larger than $\Lambda$; on the other hand, chiral symmetry is still
quite good for the $u$ and $d$ quarks, whose masses are several
$\MeV$.  For the strange quark with mass $\sim 100\MeV$, chiral
symmetry is reasonably good. We will see in tomorrow's lecture how
these masses are incorporated into the chiral Lagrangian.

\item
The symmetry could suffer small ``hard'' symmetry breaking due to a
marginal operator.  An example is the breaking of isospin in QCD by
electromagnetism. The gauge coupling to quarks is dimension 4, but
because $\alpha/4\pi$ is small, electromagnetism does not break
isospin symmetry by much.  For example, 
\beq
\frac{m_{\pi^+}^2 -m_{\pi^0}^2}{m_\rho^2}\simeq 0.002\sim
\alpha/3\ .
\eeq
(You will see later why the above ratio is a sensible one to take, rather
than dividing by a pion mass, for example).
If one is considering a very large span of length scales (for example,
from the GUT scale at $10^{16}\GeV$ down to $\Lambda_{QCD}$) one might
need to run these hard symmetry breaking terms by means of the
renormalization group, to resum the logarithmic we discussed above.

Note that symmetry breaking like this is only possible if there is no
way for the marginal symmetry breaking operators to generate relevant
symmetry breaking terms.  In the above case of case of isospin
breaking, QED radiative corrections {\it can} renormalize quark
masses, but as argued above, chiral symmetry ensures that quark
masses behave like marginal operators, and so electromagnetism and the
$u$-$d$ quark mass difference will play similar roles in isospin
breaking in the chiral Lagrangian.  However, consider a scalar field
with a shift symmetry $\phi\to \phi+f$, which would forbid a scalar
mass term.  Now suppose the symmetry is broken by a small $\lambda
\phi^4$ interaction.  This interaction will generate a mass term for
the scalar at one loop, of size $\lambda/(4\pi)^2 \Lambda^2 \phi^2$.
The moral is that if you break a symmetry with some operator, all
possible symmetry breaking operators will be generically be
radiatively generated, 
multiplied by coupling constants and the appropriate power of the
cutoff $\Lambda$. Exceptions to this rule occur in supersymmetric theories.

\item
Explicit symmetry breaking could occur in irrelevant operators.  This
may seem peculiar, but it is a common occurrence.  In this case the symmetry is better at low energies
than at high energies.  In fact, one can discover symmetries in the
low energy EFT that do not exist at all at high energy!  These are
called accidental symmetries, and I will say more about them below.
 
\end{enumerate} 

\subsubsection{Spontaneously broken symmetries}

Global symmetries can be spontaneously broken, which means that the ground
state $\ket{\Omega}$ of the theory is not invariant under the symmetry
transformation.  If $Q$ is the symmetry charge for a continuous
symmetry (such as $U(1)$, as opposed to a discrete symmetry such as
$CP$) then spontaneous symmetry breaking implies $Q\ket{\Lambda}\ne
0$, so that a symmetry rotation of the ground state produces a new
state:
\beq
e^{i\alpha Q}\ket{\Lambda} = \ket{\Lambda'}\ .
\eeq
The charge $Q$ still commutes with the Hamiltonian, $\[Q,H\]=0$,
assuming it was an exact symmetry to begin with.  Therefore the state 
$\ket{\Lambda'}$ is degenerate with the old one $\ket{\Lambda}$, and
is also a perfectly fine ground state of $H$. Thus spontaneous
symmetry breaking implies an infinitely degenerate manifold of ground
states.

The universe as a whole will choose one of those ground states at
random.  However, there will be low energy excitations where the
groundstate smoothly interpolates between $\ket{\Lambda}$ in one
region of space and $\ket{\Lambda'}$ in another; these excitations
must have vanishingly small energy at long wavelength as the
groundstates $\ket{\Lambda}$ and $\ket{\Lambda'}$ are degenerate.
Such excitations are called Goldstone bosons. It is important to
realize that since $\[Q,H\]=0$ even if there is spontaneous symmetry
breaking,  the symmetry is still exact in the sense that energy
eigenstates come in irreducible multiplets of the symmetry.  The
symmetry is just
realized in a funny way (called a ``nonlinear realization''), under which the Goldstone bosons shift, e.g,
$\phi\to\phi+\alpha$.   Symmetry multiplets are filled out with
many-particle states now, involving massless Goldstone bosons.  

If the symmetry is both explicitly  and spontaneously broken, then
$\[Q,H\]\ne 0$ and 
there is typically a unique ground state $\ket{\Lambda}$, but a
manifold of states $\ket{\Lambda'}$ which are
nearly degenerate with it.  This means that there are  massive
excitations connecting the states;  if the explicit  symmetry breaking
is small, then the excitations can be light and are called pseudo
Goldstone bosons (or PGBs).  Pions are  PGBs of spontaneously broken
chiral symmetry;  more  on
this tomorrow.

 Goldstone bosons and PGBs often play a role
in effective field theory for the obvious reason that they are
naturally light degrees of freedom.

\subsubsection{The standard model as an EFT: love those accidental symmetries}

It is profitable to consider the 
standard model to be an EFT, and ask what operators can be added to
it, consistent with  the $SU(3)\times SU(2)\times U(1)$ gauge
symmetry. Interestingly enough, there is only  one marginal operator
that could be added:  a $\theta
\widetilde G_{\mu\nu}G^{\mu\nu}$ term for the gluons.  This operator
violates $T$ and $CP$, and we know from an experimental upper bound on
the neutron electric dipole moment
 that $\theta\lesssim 10^{-9}$ (where
$\theta$ is defined in a basis where quark masses are real and
positive). The phenomenological absence of this operator in the
standard model (which is infinitely renormalized and is therefore
dependent on short distance physics) is
called the ``strong CP problem''.

  At dimension 5, a new
contribution exists, of the form
\beq
\frac{\lambda_{ij}}{\Lambda} (\ell_i H^\dagger)(\ell_j H^\dagger) +
\text{h.c.}\ ,
\eeq
where $\ell$ is the left-handed lepton doublet from family $i$, and $H$ is the Higgs doublet. When the Higgs field
is given a vev, this term gives a Majorana (lepton number violating)
mass of size $m_\nu\sim (250\GeV)^2/\Lambda$.  Light neutrino masses are
observed in nature, and the existence of this dimension 5 operator
can explain why they are so small:  $\Lambda$ is big.  For example,
with $\Lambda=10^{14}\GeV$, $m_\nu \sim 0.6\eV$.

At dimension 6, one encounters a host of four fermion operators. The
most interesting classes are those which lead to flavor changing
neutral currents (FCNC), such as those contributing to $b\to s\gamma$,
$K^0\to \mu^+\mu^-$, and $\Delta S=2$ operators; those which give rise
to $CP$-violating electric dipole moments for quarks and leptons (to
date unobserved); and those violating baryon and lepton number. The lack of
evidence for any of these processes can be nicely explained in the
standard model:  the scale $\Lambda$ where new physics kicks in must be
very high. This is a hugely appealing feature of the standard model--- even if baryon number is completely
violated at the scale $\Lambda$, it becomes an accidental symmetry in
the standard model simply because gauge invariance and the particle
content restricts the theory so much that there are no relevant or
marginal gauge invariant
baryon number violating operators.  Similarly for lepton number.  

However, the standard model has  problems in the form of two relevant
operators: the cosmological constant (observed to correspond to a
vacuum energy density $\sim (10^{-3}\eV)^4$, and  the Higgs mass
$m_H$,  which must be in the $100\GeV$ range for the 
model to work.  As discussed earlier, the natural size for $m_H^2$ will
be $\lambda/(4\pi)^2 \Lambda^2$, and so raising $\Lambda$ to explain
the absence of rare processes exacerbates the fine tuning of the Higgs mass
(a required precise cancellation  between long
distance and short distance contributions to the Higgs mass). New physics at the
scale $\Lambda=1\TeV$ (such as supersymmetry or technicolor), makes it
possible to introduce symmetries which forbid the introduction of a
Higgs mass term, solving the fine-tuning problem. However, with 
$\Lambda$ so low, it is no longer automatic that baryon number and
lepton number should be well conserved, nor that FCNC should be
absent. Solving the hierarchy problem has introduced a host of new
problems. This is why there are high hopes for LHC to turn up
something interesting, or even revolutionary.  As for the cosmological
constant problem, no one has yet discovered a model where it is even
remotely close to being
naturally small.

The sociological pendulum is swinging.   Since 't Hooft's emphasis on
naturalness in the late 1970's physicists became obsessed with solving
the naturalness problem of the Higgs, and then by destroying accidental symmetries,
had to confront  model building
conundrums which aren't present  in the standard
model. Now people are questioning whether fine tuning is bad after all,
or whether only in an apparently  fine tuned universe would galaxies and a world
like ours be possible.  For example, if the cosmological constant were
much bigger than it is, the universe would fly apart and galaxies
would never form.  So instead of asking:  ``is it unlikely that the
cosmological constant should be small?'' (whose answer is presumably
``Yes'') one asks: `` is it unlikely that the cosmological should be
small in our universe, given that it has galaxies?'', to which the
answer is ``No'').  This goes under the rubric ``anthropic
principle'' although the proponents are trying making it sound more
respectable with the title ``galactic principle''.  It is not as
stupid as it sounds on first acquaintance, but I must confess that I
find it a supremely uninteresting development in physics.  I encourage you to read some of the
literature on the subject if you are interested, such as ref. \cite{Susskind:2003kw}.

\subsubsection{Quantum field theory on the lattice: more accidental symmetries}

Accidental symmetries also play an important role in lattice field
theory, where one formulates a field theory on a lattice spacetime,
and computes properties numerically, approaching the continuum limit
by sending the lattice spacing to zero. Why should lattice QCD formulated on
a 
hypercubic lattice yield  Lorentz invariant theory in this limit?
The reason lattice field theory works is because of  accidental symmetry:
Operators on the lattice are constrained by gauge invariance and the
hypercubic symmetry of the lattice.  While it is possible to write
down operators which are invariant under these symmetries, while
simultaneously violating Lorentz symmetry, such operators have high dimension and
are not relevant.   For example, if $A_\mu$ is
a vector field defined on the lattice, then the 
marginal operator  $A_1A_2A_3A_4$ is 
hypercubic invariant, and violates
Lorentz symmetry;  however,
the only vector field in lattice QCD is the gauge potential, and such
an operator is forbidden because it is not gauge invariant. Since
irrelevant operators vanish in the continuum limit where the momentum
cutoff offered by the lattice goes to infinity, Lorentz symmetry is
achieved, even though it was never a symmetry of the lattice action.

The same principle has been recently exploited in constructing
supersymmetric gauge theories on the lattice
\cite{Kaplan:2002wv,Cohen:2003xe,Cohen:2003qw,Kaplan:2005ta,Catterall:2004zy,Catterall:2005fd}.
Supersymmetry is related to 
Lorentz symmetry, and one cannot have the full supersymmetry be exact
on the lattice. The idea is to maintain 
enough exact supersymmetry, however, to ensure that operators that do
not respect the full continuum supersymmetry are irrelevant.  

Another example of how effective field theory ideas play a big role in
the lattice lattice field theory is afforded by
domain wall fermions.  For several decades following Wilson's
development of lattice gauge theory it seemed impossible to
formulate lattice fermions invariant under a chiral symmetry, and this
inability 
made it very difficult to compute realistic QCD processes.
However, it is possible  to show that on a five dimensional lattice one can
formulate a theory of massive fermions without any chiral symmetry
which has 
massless fermion modes bound to the four-dimensional edge of the lattice \cite{Kaplan:1992bt}.
It is immediately evident that at long wavelength, with the heavy
5-dimensional modes integrated out, one is left with a four
dimensional lattice theory of massless fermions with an accidental chiral
symmetry.  A little less obvious but easy to compute is the fact that a subtle residual
chiral symmetry breaking effect remains after the heavy modes are
integrated out, which accounts for what is known as the chiral
anomaly.  The kinetic term of the effective four dimensional theory
is know as the ``overlap operator'' \cite{Neuberger:1997bg}. Because
of their chiral properties, domain wall fermions (the five dimensional
formulation) or overlap fermions (the effective four dimensional
description) are presently the subject of much computational effort in the
lattice community and will probably become the standard lattice
fermions for simulating QCD as computers become faster.

\vfill
\eject
\subsection{Problems}

\bigskip
\hrule
\bigskip
\noindent
{\bf II.1)} Consider a repulsive $\delta$ function potential in
$d=2$ and $d=3$ spatial dimensions.  Regulate the $\delta$ function as a spherical
barrier of height $V_0$ and radius $r_0$ inside a box of fixed radius
$R_0$, and look at the lowest energy eigenvalue as $r_0/R_0\to 0$,
scaling $V_0$ appropriately so that $\int d^dr\, \delta_{r_0}(\vec r)
= 1$.
Compare your result with that of the free theory, $V_0=0$. Make sense
of your answer given our analysis that $d=2$ is the critical dimension
for the $\delta$-function potential for nonrelativistic particles.

\bigskip
\noindent
{\bf II.2)}
Derive the RG equation \eq{rg2d} by more conventional techniques by
considering the one loop diagram for two particle scattering using
dimensional regularization and an MS (minimal subtraction)  scheme.  You may
find some of the formalism presented in the fourth lecture to be
useful here, as well as the appendix B.

\bigskip
\noindent
{\bf II.3)} Write down a dimension 6 operator in the standard model
which violates baryon number, and estimate the lifetime of a proton in
terms of $\Lambda$.  If the lifetime of a proton is $> 10^{34}$ years,
roughly how
big does $\Lambda$ have to be?

\bigskip
\noindent
{\bf II.4)} Show that any dimension 6 operator in the standard model which violates baryon
number preserves $(B-L)$, where $B$ is baryon number and $L$ is lepton
number.  A process that violates $B$ but not $L$ could contribute to
oscillations between neutrons and anti-neutrons, analogous to Majorana
neutrino oscillations.  What is the lowest
dimension operator 
in the standard model which violates $B-L$ symmetry and can contribute
to $n-\mybar n$ 
oscillations? Roughly what is the oscillation rate as a function
of $\Lambda$? 


\vfill\eject
\section{ Chiral perturbation theory} 
\label{sec:3}

\bigskip
\hrule
\bigskip

\subsection{Chiral symmetry in QCD}
\label{sec:3a}

QCD is the accepted theory of the strong interactions.  At large
momentum transfer, as in deep inelastic scattering processes and the
decays of heavy particles such as the $Z$, the theory is perturbative
due to asymptotic freedom. The flip side is that in the infrared, the
theory becomes nonperturbative.  This is good in the sense that we know
that the light hadrons don't look at all like a collection of quarks weakly
interacting via gluon exchange.  But it does mean that QCD is not of
much help in quantitatively understanding hadron phenomenology without
resorting to lattice QCD and a computer.
However, there does exist an effective field theory which is very
powerful for analytically treating  the interactions of the lightest hadrons, the
pseudoscalar octet, consisting of the $\pi$, $K$, $\mybar K$ and $\eta$. 

The reason that the pseudoscalar octet mesons are lighter is because
they are the
pseudo-Goldstone bosons (PGBs) that arise from the spontaneous
breaking of an approximate symmetry in QCD.

Consider the QCD Lagrangian, keeping only the three lightest quarks, $u$,
$d$ and $s$:
\beq
\CL =\sum_{i=1}^3\left( \mybar q_i i \Dsl q_i - m_i \mybar q_i q_i\right)
-\frac{1}{2}\Tr G_{\mu\nu}G^{\mu\nu}\ ,
\eeq
where $D_\mu = \partial_\mu+ig A_\mu$ is the covariant derivative,
$A_\mu = A_\mu^a T_a$ are the eight gluon fields with $T_a$ being
$SU(3)$ generators in the $3$ representation, and $G_{\mu\nu}$ being
the gluon field strength.  Note that if I write the kinetic term in
terms of right-handed and left-handed quarks, projected out by
$(1\pm\gamma_5)/2$ respectively, then the kinetic term may be written
as
\beq
\sum_i \mybar q_i i \Dsl q_i = \sum_i\left(\mybar q_{Li} i \Dsl q_{Li}
  + \mybar q_{Ri} i \Dsl q_{Ri}\right)\ .
\eqn{qcdkin}\eeq
This term by itself  evidently respects a $U(3)_L\times U(3)_R$
symmetry, where I rotate the three flavors of left-handed and
right-handed quarks by independent unitary matrices.  One combination
of these transformations, the $U(1)_A$ transformation where $q_i\to
e^{i\alpha\gamma_5} q_i$,  is in fact not a symmetry of the quantum
theory, due to anomalies; it is a symmetry of the action but not of
the measure of the path integral.  This leaves us with a $U(1)_V \times
SU(3)_L\times SU(3)_R$ symmetry.  The $U(1)_V$ is just baryon number,
under which both left- and right-handed quarks of all flavors pick up
a common phase.  The remaining $
SU(3)_L\times SU(3)_R$ symmetry, under which $q_{Li}\to
L_{ij}q_{Lj}$ and $q_{Rj}\to
R_{ij}q_{Rj}$, where $R$ and $L$ are independent $SU(3)$ matrices, is
called ``chiral symmetry''.  

$SU(3)_L\times SU(3)_R$ is
not an exact symmetry of QCD, however.  The quark mass terms may be written
as
\beq
 \sum_i m_i \mybar q_i q_i = \sum_{i,j} \mybar q_{Ri} M_{ij} q_{Lj} + h.c.\ ,\qquad
 M = \begin{pmatrix} m_u &&\cr &m_d&\cr &&m_s\cr\end{pmatrix},
\eqn{qmass}\eeq
where the quark masses $m_i$ are called ``current masses'', not to be
confused with the much bigger constituent quark masses in the quark model.
Since the mass term  couples left- and right-handed quarks, it is not
invariant under the full chiral symmetry.
Several observations:
\begin{itemize}
\item 
Note that {\it if} the mass matrix $M$ were a dynamical field,
transforming under $SU(3)_L\times SU(3)_R$ as 
\beq M\to R M L^\dagger\
,\eqn{masstrans}
\eeq
then the Lagrangian {\it would} be chirally invariant. Thinking of the
explicit breaking of chiral symmetry as being due to spontaneous
breaking due to a field $M$ which transforms as above makes it simple
to understand how $M$ must appear in the effective theory, which will
have to be chirally invariant given the above transformation.  This is
called treating $M$ as a ``spurion''.
\item 
The symmetry is broken to the extent that $M\ne RML^\dagger$. Since
$m_u$ and $m_d$ are much smaller than $m_s$, 
  $SU(2)_L\times SU(2)_R$ is not broken as badly as $SU(3)_L\times SU(3)_R$;
\item If all three quark masses were equal but nonzero, then
  QCD would respect an exact $SU(3)_V\subset SU(3)_L\times SU(3)_R$
  symmetry, where one sets $L=R$. This is the $SU(3)$ symmetry of
  Gell-Mann.   
\item Since $m_d-m_u$ is small, $SU(2)_V\subset SU(3)_V$, where $L=R$
  and they act nontrivially only on the $u$ and $d$ quarks, is quite a
  good approximate symmetry...also known as isospin symmetry. 
\item
Independent vector-like phase rotations of the three flavors of quarks
are exact symmetries...these three $U(1)$ symmetries are linear
combinations of baryon number, $I_3$ isospin symmetry, and $Y$ 
(hypercharge). The latter two are violated by the weak interactions,
but not by the strong or electromagnetic forces. 
\end{itemize}

We know that this still is not the whole story though.  An added
complication is that the QCD vacuum spontaneously breaks the chiral 
$SU(3)_L\times SU(3)_R$ symmetry down to Gell-Mann's  $SU(3)_V$
 via the quark condensate:
\beq
\expect{0}{\mybar q_{Rj} q_{Li}}{0} = \Lambda^3 \delta_{ij}\ ,
\eqn{sig1}\eeq
which transforms as a $(3,\mybar 3)$ under $SU(3)_L\times SU(3)_R$.
Here $\Lambda$ has dimensions of mass. If one redefines the quark
fields by a chiral transformation, the Kronecker $\delta$-function
above gets replaced by a general $SU(3)$ matrix,
\beq
 \delta_{ij}\to (LR^\dagger)_{ij} \equiv \Sigma_{ij}\ .
\eqn{sig2}
\eeq
If $L=R$ (an $SU(3)_V$ transformation), $\Sigma_{ij}=\delta_{ij}$
which shows that the condensate leaves unbroken the $SU(3)_V$
symmetry. For $L\ne R$, $\Sigma_{ij}$ represents a different vacuum
from \eq{sig1}, and if it wasn't for the
explicit breaking of $SU(3)_L\times SU(3)_R$ by quark masses in the QCD
Lagrangian, these vacua would be degenerate. By
Goldstone's theorem therefore,  there would have to be eight  exact Goldstone bosons ---
one for each of the eight broken generators --- corresponding to
long wavelength, spacetime dependent rotations of the condensate.  We
will parametrize these excitations  by replacing
\beq
\Sigma\to \Sigma(x)\equiv e^{2i \bfpi(x) /f}\ ,\qquad \bfpi(x) = \pi_a(x)
T_a
\eeq
where the $T_a$ are the $SU(3)$ generators ($a=1,\ldots,8$) in the
defining representation normalized to 
\beq
\Tr T_a T_b = \half\delta_{ab}\ ,
\eeq
$f$ is a parameter with dimension of mass which we will relate to the
pion decay constant $f_\pi$,  and the $\pi_a$ are eight mesons
transforming as an octet under $SU(3)_V$.  These bosons correspond to
long wavelength excitations of the vacuum.

If you are somewhat overwhelmed by this amazing mix of symmetries that are gauged, global, exact,
approximate, spontaneously broken and anomalous (and usually more than
one of these attributes at the same time), rest assured that it took a
decade and many physicists to sort it all out (the 1960's).

\subsection{Quantum numbers of the meson octet}
\label{sec:3b}
A useful  basis for $SU(3)$ generators is $T_a=\half
\lambda_a$, where $\lambda_a$ are Gell Mann's eight matrices. The
meson matrix
$\bfpi\equiv \pi_a T_a$ appearing in the exponent of $\Sigma$ is a
traceless $3\times 3$ matrix.  We know that under an $SU(3)_V$
transformation $L=R=V$, 
\beq
\Sigma\to V \Sigma V^\dagger = e^{2 i V \bfpi V^\dagger/f}\ ,
\eeq
implying that  under $SU(3)_V$ the mesons transform as an octet
should, namely
\beq
\bfpi \to V \bfpi V^\dagger\ .
\eqn{supi}
\eeq
Then by restricting $V$ to be an $I_3$ ($T_3$) or a $Y$ ($T_8$) rotation we can read
off the quantum numbers of each element of the $\bfpi$ matrix and
identify them with real particles (problem III.1):
\beq
\bfpi = \frac{1}{\sqrt{2}}\begin{pmatrix} \frac{\pi^0}{\sqrt{2}} +
  \frac{\eta}{\sqrt{6}} & \pi^+ & K^+\cr \pi^- &
  -\frac{\pi^0}{\sqrt{2}} + \frac{\eta}{\sqrt{6}} & K^0\cr K^- &
  \mybar K^0 & -\frac{2\eta}{\sqrt{6}}\end{pmatrix}
\eqn{piqn}
\eeq
The normalization is such  that
\beq
\Tr (\bfpi\bfpi) = \half\sum_a (\pi_a)^2 = \half (\pi^0)^2 +\half \eta^2 +
\pi^+\pi^- + K^+K^- + K^0\mybar K^0\ .
\eeq

\subsection{The chiral Lagrangian}
\label{sec:3c}

\subsubsection{The leading term and the meson decay constant}

We are now ready to write down the effective theory of excitations of
the chiral condensate (the chiral Lagrangian), ignoring all the other
modes of QCD.  This is analogous to
the quantization of rotational modes of a diatomic molecule, ignoring
the vibrational modes. We are guided by two basic principles of
effective field theory: (i) The chiral Lagrangian must exhibit the
same approximate chiral symmetry as QCD, which means that it must be
invariant under $\Sigma\to L\Sigma R^\dagger$ for arbitrary
$SU(3)_L\times SU(3)_R$ matrices $L$, $R$ in the ``chiral limit'',
$M\to 0$.  We will also be able to
incorporate symmetry breaking effects by including the matrix $M$,
requiring that the chiral Lagrangian be invariant under the chiral
symmetry if $M$ were to transform as in \eq{masstrans}.  (ii) The
other principle is that the   effective
theory be an expansion of local operators suppressed by powers of a
cutoff $\Lambda$, which is set by the scale of physics we are
ignoring, such as the $\rho$, $K*$, $\omega$, and $\eta'$  mesons (with masses
$m_\rho=770\MeV$,  $m_{K^*} = 892\MeV$, $m_\omega=782\MeV$ and
$m_{\eta'}=958\MeV$). In practice, the cutoff seems to be at
$\Lambda\simeq 1\GeV$ in many processes. Our calculations
will involve an expansion in powers of momenta or meson masses divided
by $\Lambda$. This cutoff is  to be compared with
$m_{\pi^\pm}=140\MeV$, $m_{K^+}=494\MeV$ and $m_\eta=548\MeV$. For
purely mesonic processes, meson masses always appear squared, which
helps.  Nevertheless, one can
surmise that chiral perturbation theory will work far better for pions
than kaons or the $\eta$. This is a reflection of the fact that
$SU(2)_L\times SU(2)_R$ is a much better symmetry of QCD than
$SU(3)_L\times SU(3)_R$.

The lowest dimension chirally symmetric operator we can write down is
\beq
\CL_0 = \frac{f^2}{4}\Tr \partial \Sigma^\dagger \partial \Sigma = \Tr
\partial \bfpi \partial \bfpi + \frac{1}{3 f^2}\Tr [\partial\bfpi,\bfpi]^2 +
\ldots
\eqn{chilag0}\eeq
Note that the $f^2/4$ prefactor is fixed by requiring that  the mesons
have
canonically normalized kinetic terms. Thus we have an infinite tower
of operators involving a single unknown parameter, $f$. From the above
Lagrangian, it would seem that the only way to determine $f$ is by
looking at $\pi\pi$ scattering.  However there is a better way: by
looking at the charged pion decay $\pi\to \mu\nu$.  This occurs through
the ``semi-leptonic'' weak interaction \eq{fermi}, namely the operator
\beq
\frac{1}{\sqrt{2}}G_F  V_{ud} \left(\mybar u \gamma^\mu (1-\gamma_5)d\right)\left(\mybar
\mu \gamma_\mu (1-\gamma_5)\nu_\mu\right) + \text{h.c.}
\eeq
The matrix element of this operator sandwiched between $\ket{\mu\nu}$
and $\bra{\pi}$ factorizes, and the leptonic part is perturbative.  We
are left with the nonperturbative part,
\beq
\expect{0}{\mybar u \gamma^\mu(1-\gamma_5)d}{\pi^-(p)} \equiv i
\sqrt{2}\,f_\pi p^\mu\ .
\eqn{fpi}\eeq 
The pion decay constant $f_\pi$ is determined from the charged pion
lifetime to be  $f_\pi = 92.4 \pm .25 \MeV$.

Even though QCD is nonperturbative, we can easily match this charged current
operator onto an operator in the chiral Lagrangian.  That is because
we can write
\beq
\mybar u \gamma^\mu(1-\gamma_5)d = 2 \left( j^\mu_{L1} + i
  j^\mu_{L2}\right)\ ,
\eeq
where $j^\mu_{La}$ are the eight $SU(3)_L$ currents
\beq
j^\mu_{La} \equiv \mybar q \gamma^\mu
\left(\frac{1-\gamma_5}{2}\right) T_a q\ .
\eeq
To compute these currents in the effective theory is easy, since we
know that under infinitesimal $SU(3)_L$ transformations the change in
$\Sigma$ is
$\delta_{La} \Sigma = i T_a \Sigma$, from which we can compute the
left-handed currents from the Lagrangian \eq{chilag0} using Noether's theorem.  The result is:
\beq
j^\mu_{La} = -i\frac{f^2}{2} \Tr T_a \Sigma^\dagger \partial^\mu
\Sigma = f \Tr T_a \partial^\mu \bfpi + O(\bfpi^2)\ .
\eeq
In particular,
\beq
2\left( j^\mu_{L1} + i
  j^\mu_{L2}\right) = 2f\,\Tr\begin{pmatrix} 0 & 1 & 0\cr 0 & 0 & 0\cr 0
  & 0 & 0\cr \end{pmatrix} \partial^\mu \bfpi + O(\bfpi^2) = \sqrt{2} f
\partial^\mu \pi^-+O(\pi^2)\ ,
\eeq
were I made use of \eq{piqn}.  Comparing this equation with \eq{fpi}
we see that to this order, 
\beq
f=f_\pi=93\MeV\ .
\eqn{fdef}
\eeq

In general it is not possible to exactly match quark operators to a unique
operator in the chiral Lagrangian;  it was possible for the
semi-leptonic decays simply because the weak operator factorized into a
leptonic matrix element and a hadronic matrix element of an $SU(3)_L$
symmetry current. For a purely hadronic weak decay, such as $K\to
\pi\pi$ the four quark operator cannot be factorized, and matching to
operators in the chiral Lagrangian involves  coefficients
which can only be computed on a lattice.  Even for these processes the
chiral Lagrangian can be predictive, relating weak decays with
different numbers of mesons in the final state.

\subsubsection{Explicit symmetry breaking}

Up to now, I have only discussed operators in the chiral Lagrangian
which are invariant under chiral symmetry. Note that that  all
  chirally invariant operators must involve  derivatives
  (other than the operator $1$). For example, one cannot write down a
  chirally invariant
  mass term for the pions. 
  Recall that without explicit chiral symmetry breaking in the QCD 
  Lagrangian, there would be an infinite number of inequivalent 
  degenerate vacua corresponding to different constant values of the 
  matrix $\Sigma$; therefore the energy (and the Lagrangian) can only 
  have operators which vanish when $\Sigma$ is constant, up to an 
  overall vacuum energy independent of $\Sigma$. In fact, rotating 
  $\Sigma\to \Sigma' = \Sigma + id\theta_a \,T_a\,\Sigma$ for constant
  $d\theta_a$ is an exact
  symmetry of the theory ($SU(3)_L$), and corresponds to {\it
    shifting} the pion 
  fields $\pi_a\to \pi_a + d\theta_a \, f/2+O(\pi^2)$. Purely  derivative
  interactions are a consequence of this shift symmetry. In the
  literature, this is called a {\it nonlinearly realized} symmetry,
  which is to say, a spontaneously broken symmetry.  A theory of massless 
   particles with nontrivial interactions at zero momentum transfer 
   (such as QCD) 
   would suffer severe infrared divergences, and so if the 
   interactions had not been purely derivative, the theory would 
   either not make sense, or would become nonperturbative like 
   QCD or else undergo spontaneous breaking of the vector $SU(3)$.  

This all changes when explicit chiral symmetry breaking is included.
Now not all vacua are equivalent, the massless Goldstone bosons become
massive ``pseudo-Goldstone bosons'' (PGBs), and acquire non-derivative
interactions.  In pure QCD, the only sources of explicit chiral
symmetry breaking are instantons (which  break the $U(1)_A$
symmetry) and  the quark mass matrix.  Electromagnetic interactions
also introduce chiral symmetry breaking, as do weak interactions.

To include the effect of quark masses, we need to include the mass
matrix $M$, recalling that if it transformed as in \eq{masstrans},
then the theory would have to be invariant.  Just as with derivatives,
each power of $M$ will be accompanied by $1/\Lambda$.  The
leading operator we can write down is
\beq
\CL_M = \Lambda^2 f^2 \left(\frac{c}{2}
  \frac{1}{\Lambda} \Tr M \Sigma + \text{h.c.}\right) \equiv
  \half  f^2\Tr({\widetilde \Lambda}  M) \Sigma + \text{h.c.}\ ,
\eqn{massterm}\eeq
where $c$ is an unknown dimensionless coefficient, and I defined 
\beq
c \Lambda\equiv {\widetilde \Lambda}  = O(\Lambda)\ .
\eeq
   Expanding to second order in the
${\boldsymbol \pi}$, I get
\beq
\CL_M = -m_\pi^2 \pi^+\pi^- - m_{K^+}^2 K^+K^- - m_{K^0}^2 K^0\mybar
K^0 - \half \begin{pmatrix}\pi^0 & \eta\cr \end{pmatrix} M_0^2
\begin{pmatrix} \pi^0\cr \eta\cr \end{pmatrix}\ ,
\eqn{chilag0M}\eeq
with
\beq
 m_\pi^2 ={\widetilde \Lambda}(m_u+m_d) \
,\qquad m_{K^+}^2 ={\widetilde \Lambda}(m_u + m_s) \ ,\qquad m_{K^0}^2 = {\widetilde \Lambda}(m_d + m_s)\ ,
\eqn{mesmass1}\eeq
and 
\beq
M_0^2 ={\widetilde \Lambda} \begin{pmatrix} (m_u + m_d) & (m_u-m_d)\cr (m_u-m_d)& \frac{1}{3}(m_u + m_d + 4
    m_s)\cr \end{pmatrix} 
\eqn{mesmass2}\eeq
Note that (i) the squares of the meson masses are proportional to
quark masses;  (ii) $\pi^0-\eta$ mixing is isospin breaking and
proportional to $(m_u-m_d)$; (iii)  expanding in powers of  $(m_u-m_d)$,  
$m_\eta^2$ and $m_{\pi^0}^2$ are given by the diagonal entries of
$M_0^2$, up to corrections of $O\left( (m_u-m_d)^2\right)$; (iv) we cannot directly relate quark and
meson masses because of the unknown coefficient ${\widetilde \Lambda}$.  

Ignoring isospin breaking, the masses obey the Gell-Mann Okuba formula
\beq
3 m_\eta^2 + m_\pi^2 = 4 m_K^2\ .
\eeq
The two sides of the above equation are satisfied experimentally to
better than 1\% accuracy. 

It is not difficult to include the effects of electromagnetism in the
chiral Lagrangian (problem III.6).  To leading order in $\alpha$,
electromagnetic corrections shift the square of the masses for the charged mesons:
\beq
m_{\pi^+}^2 = {\widetilde \Lambda}(m_u + m_d) + \frac{\alpha}{4\pi} \Delta^2\ ,\qquad
 m_{K^+}^2 = {\widetilde \Lambda}(m_u
+ m_s) + \frac{\alpha}{4\pi} \Delta^2\ ,
\eqn{emmass}\eeq
while leaving  neutral meson  masses unchanged. In the above formula,
$\Delta$ has mass dimension 1, and is $O(\Lambda)$ in size;  the
prefactor of $\alpha/4\pi$ arises since the splitting must arise from
a loop diagram involving a photon.
Following Weinberg, we can also use the above formula to calculate the
ratios of quark masses via the formulas
\beq
\frac{(m^2_{K^+} - m^2_{K^0}) - (m^2_{\pi^+} -
  m^2_{\pi^0})}{m^2_{\pi^0}} = \frac{m_u-m_d}{m_u+m_d}\ ,\qquad
\frac{3 m_\eta^2 - m_{\pi^0}^2}{m_{\pi^0}^2} = \frac{4 m_s}{m_u+m_d}\
.
\eeq
Plugging in the measured meson masses, the result is
\beq
\frac{m_u}{m_d}\simeq \half\ ,\qquad \frac{m_d}{m_s}\simeq
\frac{1}{20}\ . 
\eeq
To specify the quark masses themselves, one must perform a lattice QCD
calculation and designate a renormalization scheme. Lattice
simulations typically find $m_s$ renormalized at $\mu=2\GeV$ in the
$\overline{MS}$ scheme lies in the $80-100\MeV$ range, from which one
infers from the above ratios $m_d\sim 5\MeV$, $m_u\sim 2.5\MeV$ in the
same scheme. Evidently most of the mass of baryons and vector mesons
does {\it not} come from the intrinsic masses of the quarks.

\subsection{Loops and power counting}
\label{sec:3d}

What makes the chiral Lagrangian and EFT equivalent to low energy QCD and not just another model of
the strong interactions is that it consists of all local operators
consistent with the symmetries of QCD.  It is useful because there exists a power
counting scheme that allows one to work to a given order in a small parameter, and to be
able to make a reliable estimate of the errors arising from neglecting
the subsequent order.  As discussed in the second lecture, the power
counting scheme is intimately related to how one computes radiative
corrections in the theory.

Beyond the leading term is an infinite number of chirally invariant
operators one can write down which are higher powers in
derivatives, as well as operators with more insertions of the quark
mass matrix $M$. The derivative expansion is in powers of
$\partial/\Lambda$.  This power
counting is consistent with the leading operator \eq{chilag0}, if you consider the
chiral Lagrangian to have an overall prefactor of $\Lambda^2 f^2$,
then even in the leading operator derivatives enter as $
\partial/\Lambda$. 
 Since we have found that meson octet masses scale as
$m_{\boldsymbol\pi}^2\simeq ({\widetilde \Lambda} M)$, and since for on-shell pions $p^2\sim
m^2$, it follows that one insertion of the quark mass matrix is
equivalent to two derivatives in the effective field theory
expansion. This leads us to write the chiral Lagrangian as a function
of $(\partial/\Lambda)$ and ${\widetilde \Lambda} M/\Lambda^2$.  Including
electromagnetism is straightforward as well: since a derivative
$\partial \Sigma$ becomes a covariant derivative $D_\mu\Sigma =
\partial_\mu\Sigma-i eA_\mu\left[Q,\Sigma\right] 
$, $Q$ being the quark charge matrix ${\rm diag}(2/3,-1/3,-1/3)$,  the
photon field enters as $e A_\mu/\Lambda$.  Operators 
arising from electromagnetic loops involve two insertions of the quark
charge matrix $Q$ in the proper way (see problem (III.6)), along with a loop factor $\alpha/(4\pi)$.
Therefore the chiral Lagrangian takes the form
\beq
\CL = \Lambda^2 f^2 \widehat\CL \left[\Sigma,\, \partial/\Lambda,\, {\widetilde \Lambda} M/\Lambda^2,\, eA/\Lambda,\,
(\alpha/4\pi)Q^2 \right]\ ,
\eeq
where $\widehat\CL $ is a dimensionless sum of all local, chirally
invariant  operators (treating $M$ and $Q$ as spurions),
where the coefficient of each term (except $\CL_0$) is preceded by a
dimensionless coefficient to be fit to experiment.  These coefficients
are
  expected to be $O(1)$, but may occasionally surprise us!  This
last assumption is what allows one to 
  estimate the size of higher order corrections.  

It should be clear now in what sense the $u$, $d$ and  $s$ are light
quarks and can be treated in chiral perturbation theory, while the
$c$, $b$ and $t$ quarks are not:  whether the quarks are light or
heavy is relative to the scale $\Lambda$, namely the mass scale of
resonances in QCD.  Since the $c$ has a mass $\sim 1.5\GeV$ there is
no sensible way to talk about an approximate $SU(4)\times SU(4)$
chiral symmetry and include $D$, $D_s$ and $\eta_c$  mesons in our
theory of pseudo-Goldstone bosons\footnote{This does not mean that an
effective theory for $D-\pi$ interactions is impossible.  However, the
$D$ mesons must be introduced as heavy matter fields, similar to the
way we will introduce baryon fields later, as opposed to approximate
Goldstone bosons.}. 
Of course, you might argue that the 
strange quark is sort of heavy and should be left out as well, but if
we don't live dangerously sometimes, life is too boring.

\subsubsection{Sub-leading order: the  $O(p^4)$ chiral Lagrangian}

It is a straightforward exercise to write down sub-leading
operators of the chiral Lagrangian.  These are operators of $O(p^4)$,
$O(p^2 M)$ and $O(M^2)$, where $M$ is the quark mass matrix. This was first done by Gasser and
Leutwyler, and their choice for the set of operators has become
standard:
\beq
\CL_{p^4} &=& L_1 \left(\Tr\left(\partial_\mu
  \Sigma^\dagger\partial^\mu\Sigma\right)\right)^2 \cr &&+ L_2\Tr\left(\partial_\mu
  \Sigma^\dagger\partial_\nu\Sigma\right)\,\Tr\left(\partial^\mu
  \Sigma^\dagger\partial^\nu\Sigma\right)\cr
&& +
L_3
\Tr\left(\partial_\mu\Sigma^\dagger\partial^\mu\Sigma\partial_\nu\Sigma^\dagger\partial^\nu\Sigma\right)
\cr
&&+ L_4\Tr\left(\partial_\mu
  \Sigma^\dagger\partial^\mu\Sigma\right)\Tr\left(\chi \Sigma +
  \text{h.c.}\right)
\cr && +
L_5\Tr\left(\left(\partial_\mu
  \Sigma^\dagger\partial^\mu\Sigma\right)\left(\chi \Sigma +
  \text{h.c.}\right)\right)
\cr &&+
L_6\left(\Tr \left(\chi\Sigma+\text{h.c.}  \right)\right)^2\cr
&& +
L_7 \left(\Tr \left(\chi\Sigma-\text{h.c.}  \right)\right)^2\cr
&&+
L_8 \Tr\left(\chi \Sigma \chi\Sigma +\text{h.c.}\right)\ ,
\eeq
where $\chi\equiv 2\widetilde \Lambda M$, where $\widetilde
\Lambda$ entered in \eq{massterm}.  Additional operators
involving $F_{\mu\nu}$ need be considered when including
electromagnetism.

Note that according to our power counting, we expect the $L_i$ to be
of size
\beq
L_i\sim \frac{\Lambda^2 f^2}{\Lambda^4} = \frac{f^2}{\Lambda^2}\sim
10^{-2}\ .
\eeq

\subsubsection{Calculating loop effects}

Now consider loop diagrams in the effective theory. These are often
divergent, and so the first issue is
how to regulate them.  It is easy to show that a momentum cutoff
applied naively violates chiral symmetry;  and while it is possible to
fix that, by far the simplest regularization method
is dimensional regularization with a mass independent subtraction
scheme, such as $\overline{MS}$.

The $\overline{MS}$ scheme  introduces a renormalization scale $\mu$, usually chosen to be
$\mu=\Lambda$. However, unlike with cutoff regularization,  one
never gets powers of  the renormalization scale $\mu$ when computing a
diagram; $\mu$ can only appear in logarithms. Consider, for example, the $O( \pi^4)$ operator from
$\CL_0$, of the form $\frac{1}{f^2} (\partial\pi)^2 \pi^2$, and
contract the two pions in  $ (\partial\pi)^2$; this one-loop graph
will renormalize the pion mass.  However, since the diagram  is
proportional to $1/f^2$, and no powers of the renormalization scale
$\mu$ can appear,  dimensional analysis implies that any shift in the pion mass from this graph must be
proportional to $\delta m_\pi^2 \sim (m^0_\pi)^4/(4 \pi f)^2$, times a
possible factor of $\ln (m_\pi/\mu)$, where $(m_\pi^0)^2 \sim
\tilde\Lambda M$ is the  mass squared of
the meson at leading
order. Here I have included the factor of
$1/(4 \pi)^2$ that typically arises from a loop diagram. Ignoring the
logarithm, compare  this contribution to the pion mass contribution
from the $O(p^4)$ chiral Lagrangian, which yields $\delta m_\pi^2
\sim (\tilde \Lambda  M)^2/\Lambda^2$.  We see that so long as 
\beq
4 \pi f_\pi\gtrsim \Lambda\ ,
\eqn{flam}
\eeq
then the contribution from the radiative correction from the lowest
order operator is comparable to or smaller than the
second  order tree-level contribution, up to $\ln m_\pi^2/\mu^2$
corrections. This is satisfied for $f_\pi=93\MeV$ and $\Lambda\simeq 1\GeV$.

What about the logarithm?  Note that $\ln(\Lambda^2/m_\pi^2)
\simeq 4$. Therefore a term with a logarithm is
somewhat enhanced relative to the higher order tree-level
contributions.  It is therefore common to see in the literature a
power counting scheme of the form
\beq
p^2 > p^4 \ln (\mu^2/p^2) > p^4 > p^6 \ln (\mu^2/p^2) < p^6 \ldots 
\eqn{pcs}\eeq
which means that in order of importance, one computes processes in
the following order:
\begin{enumerate}
\item Tree level contributions from the $O(p^2)$ chiral Lagrangian;
\item Radiative corrections to the $O(p^2)$ chiral Lagrangian, keeping
  only $O(p^4\ln p^2)$ terms;
\item Tree level terms from the $O(p^4)$ chiral Lagrangian, as well as
  $O(p^4)$ radiative contributions from the $O(p^2)$ chiral
  Lagrangian;
\end{enumerate}
and so forth.  Keeping the logs and throwing out the analytic terms in
step \#2 is equivalent to saying that most of the $O(p^4)$ chiral
Lagrangian renormalized at $\mu=m_\pi$ would come from running induced
by the $O(p^2)$ Lagrangian in scaling down from $\mu=\Lambda$ to
$\mu=m_\pi$, and not from the initial values of couplings in the
$O(p^4)$ Lagrangian renormalized at $\mu=\Lambda$.  This procedure would {\it
  not} be reasonable in the large $N_c$ limit (see problem (III.7))
but seems to work reasonably well in the real world.

\subsubsection{Renormalization of $\expect{0}{\mybar q q}{0}$}

As an example of a simple calculation, consider the computation of the
ratios of the quark condensates,
\beq
x = \frac{\expect{0}{\mybar u u}{0}}{\expect{0}{\mybar ss }{0}}
\ .
\eeq
 Since the operator
$\mybar q q$ gets multiplicatively renormalized, $\expect{0}{\mybar
  q_i q_i}{0}$ is scheme dependent, but the ratio $x$ 
is not.  The QCD Hamiltonian density is given by $\CH=\ldots + \mybar q M q +
\ldots$, and so it follows from the Feynman-Hellman theorem\footnote{The
  substance of the Feynman-Hellman theorem is  that in first order
  perturbation theory, the wave function doesn't change while the
  energy does.} that
\beq
\expect{0}{\mybar q_i q_i}{0} = \frac{\partial\ }{\partial m_i} \expect
{0}{\CH}{0} = \frac{\partial\CE_0}{\partial m_i}\ ,
\eeq
where $\CE_0$ is the vacuum energy density. We do not know the value of
$\CE_0$, but we do know its dependence one the quark mass matrix; from
\eq{chilag0M}
\beq
\CE_0 = \text{const.} - 
  \half  f^2\Tr({\widetilde\Lambda}  M) \Sigma + \text{h.c.} + O(M^2\ln M) \
  \Biggl\vert_{\Sigma_{ij}=\delta_{ij}} =f^2 \widetilde\Lambda\Tr M + \ldots \ ,
\eqn{emass}\eeq
from which it follows that in this scheme
\beq
\expect{0}{\mybar q_iq_i}{0} = \widetilde \Lambda f^2\ ,
\eeq
and that in any scheme the leading result for  $x$ 
is
\beq
x =1\ .
\eeq
Well good --- this is what we started with for massless QCD in
\eq{sig1}!  To get the sub-leading logarithmic corrections, we need to
compute the $O(m^2\ln m^2)$ one-loop correction to the vacuum energy. This loop with no
  vertices's is the Feynman diagram for which Feynman rules don't work!
  As easily seen in a Euclidean  path integral, the vacuum energy
  density in a box of 4-volume $VT$ for a real, noninteracting scalar is just 
\beq
\CE_0  = -\frac{1}{VT}\ln
  \left(\det(-\square + m^2)\right)^{-1/2} = \frac{1}{VT} \half \Tr \ln (-\square +
  m^2)\ .
\eeq
In $d=(4-2\epsilon)$ Euclidean dimensions
this just involves evaluating for each mass eigenstate the integral
\beq
\frac{\mu^{4-d}}{2}\,\int \frac{d^dk}{(2\pi)^d} \ln( k^2 + m^2 ) \ .
\eeq
where the prefactor of $\mu^{4-d}$ was included to keep the mass dimension
to equal 4. 

Let us first perform the differentiation with respect to quark mass.
Then in this scheme we get the correction
\beq
\delta\,\expect{0}{\mybar q_i q_i}{0} = \half\sum_a \frac{\partial
  m^2_a}{\partial m_i}
\mu^{4-d}\,\int  \frac{d^dk}{(2\pi)^d}\frac{1}{k^2 + m^2}
\quad
\xrightarrow[\quad \overline {MS}\quad] \quad 
-\sum_a \frac{\partial
  m^2_a}{\partial m_i} \left(\frac{m_a^2\ln m_a^2/\mu^2}{32\pi^2}\right)\ ,
\eeq
where $a$ is summed over the meson mass eigenstates, and $m_i$ is the
mass of the $i^{th}$ flavor of quark.  The final result was arrived at
after performing the $\overline{MS}$ subtraction (where you only keep
the $\ln m^2$ term in the $\epsilon\to 0$ limit; see the appendix \ref{sec:7}
for dimensional regularization formulas).

To the order we are working, the quark condensate ratios are therefore given by
\beq
\frac{\expect{0}{\mybar q_i q_i}{0} }{\expect{0}{\mybar q_j q_j}{0} }=
1 -\frac{1}{32 \pi^2 \widetilde \Lambda f^2}\sum_a m_a^2\ln
m_a^2/\mu^2 \left(\frac{\partial m_a^2}{\partial m_i}- \frac{ \partial
    m_a^2}{\partial m_j}\right)\ .
\eeq
Using the masses given in \eq{mesmass1} and \eq{mesmass2}, ignoring
$\pi^0-\eta$ mixing,  we find
\beq
x &=&\frac{\expect{0}{\mybar u u}{0}}{\expect{0}{\mybar ss }{0}}=
 1 -3 g_\pi + 2 g_{K^0} + g_\eta + O(m^4) \ ,
\eeq
where 
\beq
g_P \equiv \frac{1}{32 \pi^2 f^2} m_P^2 \ln
\left(\frac{m_P^2}{\mu^2}\right)
\eeq
with $P=\pi,\,K^+,\,K^0,\,\eta$.  The answer is $\mu$ dependent, since
I have neglected to include the $O(p^4)$ Lagrangian contributions at
tree-level, and in fact it is precisely those operators that serve as
counterterms for the $1/\epsilon$ poles subtracted in
$\overline{MS}$.  However, in the usual practice of chiral
perturbation theory, I have assumed that with $\mu=\Lambda$, the
contributions from the $O(p^4)$ Lagrangian are small compared to the
chiral logs I have included.  Plugging in numbers with $\mu=1\GeV$ I
find
\beq
g_\pi\simeq -0.028\ ,\quad g_K\simeq -0.13\ ,\quad g_\eta\simeq -0.13
\eeq
implying that $x\simeq 0.70$ ---  a  30\% correction 
from the leading result $x=1$.  This is typical of any chiral correction
that involves the strange quark, since $m_K^2/\Lambda^2\simeq 25\%$.
Corrections to $\vev{\mybar u u}/\vev{\mybar d d}$ will be {\it much}
smaller, since they depend on isospin breaking, of which a typical measure
is $(m_{K^0}^2 - m_{K^+}^2)/\Lambda^2 \simeq 0.004$.

\subsection{Including baryons}
\label{sec:3e}

\subsubsection{Transformation properties and couplings}

It is interesting to include the baryon octet into the mix.  This is
reasonable so long as we consider processes with momentum transfer
$\ll \Lambda$.  Thus we might consider the weak decay $\Lambda\to N
\pi$, but not the annihilation $N\mybar N\to \pi\pi$. There are two
separate issues here:  (i) How do we figure out how baryons transform
under the chiral  $SU(3)\times SU(3)$ symmetry so that we can couple
them to $\Sigma$, and (ii) do we need or desire the Dirac spinor
formulation if we are only going to consider low momentum transfer
processes?   I will address the first question first, using Dirac
spinors.  Then I will introduce the heavy baryon formalism of Jenkins
and Manohar, replacing the Dirac spinors.

First consider a world where the $u$, $d$ and $s$ are massless. We
know that the baryons transform as an octet under the unbroken
$SU(3)_V$ symmetry, but how do they transform under $SU(3)\times
SU(3)$?  The answer is: just about any way you want.  To see this,
consider a baryon field $B$ written as a $3\times 3$ traceless matrix
of Dirac spinors, transforming as an octet under $SU(3)_V$:  
\beq
B\to VBV^\dagger\ .
\eqn{b3}\eeq
By considering $T_3$ and $T_8$ transformations it is possible to
determine the form of the matrix $B$, just as we did for the meson
matrix $\bfpi$:
\beq
B = \begin{pmatrix}
\frac{\Sigma^0}{\sqrt{2}} + \frac{\Lambda}{\sqrt{6}} & \Sigma^+ & p\cr
\Sigma^- &  -\frac{\Sigma^0}{\sqrt{2}} + \frac{\Lambda}{\sqrt{6}} &
n\cr
\Xi^- & \Xi^0 & -\frac{2 \Lambda}{\sqrt{6}}\end{pmatrix}\ .
\eeq

Now construct the left- and right-handed baryons $B_{R,L} =
\half(1\pm\gamma_5)B$. Suppose that  $B$ transformed
as the $(8,1)\oplus (1,8)$ representation under $SU(3)\times SU(3)$,
namely
\beq
B_R\to RB_R R^\dagger\ ,\qquad B_L\to LB_LL^\dagger\ .
\eqn{brl}\eeq
But then I could define
\beq
B'_R \equiv \Sigma B_R \ ,\qquad B_L' =
B_L \Sigma\ .
\eeq
The new field $B'$  works equally well as a baryon field.  However
$B'$ transforms as a $(3,\mybar 3)$ under $SU(3)\times SU(3)$:
\beq
B' \to L B' R^\dagger\ .
\eeq
Note that both $B$ and $B'$ both transform properly as an $SU(3)_V$
octet when $R=L\equiv V$, as in \eq{b3}.  What we are seeing is that
when you have massless pions around, you can't tell the difference
between a baryon, and a superposition of that baryon with a bunch of
zero momentum massless pions, and yet the two will have different
$SU(3)\times SU(3)$ transformation properties.  This is not a problem
--- rather it is liberating.  It means we can choose whatever
$SU(3)\times SU(3)$ transformation rule we wish for the baryons, so
long as \eq{b3} still holds.

While the basis \eq{brl} looks appealing, it has its drawbacks.  For
example it allows the interaction 
\beq
m_0 \Tr \mybar B_L  \Sigma B_R +\text{h.c.} = m_0 \Tr \mybar B B +
\frac{2m_0}{f}\Tr \mybar B i\gamma_5  \bfpi B + \ldots
\eeq
which makes it {\it look} like the pion can have non-derivative
couplings...which ought to be impossible for a Goldstone boson.
In fact, Dirac's analysis for nonrelativistic spinors shows that the
$\gamma_5$ coupling is in fact a derivative coupling at 
low momentum transfer, made obscure.

A better basis is the following. Define
\beq
\xi = e^{i\bfpi/f} = \sqrt{\Sigma}\ .
\eeq
One can show that under an $SU(3)\times SU(3)$ transformation
\beq
\xi\to L \xi U^\dagger = U \xi R^\dagger\ ,
\eeq
where $U$ is a uniquely defined matrix which is in general a function
of the constant $L$ and $R$ matrices characterizing the  $SU(3)\times
SU(3)$ transformation, as well as the $\bfpi(x)$ field.  For the
special case of $SU(3)_V$ transformations, $R=L=V$ and it is easy to
show that $U=V$ as well, independent of $\bfpi$.  For more general
 $SU(3)\times SU(3)$ transformations $U$ is a mess, but we will not
need to know its exact form.  Now if we take the basis \eq{brl} and
replace $B_L\to \xi^\dagger B_L \xi$, $B_R\to \xi B_R \xi^\dagger$, we
get a new basis where left- and right-handed components of $B$
transform the same way, namely
\beq 
B \to U B U^\dagger\ .
\eeq
Given the above transformation rule, $B$  cannot couple directly to
$\Sigma$, but we can define the axial and vector currents and chiral
covariant derivative:
\beq
A_\mu &\equiv& \frac{i}{2}\left(\xi^\dagger \partial_\mu \xi -
  \xi\partial_\mu \xi^\dagger\right) \xrightarrow[\  SU(3)\times
SU(3)\ ]{} U A_\mu U^\dagger\ ,\cr &&\cr
V_\mu &=& \half\left(\xi^\dagger \partial_\mu \xi +
  \xi\partial_\mu \xi^\dagger\right) \xrightarrow[\  SU(3)\times
SU(3)\ ]{} U V_\mu U^\dagger + U\partial_\mu U^\dagger\ ,\cr &&\cr
D_\mu B &\equiv& (\partial_\mu B + [V_\mu,B])  \xrightarrow[\  SU(3)\times
SU(3)\ ]{} U (D_\mu B)U^\dagger\ .
\eeq

Armed with this formalism we can write down an effective theory for
meson-baryon interactions, whose first few chirally symmetric terms are
\beq
\CL_1 =
\Tr \mybar B(i\gamma^\mu D_\mu - m_0) B 
-   D\Tr\mybar B \gamma^\mu \gamma_5 \left\{ A_\mu,B\right\} - F \Tr
\mybar B \gamma^\mu \gamma_5 \left[ A_\mu,B\right]\ .
\eeq
Note that the common octet mass $m_0$ is chirally symmetric, independent of
the quark masses.  

Expanding $A_\mu$ and $V_\mu$ in the meson fields,
\beq
A_\mu = -\frac{1}{f}\partial_\mu \bfpi + O(\bfpi^3)\ ,\qquad V_\mu =
\frac{1}{2f^2}\left(\bfpi \partial_\mu \bfpi-( \partial_\mu \bfpi)
  \bfpi\right) + O(\bfpi^4)\ .
\eeq
  For non-relativistic baryons, the Dirac analysis implies that  $\mybar B
\gamma^0 B$ and $\mybar B \vec\gamma\gamma_5 B$ are big, equal to $1$
and $\vec S$ (the baryon spin) respectively; in contrast, the bilinears $\mybar B
\vec\gamma B$ and $\mybar B \gamma^0\gamma_5 B$  are small,
given by $\vec q/m_0$ and $(\vec S\cdot\vec q)/m_0$ respectively,
where $\vec q$ is the 3-momentum transfer.
Therefore the leading meson-baryon interactions for nonrelativistic 
baryons (written as 2-component spinors) is
\beq
\frac{1}{2f^2}\,\Tr \left( B^\dagger
[\bfpi  \dot \bfpi - \dot \bfpi \bfpi,B]\right) + D
\Tr\left(B^\dagger  \vec \sigma\cdot \{\vec\nabla \bfpi, B\}\right) +
F \Tr\left(B^\dagger \vec \sigma\cdot \left[\vec\nabla \bfpi, B\right]\right)\ .
\eeq
The vector current interaction is also called the Weinberg-Tomazawa
term; it does not involve any unknown parameters and is required by
chiral symmetry.  The axial current
interaction involves two new couplings $D$ and $F$ that  may be fit to semi-leptonic baryon decay,
using the same wonderful fact exploited in relating $f$ to
$f_\pi$: namely that weak charged currents happen to be $SU(3)$
currents, and so can be unambiguously computed in the effective
theory. The combination $(D+F)=g_A=1.25$ is derived from neutron
decay.  From hyperon decay, one determines $F\simeq 0.44$, $D\simeq
0.81$. Because the axial interactions involve derivatives of the
mesons, they contribute to $p$-wave scattering, but not $s$-wave; in
contrast, the vector interaction contributes to $s$-wave scattering.

The above Lagrangian is $O(p)$, involving single derivatives on the
meson fields.  Therefore symmetry breaking terms involving the quark mass matrix $M$ are
subleading, as we have seen that $M\sim p^2$ in our power counting.
There are three such terms,
are
\beq
\CL_2 =  a_1 \Tr\mybar B (\xi^\dagger M \xi^\dagger + \text{h.c.})
  B + a_2 \Tr\mybar B B(\xi^\dagger M \xi^\dagger +
  \text{h.c.})+ a_3 \Tr\left(M \Sigma + \text{h.c.}\right)\Tr
\mybar B B\ .
\eeq
 A combination of three of the mass parameters $m_0$, $a_1$, $a_2$,
and $a_3$ may be determined from  the octet masses;  to fit the fourth combination
requires additional data, but can be done (see problem (III.9)). At
the same order one should consider chirally symmetric terms with two
derivatives acting on the mesons, such as subleading terms in the
nonrelativistic expansion of $\CL_1$, as well as new terms such as $\Tr \mybar B A_\mu A^\mu B$.
However,  I believe that there are too many such subleading  terms to fit to
low energy meson-baryon scattering data.

\subsubsection{The heavy baryon formulation}

It was clearly awkward to have to start with Dirac spinors, and then
take the nonrelativistic limit in order to disentangle the
order of each contribution in the chiral expansion. Furthermore, the  EFT expansion for
meson-baryon interactions is complicated by the 
fact that for on-shell baryons, $i\partial_t B\sim m_0 B$, and $m_0\sim
\Lambda$.  So how can there be a derivative expansion since
$\partial_t/\Lambda$ and $m_0/\Lambda$ are both $O(1)$?  The answer is
that for small momentum transfer, all one ever sees is the combination
$(i\partial_t-m_0)$, which is small.  However, a cancellation between
two large parameters is the bane of effective field theory...it makes
power counting obscure.

Another issue is that we have neglected the baryon decuplet, which
might not be justified since the $\Delta$ resonance lies not far
above threshold in $\pi N$
scattering.

Both complications were addressed in the Jenkins-Manohar approach \cite{Jenkins:1990jv,Jenkins:1991es},
which applied the formalism developed for HQET, which I discussed
briefly in my first lecture. One defines the baryon field $B_{\rm v}$ for
baryons with velocity $v_\mu$ (a
$3\times 3$ traceless matrix)
\beq
B_{\rm v}(x) = e^{i m_0 \sla{v} v_\mu x^\mu} B(x)\ ,
\eeq
where $B$ is the baryon field I introduced above. They also introduce
spin operators $S^\mu_{\rm v}$ satisfying
\beq
v_\mu S^\mu_{\rm v} =0\ ,\quad
S^2_{\rm v} B_{\rm v} = -\frac{3}{4} B_{\rm v}\ ,\quad
\{S_{\rm v}^\alpha,\, S_{\rm v}^\beta\} =
\half(v^\alpha v^\beta-\eta^{\alpha\beta})\ ,\quad
\[S^\alpha_{\rm v},\,S^\beta_{\rm v}\] = i \epsilon^{\alpha\beta\mu\nu}v_\mu
S_{v\nu}\ .
\eqn{l1b}
\cr\eeq
In terms of these operators one can work out the Dirac bilinears
such as
\beq
\mybar B_{\rm v} \gamma^\mu B_{\rm v} = v^\mu \mybar B_{\rm v} B_{\rm v}\ ,\qquad \mybar B_{\rm v}
\gamma^\mu \gamma_5 B_{\rm v} = 2\mybar B_{\rm v} S^\mu_{\rm v} B_{\rm v}\ ,
\eeq
etc.  It is then possible to rewrite the leading terms in the baryon
Lagrangian of \eq{l1b} as
\beq
\CL_1 = i\Tr \mybar B_{\rm v}(v_\mu D^\mu)B_{\rm v} + 2D\Tr \mybar B_{\rm v}
S^\mu_{\rm v}\{A_\mu,B_{\rm v}\} + 2 F \Tr \mybar B_{\rm v} S^\mu_{\rm v} \left[A_\mu,
B_{\rm v}\right]\ .
\eeq
Note that the large baryon mass $m_0$ has disappeared from the expression.

The decuplet field $T_{\rm v}^\mu$ consisting of the $\Delta$, $\Sigma^*$,
$\Xi^*$ and $\Omega$ may be introduced in a similar way, with a few
additional complications coming from the fact that they possess  spin
$\frac{3}{2}$, and therefore have to be represented as
Rarita-Schwinger fields instead of Dirac spinors, and because, as a
decuplet of $SU(3)_V$, they form a 3-index symmetric tensor, rather
than a $3\times 3$ matrix like the baryon octet.

\vfill
\eject
\subsection{Problems}

\bigskip
\hrule
\bigskip

\noindent
{\bf III.1)}
Verify that \eq{piqn} follows from \eq{supi}.

\bigskip
\noindent
{\bf III.2)}
How does $\Sigma$ transform under $P$ (parity)?  What does this
transformation imply for the intrinsic parity of the $\pi_a$ mesons?
How does $\Sigma$ transform under $C$ (charge conjugation)? Which 
of the mesons are eigenstates of $CP$, and are they $CP$ even or odd?
Recall that under $P$ and $C$ the quarks transform as 
\begin{equation}
\begin{aligned}
P:&\quad q\to \gamma^0 q\ ,\cr
C:&\quad q\to C \mybar q^T\ ,\qquad C=C^\dagger=C^{-1}=-C^T\ ,\quad
C\gamma_\mu C=-\gamma_\mu^T\ ,\qquad C\gamma_5C=\gamma_5\ .
\end{aligned}
\end{equation}

\bigskip
\noindent
{\bf III.3)} How do we know that $c$, and hence $\widetilde \Lambda$, is positive in
\eq{massterm}? How would the world look different if it were negative?
Hint: consider what $\Sigma$ matrix would minimize the vacuum energy,
and its implications for the spectrum of the theory.

\bigskip
\noindent
{\bf III.4)} 
An axion is a hypothetical particle proposed to explain why the electric dipole
moment of the neutron is so small (the strong $CP$ problem).  It
couples to quarks through the quark mass matrix, where one makes the
substitution
\beq
M \to M e^{i a X/f_a}
\eeq
in \eq{qmass}, where $a$ is the axion field, $f_a$ is the axion decay
constant, and $X$ is a $3\times 3$ diagonal matrix constrained to have $\Tr X =
1$. Compute the axion mass in terms of $m_\pi$, $f_\pi$ and $f_a$,
dropping terms of size $m_{u,d}/m_s$.  Hint: use the remaining freedom
in choosing $X$ to ensure that the axion does not mix with the $\pi^0$
or the $\eta$ mesons.

\bigskip
\noindent
{\bf III.5)} Compute the current left-handed current $j^\mu_{La}$
  through $O(\pi^3)$. Draw the one-loop diagrams that alter the
  relation between $f$ and $f_\pi$.  Estimate the contribution
  proportional to $\ln (m_\pi^2/\Lambda^2)  $

\bigskip
\noindent
{\bf III.6)} Include electromagnetism in the QCD Lagrangian
\eq{qcdkin} by having the photon couple to left- and right-handed
quarks through 
the charge matrix $Q_L$ and $Q_R$ respectively.  What are these two
matrices?  How would they have to transform under $SU(3)_L\times
SU(3)_R$ if the theory was to remain chirally invariant? Treating
$Q_{L,R}$ as spurions, show how to
modify the leading term in the chiral Lagrangian to ensure that it is
gauge invariant under $U(1)_{\text{em}}$?  Write down the leading
non-derivative operator involving powers of $\Sigma$, $\Sigma^\dagger$,
and a pair of $Q$'s. Show that this operator contributes to the masses
of the charged pseudoscalars as given in \eq{emmass}. Does 
$\Delta^2$ in \eq{emmass} arise from a contact interaction, or a
photon loop on a 
meson line? (Hint:
think about renormalization schemes.) Approximately how big would the
contribution to
$\Delta$ be from a 1-loop diagram with momentum cutoff $\Lambda$?

\bigskip
\noindent
{\bf III.7)}
Show that $4\pi f_\pi \gg \Lambda$  is obeyed in QCD for large  $N_c$ (where $N_c$
is the number of colors, and $N_c=3$ in the real world), satisfying
\eq{flam}.  However, show that the power counting scheme below \eq{pcs} doesn't
make sense in this limit.  What is the correct scheme?

\bigskip
\noindent
{\bf III.8)}
Draw the Feynman diagrams that lead to $B\pi\to B\pi$ scattering at
leading order in chiral perturbation theory, where $B$ and $\pi$ are
the baryon and pion octets respectively.  

\bigskip
\noindent
{\bf III.9)}
Assuming no isospin breaking ($m_u=m_d=\mybar m$), compute the baryon octet masses in terms
of $m_0$, $a_1$, $a_2$ and $a_3$. Why can't one determine these four
parameters from the four masses $M_N$, $M_\Sigma$, $M_\Lambda$ and
$M_\Xi$? The failure to determine four parameters from four masses
implies that there must be a relationship predicted for the baryon masses --- the
Gell-Mann Okuba formula for baryons.  What is it?  What sort of
additional data would one need to uniquely 
determine all four constants?

\vfill\eject

\section{ Effective theories of nucleons}
\label{sec:4}

\bigskip
\hrule
\bigskip

This is a nuclear physics school, and in the  last lecture we finally
mentioned nucleons!  So the obvious question is: can the effective field
theory for nucleons cast any light on the properties of nuclei or
nuclear matter?  Today I will tell you about a couple of approaches in
that direction.  The first uses the chiral Lagrangian derived
yesterday in the mean field approximation, and shows how Bose-Einstein
condensation of pions or kaons might occur.  The second goes beyond
mean field theory, and applies effective field theory to low energy few-body
scattering.  I will mention some issues in trying to push these
technologies farther. An amusing spin-off is that the techniques we
develop can be profitably used to analyze systems of trapped atoms.  

\subsection{Kaon condensation}
\label{sec:4a}

In the early seventies it was suggested by A. B. Migdal that pions might
Bose condense in nuclear matter.  His reason for thinking
this is that pions have attractive $p$-wave interactions with nucleons
(although repulsive $s$-wave interactions). A decade of research
showed that competition between $s$-wave repulsion and $p$-wave
attraction  made pion condensation dicey at best.  About 15 years later Ann
Nelson and I proposed that kaon condensation ($K^-$ condensation, to
be precise) might be favored
instead \cite{Kaplan:1986yq}.  This seemed surprising, given that the
binding energy per 
nucleon in iron is about $9\MeV$, while the binding energy for a $ K^-$
in nuclear matter would have to exceed the kaon mass ($490\MeV$ in vacuum) in
order to lead to Bose condensation!   The consensus after another couple of decades work,
supported by analysis of energy levels in kaonic atoms, is that a
binding energy of hundreds of MeV is 
not unreasonable for a $K^-$ in dense matter.  However it remains  unclear whether
kaon condensation could occur at $3\times$ nuclear density as we
predicted,  or only at  higher densities, or possibly never due to prior
appearance of other forms of
strangeness (quark or hyperon matter).   

The mean field calculation for meson condensation is rather easy to understand:
(i) one assumes a background meson field;  (ii) one computes the
dispersion relation for the baryon octet in this background meson
field, in the presence of chemical potentials for baryon number and
electric charge;  (iii) one occupies all the fermion energy levels up
to the Fermi energy, for the baryon octet and for electrons;  (iv)
one minimizes the total free energy of the 
system as a function of the background meson field, subject the
constraint of charge neutrality.   This simplest approach ignores a host of
important effects, such as other
nuclear forces beside that of octet meson exchange, as well as
deviations from mean field such as spatial correlations.

While a full treatment along the above lines typically requires
numerical computation, determining the  critical density for the onset
of condensation can be done analytically.  For this one need only expand the free
energy to second order in the meson field, and look for an instability
--- that is, a negative mass squared for the meson. This calculation
was performed by Politzer and Wise in ref.~\cite{Politzer:1991ev}, in
greater generality than I will do here.

Consider a system with neutrons, protons, electrons and a spatially homogeneous
kaon condensate, with a chemical potential for electrons, $\mu_e$, to
constrain the system to be charge neutral.  The kaon condensate  corresponds to
a classical field
\beq
K^- = v e^{-i \mu  t}\ ,
\eeq
where the kaon chemical potential $\mu = \mu_e$ equals the electron chemical potential since the $K^-$
also carries electric charge $-e$.  The amplitude $v$ is unknown.
Now plug this field into the chiral Lagrangian, and calculate the
neutron mass to $O(v^2)$, setting $m_u=m_d=0$.  One finds
\beq
\Delta M_N = (M_N(v) - M_N(0))  = \frac{|v|^2}{2f^2} \left(- \mu_e +(4 a_3 + 2
  a_2) m_s   \right)\ ,
\eqn{nmass}\eeq
where the negative first term is from the attractive Weinberg-Tomazawa
$s$-wave interaction.  It is possible to determine $a_2 m_s$ by
calculating the baryon octet mass splittings (problem (III.9)) with
the result
\beq
 a_2 m_s=\half(M_\Sigma-M_N )= 134\MeV\ .
\eeq
However, $a_3 m_s$ cannot be determined from the baryon spectrum, but
rather from pion-nucleon scattering.  The result is very uncertain.
The $a_3$ operator can be related to the dependence of the nucleon
mass on the strange quark mass:
\beq
m_s\frac{\partial M_N}{\partial m_s} = -2(a_2 + a_3) m_s\ .
\eqn{scont}\eeq
Some extractions have given this quantity to be $+350\MeV$,
corresponding to $(2 a_3 + a_2)m_s  = -450\MeV$. On the other hand, if the strange quark were very
heavy, then one could calculate the above quantity in perturbative
QCD, with the result $m_s\frac{\partial M_N}{\partial m_s}= (2/29)M_N
= +70\MeV$ --- which would correspond to $(2 a_3 + a_2)m_s  = -205\MeV$.
As for the size of $\mu_e$, it is typically $\sim 200\MeV$ in the
cores of neutron stars in conventional calculations.

So turning on a kaon vev $v$ lowers the neutron mass, an effect that
save more energy density the denser the neutrons.  On the other hand
it costs energy to make kaons because of their mass.  The net energy density
is then
\beq
m_K^2|v|^2 - \rho \Delta M_N\  = |v|^2\left( m_K^2 -\frac{\rho}{2f^2}\left(\mu_e -(4 a_3 + 2
  a_2) m_s  \right)\right)\ ,
\eeq
 and we see an instability at the critical neutron density
\beq
\rho_c = f^2\frac{m_K^2}{\mu_e/2-(2 a_3 + a_2) m_s}\ .
\eeq
For a crude calculation you can take nuclear density to equal
$\rho_0\simeq f^2 m_\pi$, so we get, if we ignore the $\mu_e$ term,
\beq
\frac {\rho_c}{\rho_0} \simeq -\frac{m_K^2}{m_\pi(2 a_3 + a_2) m_s}\ ,
\eeq
which ranges between $3.5$ and $8.5$ depending on what we take for
\eq{scont}.  However this simple calculation drops a lot of important
effects, such as (i) including the effects of the chemical potential
$\mu_e$, which lower $\rho_c$; (ii) 
with $K^-$ one can have nearly equal numbers of protons and neutrons
and still have charge neutrality --- this is energetically favorable due to the nuclear
symmetry energy, and will also lower $\rho_c$; (iii) spatial
correlations which we have neglected here reduce the kaon-nucleon
attraction and raise $\rho_c$.
  I think  
progress on the subject in the near term will come from a lattice
calculation of $a_3 m_s$.

\subsection{An EFT for pion-less nucleon-nucleon scattering}
\label{sec:4b}

We would like to go beyond mean field theory and at least set up an
honest, effective field theory formulation for nucleon-nucleon
interactions which could be used to compute properties of nuclear
matter in terms of a few phenomenological parameters which could be
fit to data.  This contrasts with the traditional approach of
constructing a nucleon-nucleon potential, tweaking it until it is
capable of fitting all of the nucleon-nucleon scattering phase shift
data, and then using that potential to perform an $N$-body
calculation, adding three-body interactions as needed to explain the
data. Both methods make use of phenomenological parameters to fit the
data, both methods are predictive in the sense that there are far more
data than free parameters.  So how would an effective field theory
treatment be any better than using a potential model? There are
several ways:
\begin{enumerate}[i.]
\item The EFT provides a natural framework for including known
  long-distance physics, such as one- and two-pion exchange,
  consistent with chiral symmetry;
\item Working in an EFT to a given order is equivalent to  including
  all the operators necessary to reproduce QCD to a given accuracy in
  a $p/\Lambda$ expansion.  In contrast, in a potential model one
  never knows whether the next observable one encounters will be
  calculable to the same accuracy with which the model fits previous
  observables.  If the effects of an operator in the EFT have been
  left out of the potential models, then the different potential
  models will disagree on observables, but they will fall on a curve 
the potential models have omitted a free parameter corresponding to a
certain operator in the EFT, then they will disagree among themselves,
but their predictions will fall upon a curve.  This curve represents
the arbitrary value the model assigns to a free parameter in the EFT.
An example is given below.

\item Unlike potential models, EFT never suffers from any ambiguity
  about ``on-shell'' and ``off-shell'' interactions.

\item Relativistic effects, such as time retardation are simple to
  include in an EFT, but not in a potential model.

\item Dynamical processes, scattering, inelastic collisions --- all
  of these are more simply treated in an EFT.

\end{enumerate}

For an effective field theory to be useful, however, it is necessary that
there exist a gap in the spectrum.  For momentum transfer far below
the pion mass, this is clearly the case:  then one can use a pion-less
EFT, where the only interactions are $n$-body nucleon contact
interactions (as well as electromagnetism).   I will start by
describing such a theory, which has received a lot of attention over
the past decade.

\subsection{The pion-less EFT for nucleon-nucleon interactions}
\label{sec:4c}

Given that the pion-less theory consists only of nonrelativistic
nucleons, the Lagrangian is quite simple. In $(4-D)$ dimensions it
takes the form
\beq
 \CL_{eff}&=&
{{  N}^\dagger} \left( i\partial_t + \nabla^2/2M\right)   N
\nonumber\\ &&+(\mu/2)^{4-D}\left[C_0 ({{  N}^\dagger}   N)^2
+ \frac{C_2}{ 8} \left[(  N   N)^\dagger( N {
\mathop\nabla^\leftrightarrow}{ }^2   N) +h.c\right]
+...\right]\ ,
\eqn{yukeff}
\eeq
where
\beq
 {  \mathop\nabla^\leftrightarrow}^2 \equiv  {  \mathop\nabla^\leftarrow}^2 -2
{ \mathop\nabla^\leftarrow \cdot \mathop\nabla^\rightarrow}+ {
\mathop\nabla^\rightarrow}^2\ .
\eeq
I have suppressed everywhere the spin and isospin indices and their contractions 
which project these 
interactions onto the various scattering channels ($\si$, $\siii$,
etc.).  The ellipsis indicates higher derivative operators,
and $(\mu/2)$ is an
arbitrary mass scale introduced to allow the couplings $C_{2n}$
multiplying operators containing $\nabla^{2n}$  to have the same dimension for
any $D$.
 I focus on  the $s$-wave channel
(generalization to higher partial waves is straightforward), and assume
that $M$ is very large so that relativistic effects can be ignored. The form of
the
$C_2$ operator is fixed by Galilean invariance, which implies that when all
particle
momenta are boosted $\bfp \to \bfp + M\bfv$, the Lagrangian must remain
invariant.
There exists another two derivative operator for $p$-wave scattering which I
will not be discussing. 


The usual effective field theory expansion requires one to
identify a class of diagrams to sum 
 which gives the amplitude $i\CA$ to the desired order in a
 $p/\Lambda$ expansion.  For nonrelativistic scattering, the
 scattering amplitude is related to the $S$-matrix by 
\beq
S= 1 + i \frac{Mp}{ 2\pi}\CA\ ,
\eeq
where $p=\sqrt{ME_\text{cm}}$ is the magnitude of the momentum that
each nucleon has in the center of momentum frame. For $s$-wave
scattering, $\CA$ is related to the phase shift $\delta$ 
by
\beq
\CA = \frac{4\pi}{ M}\frac{1}{ p\cot\delta -i p}\ .
\eqn{amp}
\eeq
However, it is well known that for a short-range two-body potential
$V(r)$ that dies
off exponentially fast for $r \Lambda>1$,
  it is not $\CA$ which in general has a good Taylor expansion in $p/\Lambda$, but rather the
quantity
$p\cot\delta$ which is expanded as:
\beq
p\cot\delta = -\frac{1}{ a} + \frac{1}{ 2}\Lambda^2\sum_{n=0}^\infty {r}_n
\(\frac{p^2}{ \Lambda^2}\)^{n+1}\ .
\eqn{erexp}
\eeq
This is called the effective range expansion, were $a$ is the scattering length, and
$r_0$ is the effective range. At best, our effective theory for
nucleon-nucleon scattering should reproduce the effective range
expansion.  Of course, if this were the whole story, it would be boring,
reproducing well known results! What will make it more interesting is
when we incorporate electromagnetic and weak interactions into the
theory. But first we need to understand the power counting of our
EFT, since Feynman diagrams give one the amplitude $\CA$ and not the quantity $p\cot\delta$.
A Taylor expansion of $\CA$ in powers of $p$ yields
\beq
\CA = -\frac{4\pi a}{ M }\[ 1 - i a p +(ar_0/2-a^2) p^2 + O(p^3/\Lambda^3)\]\ ,
\eqn{aexp}
\eeq%

For a generic short-range potential,
the coefficients  $r_n$ are generally $O(1/\Lambda)$ for all $n$. However,  $
a$ can
take on any value, which is problematic since for $1/|a|>\Lambda$, the
above momentum expansion of the amplitude has a radius of convergence set by
$1/|a|$ and not by $\Lambda$.
A general property of the scattering length is that  $1/a$ is negative for a weakly
attractive potential, vanishes for a more attractive potential which
possesses a 
bound state at threshold, and becomes positive for an even more
attractive potential with a deep bound state. (For example: if one considers an attractive Yukawa
potential for the form
\beq
V(r) = -\frac{g^2}{4\pi} \frac{e^{-\Lambda r}}{r}
\eeq
then a bound state at threshold appears for the critical coupling
$\eta \equiv g^2 M/(4 \pi \Lambda)\simeq 1.7$, at which point the
scattering length $a$ diverges.)
 First I consider the situation where the
scattering length is of natural
size $|a|\sim 1/\Lambda$, and then I  discuss the case
$|a|\gg 1/\Lambda$, which is   relevant for realistic $NN$ scattering.

\subsubsection{The case of a  ``natural'' scattering length:
  $1/|a|\simeq \Lambda$}

In the regime $|a|\sim  1/\Lambda$ and $|r_n|\sim  1/\Lambda$,
the expansion  of the amplitude $\CA$  in \eq{aexp}
converges up to momenta $p\sim \Lambda$, and it  is this
expansion that we 
wish to reproduce in an effective field theory.

The complete tree level $s$  partial wave amplitude in the center of mass
frame arising from $\CL_{eff}$ is
\beq
i \CA_{\rm tree}^{(cm)} = -i(\mu/2)^{4-D} \sum_{n=0}^\infty C_{2n}(\mu) p^{2n}\
,
\eqn{tree}
\eeq
where the coefficients $C_{2n}(\mu)$ are  the
couplings in the Lagrangian of operators with $2n$ gradients contributing to
$s$-wave scattering.  One may always
trade time derivatives for spatial gradients, using the equations of motion
when computing
$S$-matrix elements, and so I will ignore such operators.

\begin{figure}[t]
\centerline{\epsfxsize=4.5 in \epsfbox{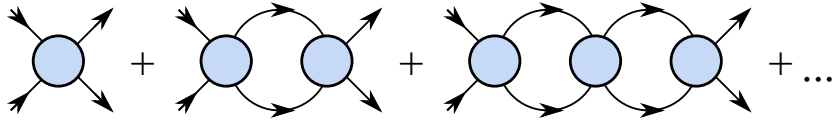}}
\noindent
\caption{\it The bubble chain arising from local operators. 
The vertex is given by the tree
level amplitude, \eq{tree}.}
\label{bubbles}
\vskip .2in
\end{figure}

Beyond tree level one encounters the loop diagrams  shown in
Fig.~\ref{bubbles}.  Formally, these are {\it all} the diagrams one
encounters in a nonrelativistic theory...if you cut the diagrams in
half somewhere in the middle, you can only encounter the two original
particles and no additional  particle-antiparticle pairs.  The loop
integrals one encounters are all of the form

\beq
\openup3\jot
I_n&\equiv& i(\mu/2)^{4-D} \int \frac{{\rm d}^D q}{ (2\pi)^D}\, \frac{{\bf q}^{2n}
}{\( E/2 + q_0 -\frac{{\bf q}^2}{ 2M} + i\epsilon\)\( E/2 - q_0
-\frac{{\bf q}^2}{ 2M} + i\epsilon\)}
\nonumber\\
&=& (\mu/2)^{4-D} \int \frac{{{\rm d}}^{(D-1)}{\bf  q}}{ (2\pi)^{(D-1)}}\,
{\bf q}^{2n} \(\frac{1}{ E  -{\bf q}^2/M + i\epsilon}\)
\nonumber\\
&=& -M (ME)^n (-ME-i\epsilon)^{(D-3)/2 } \Gamma\(\frac{3-D}{ 2}\)
\frac{(\mu/2)^{4-D}}{  (4\pi)^{(D-1)/2}}\ .
\eqn{loopi}
\eeq

In order to define the theory, one must specify a subtraction scheme; different
subtraction
schemes amount to a reshuffling between contributions from
the vertices and contributions from the the UV part of the
loop integration.  How does one choose a subtraction scheme that is useful?
I am considering the case $|a|, |r_n|\sim 1/\Lambda$, and wish to reproduce the
expansion of
the amplitude \eq{aexp}.  In order to do this via Feynman diagrams, it is
convenient if any Feynman
graph with a particular set of operators at the vertices only contributes to
the expansion of the
amplitude at a particular order.  Since the the expansion \eq{aexp} is a strict
Taylor expansion in $p$, it is
it is therefore very convenient if each Feynman graph yields a simple
monomial in $p$.  Obviously,
this won't be true in a random subtraction scheme.
 A scheme that fulfills this criterion is the minimal subtraction
scheme ($\ms$) which
amounts to subtracting any
$1/(D-4)$ pole before taking the $D\to 4$ limit.
As the integral \eq{loopi} doesn't exhibit any such poles, the result in $\ms$
is simply
\beq
I_n^{\ms}=  (ME)^n \(\frac{M}{ 4\pi}\)\sqrt{-ME-i\epsilon}=
-i\(\frac{M}{ 4\pi}\) p^{2n+1}  \ .
\eqn{inval}
\eeq
Note the  nice feature of this scheme that the factors of $q$ inside the loop
get converted
to factors of $p$, the external momentum.  Similarly, a factor of the equations
of motion,
$i\partial_t +\nabla^2/2M$, acting on one of the internal legs at the vertex,
causes the
loop integral to vanish.
Therefore one can use the on-shell,
tree level amplitude \eq{tree} as the internal  vertex in loop diagrams.
Summing the bubble diagrams in the center of mass frame gives
\begin{eqnarray}
\CA & = & -\frac{  \sum C_{2n} p^{2n}
}{
1 + i(M p/4\pi) \sum C_{2n} p^{2n} }
\ .
\end{eqnarray}
 Since for this process there are no poles at $D=4$ in the $\ms$
 scheme, the coefficients 
$C_{2n}$ are
independent of the subtraction
point $\mu$.
The power counting in the $\ms$ scheme is particularly simple, as promised:
\begin{enumerate}
\item{Each propagator counts as $1/p^2$;}
\item{Each loop integration $\int {\rm d}^4q$ counts as $p^5$ (since $q_0\sim
\bfq^2/2M$);}
\item{Each vertex $C_{2n}\nabla^{2n}$ contributes $p^{2n}$.}
\end{enumerate}
The amplitude may be expanded in powers of $p$ as
\beq
\CA=\sum_{n=0}^\infty \CA_n
\  ,\qquad
\CA_n\sim O(p^n)
\eqn{ampexpand}
\eeq
where the $\CA_n$ each arise from graphs with $L\le n$ loops and can be equated 
to the low
energy scattering data \eq{aexp} in order to fit the $C_{2n}$ couplings.  In
particular,
$\CA_0$ arises from the tree graph with $C_0$ at the vertex; $\CA_1$ is given
by the
1-loop diagram with two $C_0$ vertices; $\CA_2$ is gets contributions from both
 the 2-loop diagram with
three $C_0$ vertices, as well as the tree diagram with one $C_2$ vertex, and so
forth.
Thus the first three terms are
\beq
\CA_0= -C_0\ ,\qquad
\CA_1= i C_0^2\frac{Mp}{ 4\pi}\ ,\qquad
\CA_2=  C_0^3\(\frac{Mp }{4\pi}\)^2-C_2p^2
\ .
\eqn{effexp}
\eeq
Comparing \eqsii{aexp}{effexp} I find for the first two couplings of the
effective theory
\beq
C_0 = \frac{4\pi a}{ M}\ ,\qquad C_2 = C_0 \frac{a r_0}{ 2}
\ .
\eqn{cfit}
\eeq
In general, when the scattering length has natural size,
\beq
C_{2n} \sim \frac{4\pi}{ M\Lambda} \frac{1}{ \Lambda^{2n}}
\ .
\eeq
Note that the effective field theory calculation in this scheme
is completely perturbative even though the underlying short-distance
physics need not be. Also note that our choice of subtraction scheme
 ($\ms$), while not changing the physics, made the power counting
 transparent. A feature of the fact that we are computing consistently
 to a given order in momentum is that fact that our results are
 independent of the renormalization scale $\mu$.

\subsubsection{The realistic case of an  ``unnatural'' scattering length}

One might guess that the results of the previous section would apply
to low energy $NN$ scattering, with role of $\Lambda$ played by $m_\pi$ or
$m_\pi/2$.  However, while it is true that the pion is the lightest
hadron exchanged between nucleons, the EFT is much more interesting
than the above scenario, as the $NN$ scattering lengths are
unnaturally large.  For example, the $\si$ scattering length is
$a_0=-23.714\pm.013\fm\simeq 1/(8\MeV) $, which is {\it much} bigger
than $1/m_\pi \simeq 1/(140\MeV)$.

For a nonperturbative interaction  with
a bound state near threshold, the  expansion of $\CA$ in powers of $p$
is of little practical value, as it breaks down for momenta $p\gtrsim 1/|a|$,
far below $\Lambda$.
In the above effective theory, this occurs because the couplings $C_{2n}$
are anomalously large, $C_{2n}\sim 4\pi a^{n+1} / M\Lambda^n$.
However, the problem is
not with the effective field theory method,
but rather with the subtraction scheme chosen.

Instead of reproducing the expansion of the amplitude shown in \eq{aexp}, one
needs
to expand in powers of $p/\Lambda$ while retaining $ap$ to all orders:
\beq
\CA = -\frac{4\pi}{ M}\frac{1}{ (1/a + i p)}\[ 1 + \frac{r_0/2}{ (1/a + ip)}p^2 +
\frac{(r_0/2)^2}{ (1/a + ip)^2} p^4 + \frac{(r_1/2\Lambda^2)}{ (1/a + ip) } p^4
+\ldots\]
\eqn{aexp2}
\eeq
Note that  for $p>1/|a|$ the terms  in this expansion scale as $\{p^{-1},
p^0,p^1,\ldots\}$.
Therefore, the expansion in the effective theory should take the form
\beq
\CA=\sum_{n=-1}^\infty \CA_n
\ ,\qquad
\CA_n\sim O(p^n)
\eqn{ampbiga}
\eeq
beginning at $n=-1$ instead of $n=0$, as in the expansion \eq{ampexpand}.
Comparing with \eq{aexp2}, we see that
\beq
\CA_{-1} &=&  -\frac{4\pi}{ M}\frac{1}{ (1/a + i p)}\ ,\nonumber\\
\CA_{0} &=&  -\frac{4\pi}{ M}\frac{r_0p^2/2}{ (1/a + i p)^2}\ ,
\eqn{ami}
\eeq
and so forth.
Again, the task is to compute the $\CA_n$ in the
effective theory, and  equate to the appropriate expression
 above, thereby fixing the
$C_{2n}$ coefficients.  As before, the goal is actually more ambitious:  each
particular
graph contributing to $\CA_n$ should be $O(p^n)$, so that the power counting is
transparent.

 As any single diagram in the effective theory
is proportional to positive powers of $p$, computing the leading term
$\CA_{-1}$ must involve
summing an infinite set of diagrams. It is easy to see that the leading term
$\CA_{-1}$
can be reproduced by the sum of bubble diagrams with $C_0$ vertices
which yields in the $\ms$ scheme
\beq
{\cal A}_{-1} = \frac{ -C_0}{ \left[1 + \frac{C_0 M}{ 4\pi} ip\right]}\ .
\eqn{aminus1}\eeq
Comparing this with \eq{ami} gives $C_0=4\pi a/M$, as in the
previous section.  However, there is no  expansion parameter that justifies
this summation:
each individual graph in the bubble sum goes as $C_0 (C_0 M p)^L\sim(4\pi
a/M)(i a p)^L$, where $L$ is the number
of loops. Therefore each graph in the bubble sum is bigger than the preceding
one, for $|ap|>1$,
while they sum up to something small.

This is an unpleasant situation for an effective field theory;  it is important
to have an
expansion parameter so that one can  identify the order of any particular
graph, and sum the
graphs consistently.
Without such an expansion parameter, one cannot determine the size of omitted
contributions,
and one can end up retaining certain graphs  while dropping operators
needed to renormalize those graphs.  This results
in a model-dependent description of the short distance physics,
as opposed to a proper effective field theory calculation.

Since the sizes
of the contact interactions depend on the renormalization scheme one uses, the
task becomes one of identifying the appropriate subtraction scheme that makes
the
power counting simple and manifest.
The $\ms$ scheme fails on this point;  however  this is not a
problem with dimensional regularization, but rather a problem with the
minimal subtraction scheme itself. A momentum space subtraction at threshold
behaves similarly.

Consider an alternative regularization and renormalization
scheme, namely to using a momentum cutoff equal to $\Lambda$.
Then for large $a$ one finds $C_0 \sim (4\pi/M\Lambda)$, and each additional
loop contributes
a factor of  $C_0(\Lambda + ip)M/4\pi \sim (1+ip/\Lambda)$.
The problem with this scheme is that for $\Lambda\gg p$ the term $ip/\Lambda$
from the loop is small relative to the $1$,
and ought to be ignorable;
however, neglecting it would fail to reproduce the desired result \eq{ami}.
This scheme suffers from
significant cancellations between terms, and so once again the power counting
is not manifest.

Evidently, since $\CA_{-1}$ scales as $1/p$, the desired expansion
would have each individual graph contributing to $\CA_{-1}$ scale as $1/p$.
As the tree level contribution is $C_0$, I must therefore  have
$C_0$ be of size $\propto 1/p$, and each additional loop must be $O(1)$.
This can be achieved by using  dimensional regularization and
 the \pds (power divergence  subtraction)
scheme. 
The \pds scheme involves subtracting from the
dimensionally regulated loop integrals not only the $1/(D-4)$ poles
corresponding
to log divergences, as in $\ms$, but also
poles in lower dimension which correspond to power law divergences at $D=4$.
The integral $I_n$ in
 \eq{loopi}\ has a pole in $D=3$ dimensions which can be removed by adding to
$I_n$
the counterterm
\beq
\delta I_n = -\frac{M(ME)^n \mu}{ 4\pi (D-3)},
\eeq
so that the subtracted integral in $D=4$ dimensions is
\beq
I_n^{PDS} = I_n + \delta I_n = - (ME)^n \left(\frac{M}{ 4\pi}\right) (\mu + ip).
\eqn{ipds}
\eeq
In this subtraction scheme
\begin{eqnarray}
\CA & = & -\frac{M}{ 4\pi}\[
\frac{4\pi}{ M  \sum C_{2n} p^{2n} } + \mu+ip\]^{-1}
\ .
\eqn{answer}
\end{eqnarray}
By performing a Taylor expansion  of the above expression,
and comparing with \eq{aexp2}, one finds that
for $\mu\gg 1/|a|$,
the couplings $C_{2n}(\mu)$ scale as
\beq
C_{2n}(\mu) \sim \frac{4\pi}{ M \Lambda^n \mu^{n+1}}\ .
\eqn{cscale}
\eeq
Eqs.~(\ref{eq:answer},\ref{eq:cscale}) imply that the appropriate
power counting entails   $\mu \sim p$, $C_{2n}(\mu)\sim 1/p^{n+1}$.
This is very different than the example of the ``natural'' scattering
length discussed in the previous section; the strong interactions that
give rise to a large scattering length have significantly altered the
scaling of all the operators in the theory. 
A factor of $\nabla^{2n}$ at a vertex scales as $p^{2n}$, while each loop
contributes a factor of
$p$. The power counting rules for the case of large scattering length are
therefore:
\begin{enumerate}
\item{Each propagator counts as $1/p^2$;}
\item{Each loop integration $\int {\rm d}^4q$ counts as $p^5$;}
\item{Each vertex $C_{2n}\nabla^{2n}$ contributes $p^{n-1}$.}
\end{enumerate}
We see that this scheme avoids the problems encountered with the choices of the
$\ms$ ($\mu=0$) or momentum
cutoff ($\mu\sim \Lambda$) schemes.
First of all, a tree level diagram with a $C_0$ vertex is $O(p^{-1})$, while
each loop with a $C_0$ vertex contributes $C_0(\mu) M (\mu+ip)/4\pi\sim 1$.
Therefore  each term in the bubble sum contributing
to $\CA_{-1}$ is  of order $p^{-1}$, unlike the case for $\mu=0$.  Secondly,
since $\mu\sim p$,
it makes sense keeping  both the $\mu$ and the $ip$ in \eq{ipds} as they are of
similar size, unlike
what we found in the $\mu=\Lambda$ case.
The \pds scheme  retains the nice feature of $\ms$ that
powers of $q$ inside the loop.

\begin{figure}
\centerline{\epsfysize=3 in \epsfbox{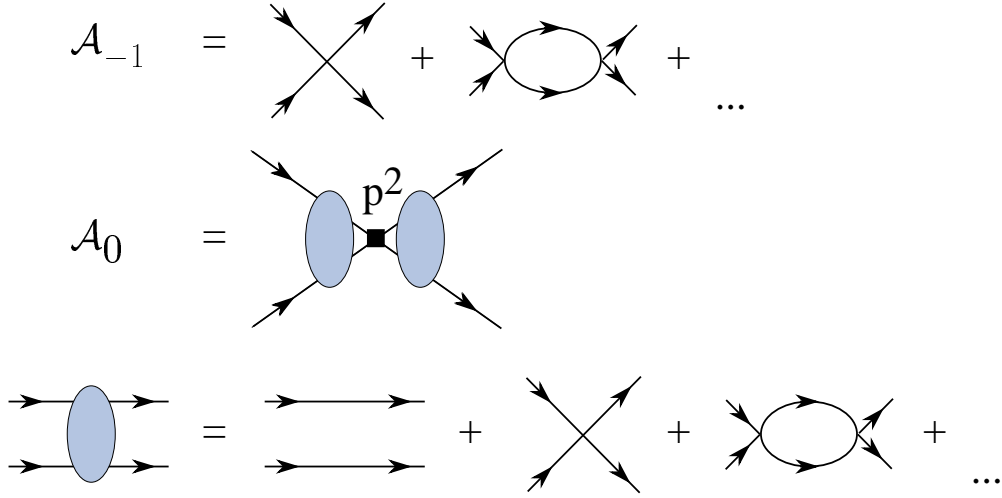}}
\noindent
\caption{\it Leading and subleading contributions 
arising from local
operators.  The unmarked vertex is the $C_0$ interaction, which is
summed to all orders; the one
marked ``$p^2$'' is the $C_2$ interaction, etc.}
\label{FG1S0_m1}
\vskip .2in
\end{figure}

Starting from the above counting rules (proposed in
\cite{Kaplan:1998tg,Kaplan:1998we} and referred to in the literature as ``KSW''
counting)  one finds that the leading order
contribution to the scattering amplitude $\CA_{-1}$
scales as $p^{-1}$ and consists of the sum of bubble diagrams with $C_0$
vertices;
contributions to the amplitude scaling as higher powers of $p$ come from
perturbative
insertions of derivative interactions, dressed to all orders by $C_0$.  The
first three
terms in the expansion are
\beq
{\cal A}_{-1}&=& \frac{ -C_0}{ \left[1 + \frac{C_0 M}{ 4\pi} (\mu + ip)\right]}\
,\nonumber\\
{\cal A}_0    &=& \frac{ -C_2 p^2}{ \[1 + \frac{C_0 M}{ 4\pi}(\mu + ip)\]^2}\
,\nonumber\\
{\cal A}_1    &=& \(\frac{ (C_2 p^2)^2M(\mu+ ip)/4\pi}{ \[1 + \frac{C_0 M}{
4\pi}(\mu + ip)\]^3}
-\frac{ C_4 p^4}{ \[1 + \frac{C_0 M}{ 4\pi}(\mu + ip)\]^2}\) \ ,
\eqn{athree}
\eeq
where the first two correspond to the Feynman diagrams
in Fig.~\ref{FG1S0_m1}.  The third term, $\CA_1$, comes from  graphs with
either one insertion of $C_4\nabla^4$ or two insertions of $C_2\nabla^2$,
dressed to all orders by the $C_0$ interaction.

Comparing \eq{athree} with the expansion of the amplitude \eq{aexp2},  the
couplings $C_{2n}$ are related
to the low energy scattering data $a$, $r_n$: 
\beq
C_0(\mu) &=& \frac{4\pi}{ M }\(\frac{1}{ -\mu+1/a}\)\ ,\nonumber\\ 
C_2(\mu) &=&  \frac{4\pi}{ M }\(\frac{1}{ -\mu+1/a}\)^2 \frac{{ 
r}_0}{ 2}\ ,\nonumber\\
C_4(\mu) &=&  \frac{4\pi}{ M }\(\frac{1}{ -\mu+1/a}\)^3 \[\frac{1}{ 4}
{ r}_0^2 + \frac{1}{ 2} \frac{{ r}_1}{\Lambda^2} \({-\mu+1/a}\)\]\ . 
\eqn{cvals}
\eeq
Note that assuming $r_n\sim 1/\Lambda$, these expressions are 
consistent with the scaling law in \eq{cscale}.

\subsubsection{Beyond the effective range expansion}

\begin{figure}[t]
\centerline{\epsfxsize=3.5 in \epsfbox{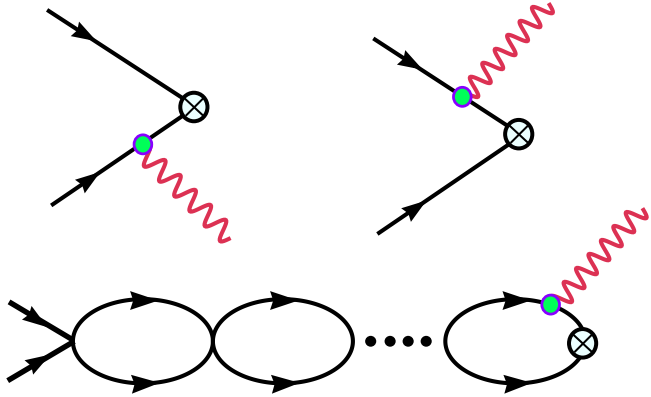}}
\noindent
\caption{\it The leading order contribution to $np\to d\gamma$.  Solid
  lines denote nucleons, wavy lines denote photons.  The photon
  coupling is through the nucleon anomalous  magnetic moment operator
  in $ \CL_B$; the resummed unmarked vertex is the $C_0$ interaction. The crossed circle represents and insertion of the
  deuteron interpolating field $\CD_i$. The bubble chain without
  photon insertions (not shown) is used to compute the wave function
  renormalization $Z$, and to fit $C_0$ to get the correction deuteron
binding energy. See \cite{Kaplan:1998sz,Kaplan:1998xi,Chen:1999tn}.}
\label{fig:lonpdg}
\vskip .2in
\end{figure}

So far, we have developed an elaborate machinery to just reproduce the
effective range expansion! The payoff comes when one includes
electromagnetic and weak interactions.  The example I will briefly describe the
application of the pion-less effective theory to 
here is the application of the pion-less effective theory to radiative
capture process $np\to d \gamma$.  At leading order, the ingredients
to the calculation are the following:
\begin{enumerate}[i.]
\item One starts with the the nucleon kinetic two-nucleon $C_0$ interaction for the
  $\siii$ channel, written as
\beq
\CL = \ldots -C_0(N^T P_i N)^\dagger (N^T P_iN)\ ,
\eqn{3sl}
\eeq
where $N$ is the nucleon doublet, and $P_i$ is the projection operator
onto the $\siii$ channel:
\beq
P_i = \frac{1}{\sqrt{8}} \sigma_2 \sigma_i \tau_2\ ,\quad \Tr P_i
P_j=\half\delta_{ij}\ ,
\eqn{projop}
\eeq
where the $\sigma_i$ act on spin and the $\tau_i$ act on isospin.
\item One uses the convenient interpolating field $\CD_i(x)\equiv N^T
  P_i N(x)$ to be the operator that creates a deuteron at the point
  $x$. The coupling $C_0$ can be fixed by ensuring that the pole in
  $\CA_{-1}$ occurs at the deuteron binding energy.  The leading order
  wave function normalization $Z$ is extracted by looking at the
  residue at the pole. ($\sqrt{Z}$ is just the amplitude for our
  operator $\CD_i$ to create a physical deuteron.)  
\item $np\to d \gamma$ occurs by emitting a magnetic photon, and so
  one needs to include in the Lagrangian the anomalous magnetic moment interaction of
  the nucleons:
\beq
\CL_B = \frac{e}{2M_N} N^\dagger(\kappa_0 + \kappa_1 \tau_3)
{\boldsymbol\sigma}\cdot {\bf B} N\ ,
\eeq
where $\kappa_0=\half (\kappa_p + \kappa_n)$ and $\kappa_1 =
\half(\kappa_p-\kappa_n)$  are the isoscalar and isovector nucleon
magnetic moments with $\kappa_p = 2.79$, $\kappa_n = -1.91$. 

\item Then at leading order one sums up the bubble chain with one
  insertion of the magnetic moment operator, as shown in
  Fig.~\ref{fig:lonpdg}.
\end{enumerate}

From these graphs one finds the capture cross section
\beq
\sigma=\frac{8\pi\alpha\gamma^5\kappa_1^2 a_0^2 }{v M_N^5}
\left(1-\frac{1 }{ \gamma a_0} \right)^2, 
\eeq
where $\alpha$ is the fine structure constant and $v$ is the magnitude of the
neutron velocity (in the proton rest frame), $a_0=-23.714\pm.013\fm$
is the $\si$ scattering length and $\gamma\equiv\sqrt{M_N B}$, where
$b$ is the deuteron binding energy.
This agrees with old results of Bethe and Longmire 
when terms in their expression involving the effective range (which
are higher order in our expansion) are
neglected.

\begin{figure}[t]
\centerline{\epsfxsize=2.5 in \epsfbox{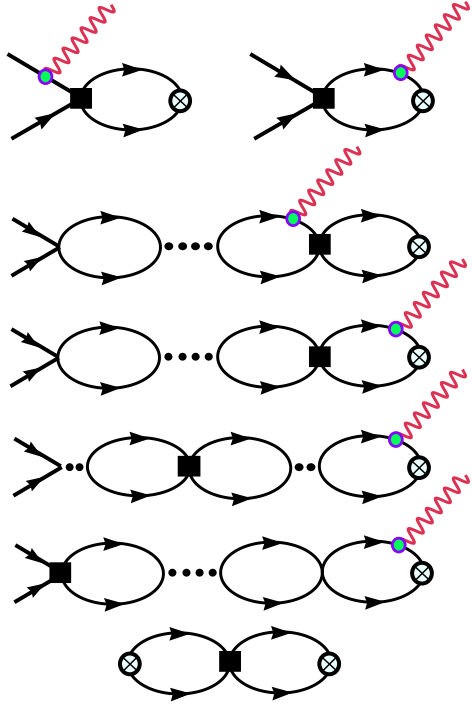}}
\noindent
\caption{\it The   graphs contributing to  $np\to
  d\gamma$ at NLO. The black square corresponds to an insertion of a $C_2$
  interaction, the circle to the nucleon anomalous magnetic moment,
  and the resummed unmarked vertex to the $C_0$ interaction.   The last graph is the contribution to wave function
  renormalization at this order. Figure from ref.~\cite{Chen:1999tn}.}
\label{fig:nlonpdg}
\end{figure}

At next-to-leading order (``NLO'') one needs to sum all relevant
diagrams involving a single insertion of a $C_2$
vertex (that is, a 2-derivative contact interaction, whose value is
fit to the experimental effective range in $NN$ scattering) for both
the $\si$ and $\siii$ channels,  as in 
Fig.~\ref{fig:nlonpdg}.

\begin{figure}[t]
\centerline{\epsfxsize=3.0 in \epsfbox{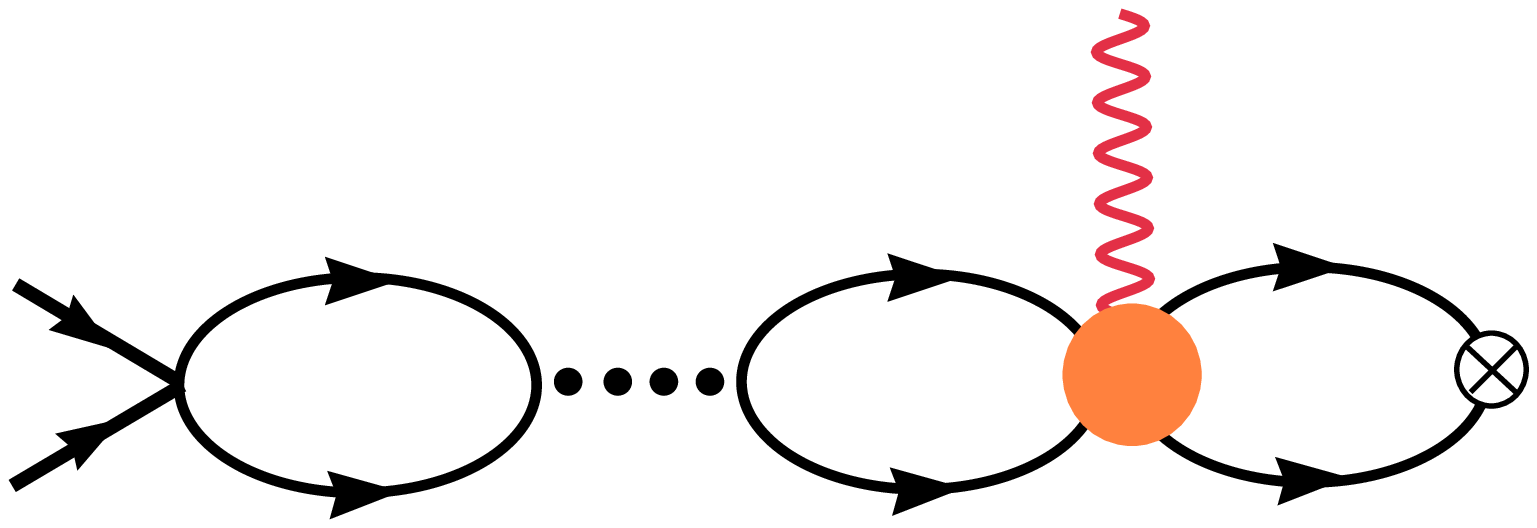}}
\noindent
\caption{\it An additional graph at NLO including an insertion of the
  $L_1$ operator. From ref.~\cite{Chen:1999tn}, courtesy of M. Savage.}
\label{fig:nlonpdgL1}
\vskip .2in
\end{figure}

 However this is not all.  At the same order one finds a new contact
 interaction which cannot be fit to $NN$ scattering data.  It is a
 2-body interaction with a magnetic photon attached, involving a new
 coupling constant $L_1$:
\beq
\CL_{L_1} = e L_1 (N^T P_i N)^\dagger (N^T  P_3 N) B_i\ .
\eeq
 A gauge
field is power counted the same as a derivative, and so the $B$ field
counts as two spatial derivatives. Thus graphs with one $L_1$
insertion and an infinite number of $C_0$ insertions comes in at the
same order as the $\kappa$ $\gamma NN$ vertex  summed with an infinite
number of $C_0$ vertices and one $C_2$ insertion. So at NLO one needs
also to include the graph in Fig.~\ref{fig:nlonpdgL1}.

\begin{figure}[t]
\centerline{\epsfxsize=3.0 in \epsfbox{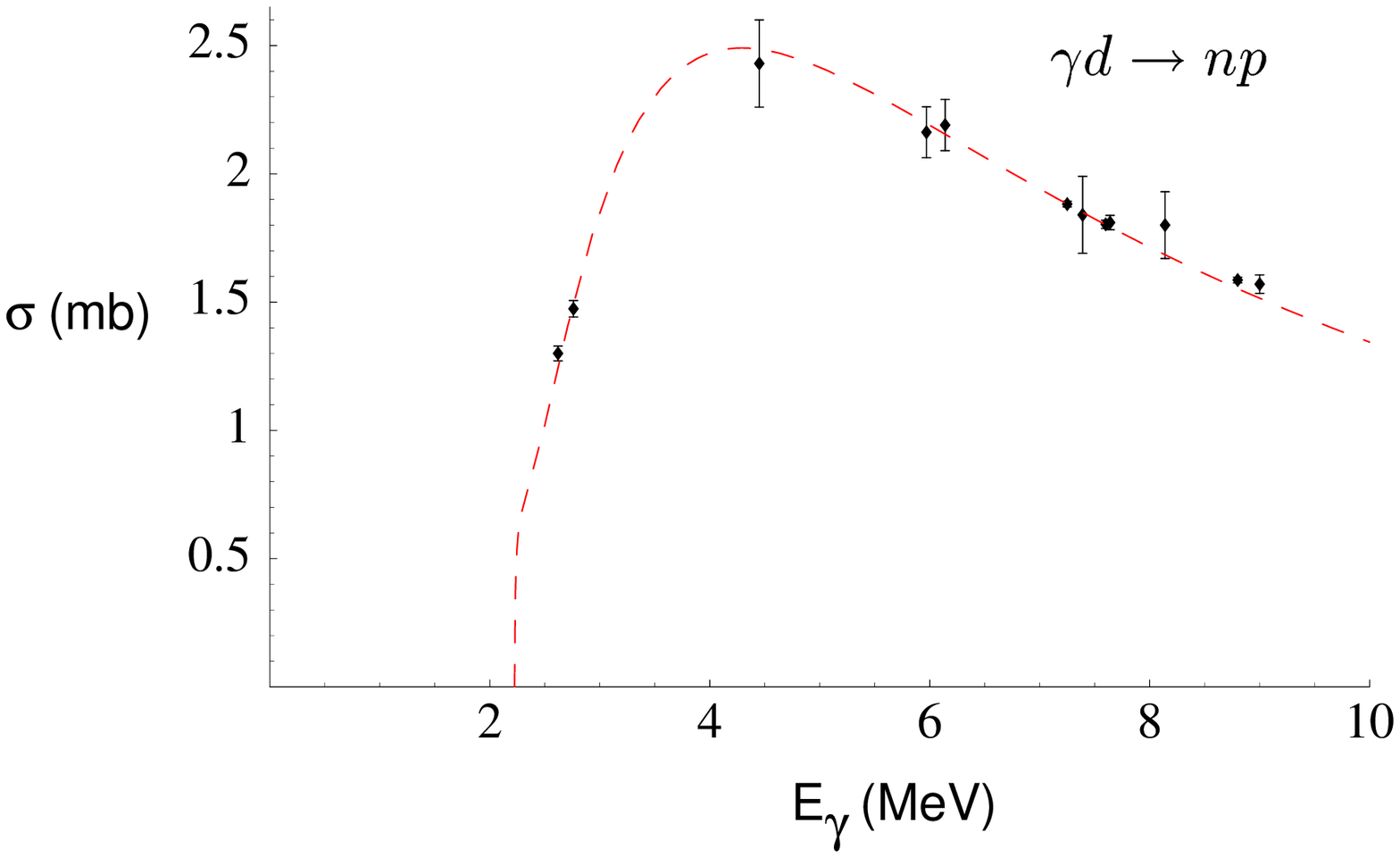}}
\noindent
\caption{\it Cross section for $\gamma d\to n p $ breakup as a
  function of photon energy $E_\gamma$.  The dashed line is the
  theoretical calculation of ref.~\cite{Rupak:1999rk}, as is this
  figure, courtesy of G. Rupak.   The data are from 
  ref. \cite{dbreakup} }
\label{fig:breakup}
\vskip .2in
\end{figure}

The NLO result deviates from the old effective range calculations,
since the $L_1$ operator is a completely new ingredient, and it also
changes the dependency of the answer upon the effective range. This new
coupling $L_1$ can be fit to data at one particular neutron
velocity, and then one has a highly accurate prediction for
neutron capture at any low velocity.  The state of the art is
presently an $N^4LO$ calculation for the related breakup process
$\gamma d\to n p$ by Gautam Rupak \cite{Rupak:1999rk}.  His results are shown in
Fig.~\ref{fig:breakup}.

\subsubsection{Few nucleon systems}

Extension of the pion-less effective  theory to systems with more than
two nucleons is a very elegant and interesting subject, pioneered by
Bedaque, Hammer and Van Kolck
\cite{Bedaque:1997qi,Bedaque:1998mb,Bedaque:1998kg} (for a more recent
review, see 
\cite{Bedaque:2002mn}).  A fascinating result of the analysis is that
$Nd$ scattering in the $j=\frac{3}{2}$ channel is well described at
leading order by summing up two-body interactions, along the lines
described in the previous section.  However in the $j=\frac{1}{2}$
channel, already at leading order a 3-body contact interaction is
needed to renormalize the scattering amplitude. Furthermore, the
strength of this interaction exhibited limit-cycle behavior as a
function of the momentum cutoff.  

\begin{figure}[t]
\centerline{\epsfysize=6cm\epsfbox[73 456 530 700]{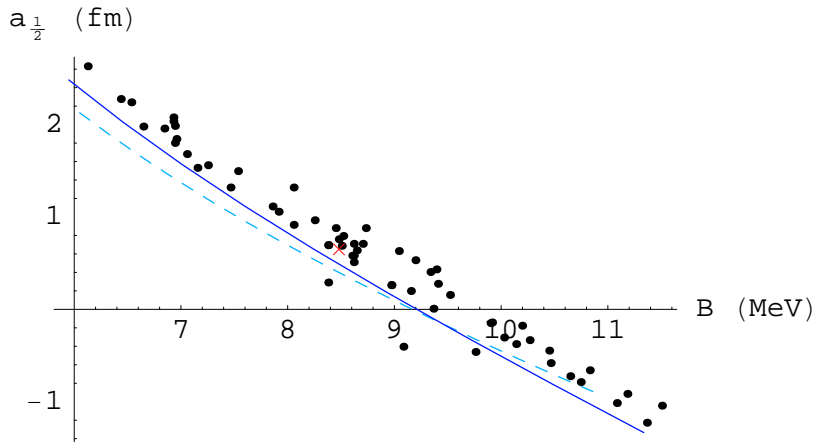}}
\caption{{\it Correlation between the $j=\half$ $s$-wave $Nd$ scattering
  length and the triton binding energy (Phillips line): predictions of
various potential models (black dots), EFT in LO (light dashed line)
and NLO (dark solid line), varying the 3-body contact interaction.
The experimental value is marked by the red cross. From
\cite{Bedaque:2002mn}, courtesy of the authors.}}
\label{fig:phillips}
\vskip .2in
\end{figure}

Unfortunately, I do not have time to discuss
it, but I did want to show one plot from the review, showing the
so-called Phillips line, in Fig.~\ref{fig:phillips}. Plotted here is a
plot of the $j=\half$ $nd$ scattering length, versus the triton
binding energy in MeV.  Plotted as black dots are the results from
numerous potential models.  They evidently fall along a rough  curve, called
the ``Phillips line''.  Also plotted are the LO and NLO results from
the pion-less effective field theory; these calculations require a
counterterm for a 3-body operator at leading order for the
scattering amplitude to be made
finite. The residual finite part of this interaction must therefore be
fit to data.  In Fig.~\ref{fig:phillips} the EFT results are shown
for a continuous range of this coupling constant for the 3-body force,
generating curves which lie close to the black dots.  The
interpretation is evident:  without realizing it, the different
potential models have assigned different, and essentially random
values to the 3-body force\footnote{Note that sequential 2-body
  interactions at short distance can be equivalent to  a 3-body
  interaction when viewed with low resolution.  So even if every model
  had included no fundamental 3-body interaction, since they all have
  different 2-body interactions at short distance, they would still be
  scattered over the Phillips line.}, hence the one-parameter spread
in results.  The figure also makes it clear that by appropriately
choosing the value for this 3-body force, the NLO EFT calculation will
agree very well 
with experiment, lying at the closest approach of the solid curve to
the red cross.  This plot is an excellent advertisement for why
effective field theory is a good tool for low energy nuclear physics.

\subsection{Including pions in the EFT for nuclear physics}
\label{sec:4d}

The original suggestion for applying effective field theory to nuclear
physics was due to Weinberg \cite{Weinberg:1990rz,Weinberg:1991um}.
His idea was use the chiral Lagrangian for meson-nucleon interactions
discussed in the previous lecture,
supplemented with multi-nucleon operators.  Realizing that the system
was nonperturbative, he advocated performing a straightforward chiral
expansion of the nucleon-nucleon potential to the desired order, and
then solving the 
Shr\"odinger equation using that potential. In Feynman diagrams,
latter step is equivalent to summing up ladder diagrams where the
rungs of the ladder are interactions between the nucleons via the
potential. 
For the expansion of the potential, the chiral power counting
Weinberg proposed was the usual one, basically just
counting inverse powers of $\Lambda$ or $f_\pi$, where $C_0\sim
1/f_\pi^2$ entered at the same order as one-pion exchange, which has
the form 
\beq
\frac{g_A^2}{f_\pi^2}\,\frac{({\bf q}\cdot\boldsymbol
  \sigma)^2}{{\bf q}^2+m_\pi^2}\ .
\eqn{ope}
\eeq

It was soon discovered that while the perturbative chiral expansion of the
potential was well defined, the ladder sum of the potential expanded
to a given order required
counterterms at arbitrarily high order in the chiral expansion
\cite{Kaplan:1996xu}.  This is equivalent to saying that Weinberg's
method sums up a subset of diagrams arbitrarily high in the chiral
expansion. This may be OK if that subset does not affect significantly
the final answer, but in that case, why sum the higher order effects
at all?  If they are important, then one has to justify the exclusion
of other terms higher order in the chiral expansion, or else what one
has is just another model for nuclear physics, and not a sensible  EFT.

To avoid this problem, 
the  KSW power counting scheme was proposed
\cite{Kaplan:1998tg,Kaplan:1998we}, which was introduced in the
previous section for the pion-less theory; one-pion
exchange enters  at $O(p^0)$ (as makes sense from the form of \eq{ope}, which
has a $q^2$ both in the numerator and the denominator) at the same
order as the $C_2$ interaction. Because this is a consistent scheme,
results are independent of renormalization scale at any order of the
expansion.  It is also a theoretically appealing scheme because pions
are treated perturbatively, and so calculations may be performed
analytically. Unfortunately, the scheme was shown to 
fail to converge well at fairly low momentum in the $\siii$ channel.

Apparently what is happening is that because of the nonperturbative
interactions, operators acquire large anomalous dimensions which
can cause them to become either much more relevant, or much less
relevant than one naively expects.  Thus the perturbative analysis of
counterterms in ref.~\cite{Kaplan:1998tg,Kaplan:1998we} can be
misleading: the need for a  counterterm at high order in the chiral
expansion could just signify that the
operator in question has a large negative anomalous dimension and is
actually very{\it  unimportant} to the calculation.  

Currently, one
needs to perform numerical calculations to understand the
nonperturbative renormalization. 
While perturbative ladder diagrams suggest an infinite number of
counterterms are needed to renormalize two nucleons interacting via
the tensor force in the
$\siii-\diii$ channels, numerical results show that only a single
counterterm is required.  That is good news for the application of Weinberg's
power counting scheme.  On the other hand, ref.~\cite{Nogga:2005hy}  that in 
arbitrarily high
partial wave channels experiencing an attractive tensor interaction, a counterterm is
required at leading order, which is not at all in agreement with Weinberg's
expansion scheme.  Some groups have adopted Weinberg's expansion scheme at face
value, but the evidence is clear that in some channels it is correct,
while in others it is not.

In conclusion, I would say that EFT for nuclear physics at momentum
transfers comparable to the pion mass and higher seems to work pretty
well if one follows a patchwork of power counting rules that are
derived from numerical experiments. I do not find this very
satisfactory though; 
 new theoretical ideas for systematizing the power
counting for this nonperturbative EFT would be welcome.

\subsection{Trapped atoms}
\label{sec:4e}

One might think that having particles with an unnaturally large
two-body scattering length would be peculiar to nuclear physics.
However, atomic physicists trapping collections of atoms can tune the
scattering length of atom-atom scattering to be very large by adjusting
an external magnetic field.  In the limit that the scattering length
diverges and the inter-particle spacing is much less than the range of
the interactions, there is no dimensionful scale for low energy scattering
other than the incoming energy.  Therefore any such system should
exhibit universal properties, up to trivial rescaling to account for
the particle masses. For example, the rescaled specific heat, or
critical temperature 
for pairing should be almost the same for tuned atoms as for a dilute neutron
gas.

This is an ideal system for  the application of the type of EFT expansion developed for
nucleon-nucleon interactions. Unfortunately, it is a nonperturbative
many-body problem (an infinite number of particles lie within a
scattering length of each other) and not amenable to analytic
calculation with any reliability.  However, knowing that a two
particle contact interaction is all that is needed to describe the system, it
is relatively simple to construct a lattice version of the problem
\cite{Chen:2003vy} which can be simulated numerically
\cite{Wingate:2005xy}. It is the beauty of effective field theory that
allows one to extract information about a complicated many-body atomic
system by analyzing fundamental fermions with a two-particle contact interaction, formulated in a
spacetime consisting of a hypercubic array of points!

\vfill
\eject

\bigskip
\noindent
{\bf Problems:}

\bigskip
\hrule
\bigskip
\noindent
{\bf IV.1)} Reproduce the result \eq{aminus1}.

\bigskip
\noindent
{\bf IV.2)} Using the lowest order interaction of \eq{3sl} and the
interpolating field $\CD_i = N^T P_i N$, where $P_i$ is the projection
operator in \eq{projop}, relate $C_0$ to the binding energy $B$ of the
deuteron, and find the wave function renormalization $Z$.  Ingredients:
\beq
G(\overline E) \,\delta_{ij}= \int {\rm d}^4x
e^{-i(Et-\bfp\cdot{\bf x})}\,\bra{0} {\rm T}\[\CD_i^{\dagger}(x)\CD_j(0)\]\ket{0}
=\delta_{ij}\frac{i\CZ(\overline E)}{  \overline E + B + i\varepsilon}\ ,
\eqn{fullprop2}
\eeq
is the sum of bubble-chain graphs with an insertion of $\CD_i^\dagger$
at one end, and  $\CD_j$ at the other.
By Lorentz invariance, the propagator only depends on the
energy in the center of mass frame, namely
\beq
\overline E\equiv  E - \frac{\bfp^2}{ 4M} +\ldots\ ,\qquad E\equiv (p^0-2M),
\eeq
where the ellipses refers to relativistic corrections to the dispersion
relation.
 The numerator $\CZ$ in \eq{fullprop2} is assumed to be smooth near
the deuteron pole, and when evaluated at the pole gives the
wave function renormalization $Z$,
\beq
\CZ(-B) \equiv Z = -i \left[ \frac{{\rm d}G^{-1}(\overline E)}{
{\rm d} E}\right]_{\overline E=-B}^{-1} .
\eeq
The results are found in the appendix of ref. \cite{Kaplan:1998sz}.


\noindent
\vfill\eject

\section{ The effective theory for color superconductivity}
\label{sec:5}

\subsection{Landau liquid versus  BCS instability}
\label{sec:5a}

A condensed matter system can be a very complicated environment; there may be
various types of ions arranged in some crystalline array, where each ion has a
complicated electron shell structure and interactions with neighboring ions
that allow electrons to wander around the lattice.  Nevertheless, the low
energy excitation spectrum for many diverse systems can be described pretty
well as a ``Landau liquid'', whose excitations are fermions with a possibly
complicated dispersion relation but no interactions.  Why this is the case can
be simply understood in terms of effective field theories, modifying the
scaling arguments to account for the existence of the Fermi surface.

Let us assume that the low energy spectrum of the condensed matter system has
fermionic excitations with arbitrary interactions above a Fermi surface
characterized by the fermi energy $\epsilon_F$ ; call them
``quasi-particles''. Ignoring interactions at first, the action can be written as
\beq
S_{free}=\int dt\,\int d^3p\, \sum_{s=\pm\half} \[
\psi_s(p)^{\dagger} i\partial_t \psi_s(p) - (\epsilon(p)-\epsilon_F)
\psi^{\dagger}_s(p)\psi_s(p)\]
\eqn{sfree}
\eeq
where an arbitrary dispersion relation $\epsilon(p)$ has been assumed.

To understand how important interactions are, we wish to repeat some
momentum space version of the scaling arguments I introduced in the first lecture. In
the present case, a low energy excitation corresponds to one for which
$(\epsilon(p)-\epsilon_F)$ is small, which means that $\bfp$ must lie
near the Fermi surface.  So in momentum space, we will want our
scaling variable to vary the distance we sit from the Fermi surface,
and not to rescale the overall momentum $\bfp$.  After all, here a
particle with $\bfp=0$ is a high energy excitation.  

\begin{figure}[t]
\centerline{\epsfxsize=2.2 in \epsfbox{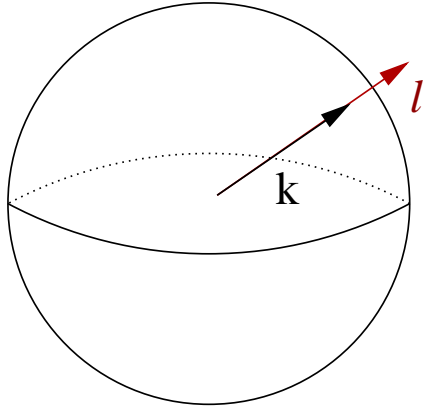}}
\noindent
\caption{\it The momentum $\bfp$ of an excitation is decomposed as $\bfp=\bfk + \bfl$,
  where $\bfk$ lies on the Fermi surface, and $\bfl$ is perpendicular
  to the Fermi surface. Small $|\bfl|$ corresponds to a small
  excitation energy.}
\label{fig:fermiscaling}
\vskip .2in
\end{figure}

This situation is
a bit
reminiscent of  HQET where we wrote  $p_\mu=m v_\mu +
k_\mu$, with $k_\mu$ being variable that is scaled, measuring the
``off-shellness'' of the heavy quark.   So in the present case we will
write the momentum as
\beq
\bfp = \bfk+ \bfl
\eeq
where $\bfk$ lies on the Fermi surface and $\bfl$ is
perpendicular to the Fermi surface
(shown in Fig.~\ref{fig:fermiscaling} for a spherical Fermi surface).
Then $\bfl$ is the quantity we vary in experiments and so we 
define the dimension of operators by how they must scale so that the theory is
unchanged when we change $\bfl \to r\bfl$.  If an object scales as
$r^n$, then we say it has dimension $n$.  Then  $[k]=0$, $[\ell]=1$, and $[\int
d^3p = \int d^2k d\ell] = 1$. And if we define the Fermi velocity as
$ {\bf v}_F(\bfk)={\boldsymbol\nabla}_k\epsilon(\bfk)$, then for   $\ell\ll k$,
\beq
\epsilon(\bfp) - \epsilon_F = \bfl\cdot {\bf v}_F(\bfk) +\CO(\ell^2)\
,
\eeq
and so $[\epsilon - \epsilon_f] = 1$ and $[\partial_t] = 1$. Given that the
action
\eq{sfree} isn't supposed to change under this scaling,
\beq
[\psi] = -\half\ .
\eeq

Now consider   an interaction of the form
\beq
S_{int} = \int dt\, \int \prod_{i=1}^4 (d^2 \bfk_i d\ell_i) \delta^3(
\bfP_{tot}) C(\bfk_1,\ldots,\bfk_4)\psi_s^\dagger(\bfp_1) \psi_{s}(\bfp_2)
\psi_{s'}^{\dagger}(\bfp_3)\psi_{s'}(\bfp_4)\ .
\eeq
This will be relevant, marginal or irrelevant depending on the dimension of
$C$. Apparently $[\delta^3(\bfP_{tot}) C] = -1$.  So how does the $\delta$
function
scale?  For generic $\bfk$ vectors, $\delta(\bfP_{tot})$ is a constraint on
the $\bfk$ vectors that doesn't change much as one changes $\ell$, so that
$[\delta^3(\bfP_{tot})]=0$.   It follows that $[C]=-1$ and that the four
fermion interaction is irrelevant...and that the system is adequately described
in terms of free fermions (with an arbitrary dispersion
relation). This is why Landau liquid theory works and
is related to why  in nuclear physics Pauli blocking allows a strongly
interacting system of nucleons to have single particle excitations.

\begin{figure}[t]
\centerline{\epsfxsize=5.0 in \epsfbox{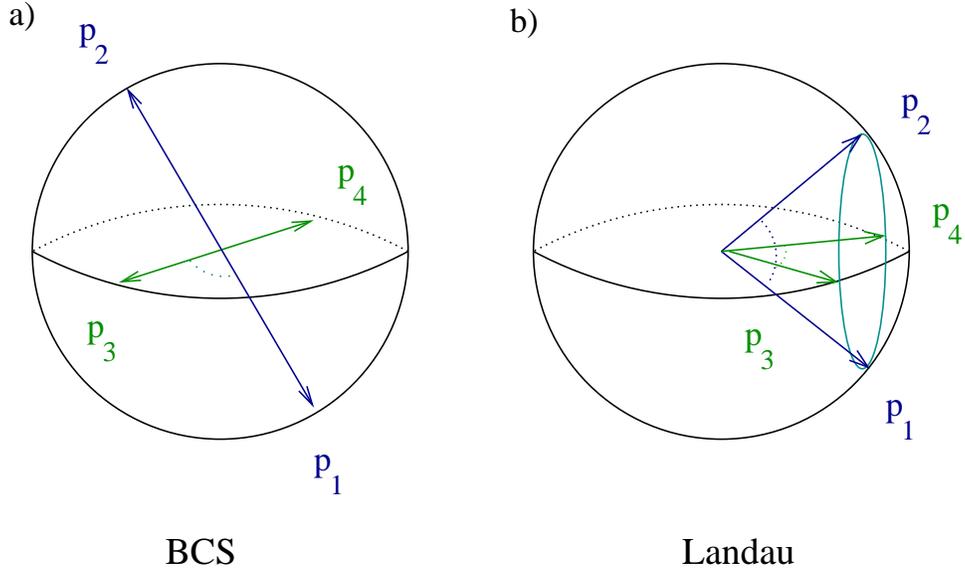}}
\noindent
\caption{\it Fermions scattering near the Fermi surface. (a) Head-on collisions:  With $\bfk_1+\bfk_2=0$, only two degrees of freedom
  in the outgoing momenta $\bfk_3$ and $\bfk_4$ are constrained, as
  they  can point to
  any two opposite points on the Fermi surface.  (b) The generic
  Landau liquid case, where the incoming particles do not collide
  head-on, and three degrees of freedom in  the outgoing momenta
  $\bfk_3$ and $\bfk_4$ are constrained, as they must point to opposite sides of
  a particular circle on the Fermi surface.  Figure from
  ref.~\cite{Schafer:2003vz}, courtesy of Thomas Sch\"afer.  }
\label{fig:fskin}
\vskip .2in
\end{figure}

This is not the whole story though, or else superconductivity would
never occur.  Let us look more closely at the assumption  above
$[\delta^3(\bfP_{tot})]=0$.   Consider the case when all the 
$\bfl_i=0$, and therefore the $\bfp_i=\bfk_i$  and lie on the Fermi
surface.  Suppose we
fix the two incoming momenta $\bfk_1$ and $\bfk_2$.    The $\delta^3(\bfP_{tot})$ then
constrains the sum  $\bfk_3+\bfk_4$ to equal $\bfk_1+\bfk_2$, which
generically means that the vectors $\bfk_3$ and $\bfk_4$ are
constrained up to point to opposite points on a circle that lies on
the Fermi surface (Fig.~\ref{fig:fskin}b).  Thus one free parameter
remains out of the four independent parameters needed to describe the
vectors $\bfp_3$ and $\bfp_4$.  So we see that in this generic case,
$\delta^3(\bfP_{tot})$ offers three constraints, even when
$\bfl_i=0$. Therefore $\delta^3(\bfP_{tot})= \delta^3(\bfK_{tot})$ is unaffected when $\bfl$
is scaled, and we find the above assumption $[\delta^3(\bfP_{tot})]=0$
to be true, and Landau liquid theory is justified.
 
However now look at the special case when the collisions of the
incoming particles are nearly head-on, $\bfk_1+\bfk_2=0$.  Now
$\delta^3(\bfP_{tot})$ constrains the outgoing momenta to satisfy
$\bfk_3+\bfk_4=0$. But as seen in Fig.~\ref{fig:fskin}a, this only
constrains $\bfk_3$ and $\bfk_4$ to lie on opposite sides of the Fermi
surface.  Thus $\delta^3(\bfP_{tot})$ seems to be only constraining
two degrees of freedom, and could be written as
$\delta^2(\bfk_3+\bfk_4)\delta(0)$. This singularity obviously arose
because the set the $\bfl_i=0$. For nonzero $\bfl$ the $\delta(0)$
becomes $\delta(\bfl_{tot})$, and as a result, the $\delta$ function
does scale with $\bfl$:  $[\delta^3(\bfP_{tot})]=-1$.  But then
$[C]=0$ for head-on collisions, and the interaction is marginal!
  Quantum corrections either make it either irrelevant or relevant;
it turns out that for an attractive interaction, the interaction becomes
relevant, and for a repulsive interaction, 
it becomes irrelevant, just as we found for a $\delta$-function
interaction in two dimensions.

Therefore, an attractive contact interaction 
between  quasiparticles becomes strong exponentially close to the
Fermi surface (since the coupling runs logarithmically), and can lead
to pairing and superconductivity just as the asymptotically free QCD
coupling leads to quark condensation and chiral symmetry breaking.
The BCS variational calculation shows that the pairing instability
does indeed occur;  the effective field theory analysis explains why
Cooper pairs are exponentially large compared to the lattice spacing
in superconductors.

\subsection{Dense quark matter}
\label{sec:5b}

With a tree-level one-gluon exchange, two quarks transforming as a
color triplet will feel an attractive interaction in the color $\mybar 3$
channel, and a repulsive interaction in the $6$ channel.
Since superconductivity is generic in any dense fermion system with an
attractive interaction, it is clear that such a phase should occur for
dense quark matter.  Just as electromagnetism is spontaneously broken
by the condensation of electron pairs in an ordinary superconductor,
color will be broken when quarks condense. Color superconductivity was
first discussed in the early 1980's \cite{Bailin:1983bm}. In the past decade there
has been a resurgence of interest in dense 
quark matter and color superconductivity, sparked by the papers
\cite{Alford:1997zt,Rapp:1997zu,Alford:1998mk,Son:1998uk}.  Two
interesting features that were discovered was that the gap is parametrically
larger in a color superconductor than in an ordinary metal
\cite{Son:1998uk}, and that the ground state of dense QCD with three
massless flavors
 spontaneously breaks
chiral symmetry, even though at high density the $\vev{\mybar q q}$
condensate is expected to vanish. As a result, there are nine
Goldstone bosons in the theory (eight for flavor, one for baryon
number).  Thus there is no gap in the spectrum of the theory in many
quantum number channels, and so the phase structure becomes very rich
when quark masses are turned on and charge neutrality is enforced. An
effective field theory is indispensable for understanding these  light
modes and the phase structure.

I will be assuming a very high quark number chemical potential,
$\mu\gg \Lambda_{QCD}$ so that perturbative QCD applies.  I will
simultaneously be assuming that $ m_c> \mu$ so that I can restrict the
discussion to three flavors.  It is certain that there is no place in
the universe where these conditions are met!  At best, one can hope
that there is a quark matter star somewhere, and that even though its
chemical potential will be $O(\Lambda_{QCD})$, qualitative features
discovered at large chemical potential  will still hold in the nonperturbative
 regime\footnote{Even should this be true, it seems not too likely
   that we will ever get enough data on such a star to really test the
   theory. My hope is that some day people will figure out how to  simulate
   lattice QCD at finite chemical potential and we can do experiments on a
   computer.  If you want to make a truly major contribution to
   nuclear physics, solving that  problem would be a fine choice!}.
 So the subject may be purely academic, but we're academics, after
 all, and it is fun.

Just as electron pairs condense in ordinary superconductivity, quark
pairs are expected to condense in the attractive color $\mybar 3$
channel.  However, if the $u$, $d$ and $s$ quarks are massless, their
Fermi surfaces match up, and quarks of different flavors can condense
with each other. A Lorentz singlet condensate of two left-handed
quarks in the most attractive color channel will transform
as a $(\mybar 3,1,\mybar 3)$ under $SU(3)_L\times SU(3)_R\times
SU(3)_c$, while a condensate of right-handed quarks will transform as
a $(1,\mybar 3,\mybar 3)$.  These are $3\times 3$ matrices, so where
are the nonzero entries?  The favored phase is to have these matrices
just be proportional to the unit matrix;  it is called the ``Color
Flavor Locked'' phase, or CFL for short.

Thus the  ``order parameter'' for this phase may be represented as
\beq
\epsilon_{abc} \epsilon^{ijk}\vev{q_{L,i}^a C q_{L,j}^b} = -\epsilon_{abc}\epsilon^{ijk} \vev{q_{R,i}^a C q_{R,j}^b} =
\Lambda^3 \delta^k_c\ ,
\eqn{cfl}\eeq
where $a,b,c$ are $SU(3)_c$ indices, and $i,j,k$  are flavor indices.
Note that flavor and color have become 
correlated as the above condensates vanish if $c\ne  k$.  (I prefer a Weyl fermion
basis for keeping track of the quantum numbers;  see problem
(V.1)). This condensate breaks $SU(3)_L\times SU(3)_R\times SU(3)_c$
down to the diagonal $SU(3)$.  

Actually the condensate above is written in the usual sloppy way we
talk about the Higgs mechanism.  In fact, gauge variant operators
always have zero expectation value unless one gauge fixes.  The actual
gauge invariant order parameters are four (and more) quark operators,
such as
\beq
(\epsilon^{ijk}\epsilon_{mnp})(\epsilon_{abc}\epsilon^{dec})\vev{q_{R,i}^a
C q_{R,j}^b\left(q_{L,m}^d C q_{L,n}^e\right)^\dagger} = \Lambda^6
\delta^k_p\ .
\eqn{cfl4}\eeq
This color neutral order parameter transforms as a $(3,\mybar 3)$
under $SU(3)_L\times SU(3)_R$, and acquires a diagonal vev, breaking
$SU(3)_L\times SU(3)_R$ down to $SU(3)_V$...just like the $\vev{\mybar
q q}$ condensate of \eq{sig1}  at zero chemical potential.  The
fermionic excitations in this ground state are 
``gapped'': it requires a minimum energy $2\Delta$ to produce a
particle-hole pair, where $\Delta$ is called the gap, and
$\Lambda^3\propto \Delta$ in the above equation.

Although sloppy, the advantage of \eq{cfl}  is that it
indicates that $SU(3)_c$ has been higgsed, and that the gluons have
become massive.  Why is this significant, since at finite density, one
already has Debye screening, which means that color electric fields
fall off from sources as if the gluon had a mass $m=g\mu$?  The difference is
that without the Higgs effect, magnetic fields at nonzero frequency
would have power law fall-off, and not be screened.  In the color
superconductor, though, magnetic fields are screened as well.  This
means that at extremely high density it is self consistent to calculate properties of
this state in perturbation theory, for then the running
coupling 
$\alpha_s$
 is small when evaluated at the large momentum scale $\mu$.

The spectrum of the theory now has gluons with mass  $g \mu$;
fermionic excitations around the Fermi surface 
with mass $\Delta$, where $\ln \Delta/\mu = O(1/g)$ \cite{Son:1998uk};
and nine  massless Goldstone bosons (an $SU(3)_V$ octet, and the
superfluid mode from broken baryon number).  This allows one to construct a
series of effective field theories.  First one integrates out degrees
of freedom at the scale $\mu$ --- for example, quark anti-quark pairs
which have  at least an energy $\mu$, since the quark has to be created above
the Fermi sea.  Next one integrates out the gluons at scale $g\mu$;
then the quasiparticle excitations at scale $\Delta$.  One is left
with a chiral Lagrangian for the Goldstone bosons, where the
coefficients are calculable.

We start with the QCD Lagrangian,
\beq
\CL = \mybar q(i\Dsl + \mu \gamma^0)q - \mybar q_L M q_R -
\mybar q_R M^\dagger q_L - \frac{1}{4} G^a_{\mu\nu}
G^{a,\mu\nu}\ ,
\eeq
where I have reintroduced the quark mass matrix $M$. Since we wish to
consider quarks near the Fermi surface, first consider free quarks
with a chemical potential $\mu$.  Then the Dirac equation reads
\beq
\left({\boldsymbol \alpha}\cdot\bfp - \mu\right) \psi_\pm = E_\pm
\psi_\pm\ ,\qquad ({\boldsymbol \alpha}\cdot\hat\bfp)\psi_\pm =\pm
\psi_{\pm}\ , 
\eeq
where ${\boldsymbol \alpha} = \gamma^0 {{\boldsymbol \gamma}}$ and
$E_\pm=-\mu\pm p$. We see that for $p\sim p_F=\mu$, the $\psi_+$ with
$E_+\sim 0$  correspond to states near
the Fermi surface and $\psi_-$, with $E_-\sim -2\mu$, to states far
from it.  We therefore follow Hong \cite{Hong:1998tn,Hong:1999ru} and define for the interacting theory
\beq
\psi_{\pm}({\bf v_F},x) = e^{i p_F v_\mu x^\mu} \left(\frac{1\pm {\boldsymbol
      \alpha}\cdot {\bf \hat v_F}}{2}\right) q\ ,
\eeq
where $v_\mu = (1,{\bf v_F})$ and ${\bf v_F}$ is the Fermi velocity we
are expanding about, with $\bfk = p_F{\bf \hat v_F}$ corresponding to
the vector $\bfk$ in Fig.~\ref{fig:fermiscaling}. The prefactor removes
the rapid phase common to all fermions in the vicinity of this patch
of the Fermi surface specified by Fermi velocity ${\bf v_F}$.

One then constructs a $1/p_F$ expansion of this theory,  integrating
out the $\psi_-$ fields and the hard gluons (whose propagators
are $1/q^2 \le 1/p_F^2$). At tree level, integrating out the $\psi_-$
fields is equivalent to replacing them by their equations of motion,
\beq
\psi_{-,L} = \frac{1}{2p_F}\left(i {\boldsymbol\alpha}_\perp \cdot {\bf
      D}\psi_{+,L} + \gamma^0 M \psi_{+,R}\right)\ .
\eqn{psieq}
\eeq
where ${\boldsymbol \gamma}_\parallel \equiv {\bf \hat v}_F ({\bf \hat
  v}_F\cdot {\boldsymbol \gamma})$ and ${\boldsymbol\gamma}_\perp=
({\boldsymbol \gamma}-{\boldsymbol \gamma}_\parallel)$.
 Integrating out the hard gluons generates a four
quark vertex which is the attraction that gives rise to quark
condensation;  this in turn introduces a ``gap'' term in the effective
theory, reflecting that we are not expanding about the perturbative
vacuum, but one with symmetry breaking;  the gap $\Delta$ is solved
for self consistently.  I highly recommend the various papers and
reviews by
Thomas Sch\"afer on the subject if you wish to understand the procedure
in detail
\cite{Schafer:2004yx,Schafer:2003yh,Schafer:2003jn,Schafer:2003vz,Schafer:2002ms}.

The effective theory to $O(1/p_F)$ takes the form (see for example \cite{Schafer:2001za,Son:1999cm})
\beq
{\CL} &=& 
 \psi_{L+}^\dagger (iv\cdot D) \psi_{L+}
  - \frac{ \Delta}{2}\left(\psi_{L+}^{ai} C \psi_{L+}^{bj}
 \left(\epsilon_{abc} \epsilon_{ijk} (X^\dagger)^{ck} \right) 
           + {\rm h.c.} \right) \nonumber \\ 
& & \hspace{0.5cm}\mbox{}
  - \frac{1}{2p_F} \psi_{L+}^\dagger \left(  (\Dsl_\perp)^2 
  + MM^\dagger \right)  \psi_{L+}  
  + \left( R\leftrightarrow L, M\leftrightarrow M^\dagger, X
    \leftrightarrow Y \right)  + \ldots ,\cr&&
\eeq
where $D_\mu=\partial_\mu+igA_\mu$.  The $SU(3)$ matrices $X$ and $Y$
are spacetime dependent, and parametrize the wiggles of the $\vev {q_L
q_L}$ and $\vev{q_R q_R}$ condensates respectively. Under
$SU(3)_L\times SU(3)_R\times SU(3)_c$ they transform as $X=(3,1,3)$
and $Y=(1,3,3)$, and  they can be
written in the form $X=e^{i(\Pi + \pi)/f}$ and $Y=e^{i(\Pi - \pi)/f}$.
 The $\Pi$ fields can be rotated away by an $SU(3)_c$ color
 transformation---this is called ``choosing unitary gauge'', and the
 gluons have a mass $gf$ in this gauge.  The $\pi$ fields are the
 octet that arises from $SU(3)_L\times SU(3)_R$ breaking, and we can
 define the $SU(3)_c$ neutral matrix $\Sigma = X Y^\dagger$ which
 transforms as $(3,\mybar 3,1)$ just like the $\Sigma$ matrix we
 discussed in lecture 3, which parametrizes the QCD groundstate at $\mu=0$.

In the above Lagrangian, the $M M^\dagger$ term is called the
Bedaque-Sch\"afer term \cite{Bedaque:2001je} and it is crucial to understanding the fate of
the CFL phase as one turns on nonzero quark masses; the discovery of
its existence and dramatic consequences  is
one of the triumphs of the effective field theory approach to color
superconductivity.  It arises from the mass dependence of $\psi_-$ in
\eq{psieq}. What Bedaque and Sch\"afer realized is that the
combination $\mu^{(BS)}_L \equiv -M M^\dagger/(2 p_f)$ enters the effective theory like the
time component of an  $SU(3)_L$ gauge field would, if one gauged
left-handed flavor symmetries.  Similarly, $\mu^{(BS)}_R\equiv -M^\dagger M/(2 p_F)$
enters like an $SU(3)_R$ gauge field. This fake gauge invariance
therefore will constrain how the Bedaque-Sch\"afer terms enter the low
energy chiral Lagrangian.

At  $O(1/p_F^2)$, the high density effective theory contains
four-fermion interactions which contain two powers of $M$ and others
that contain an $M$ and an $M^\dagger$; see \cite{Schafer:2001za}.

\subsection{The chiral Lagrangian for the CFL phase}
\label{sec:5c}

We are now ready to construct the chiral Lagrangian for the octet of
Goldstone bosons in the CFL phase. There is an excellent reason to do
so:  the CFL order parameter \eq{cfl4} is perfectly symmetric in
flavor because we assumed massless quarks.  In the real world, the
strange quark mass $m_s$
could be comparable or bigger than the gap $\Delta$ even for rather
large chemical potential $\mu$, since $\Delta$ is exponentially
smaller than $\mu$.  That means that the fermi surface of free strange
quarks would not be congruent with the fermi surface of the $u$ and
$d$ quarks, and pairing between them would be cost energy.  Thus as
the strange quark mass is turned on, the vacuum comes under stress,
and eventually phase transitions are expected.  If the phase
transitions are second order or weakly first order (with latent heat less
than $\Delta$), then they will have to involve light
degrees of freedom, and should be visible in the chiral
Lagrangian. In fact, there are a number of interesting phase
transitions one can find as $m_s$ is turned up, starting with kaon condensation.

  The cutoff of the chiral Lagrangian will be $\Delta$, the
scale of the quasiparticles we have integrated out (they are like the
baryons in the usual chiral Lagrangian discussed in the third
lecture). There several basic differences between the CFL chiral
Lagrangian, and the usual one:
\begin{enumerate}[i.]
\item Since the chemical potential violates Lorentz invariance, the
  Lagrangian will not respect Lorentz invariance either;  the speed of
  light will be replaced by the speed of sound, which to leading order
  in perturbation theory is $c/\sqrt{3}$, the result for a weakly
  interacting 
  relativistic gas.
\item Baryon number is spontaneously broken as well as $SU(3)\times
  SU(3)$, so there will be a corresponding Goldstone boson $B$, the
  ``superfluid mode''.
\item At $\mu=0$ we ignore the $U(1)_A$ symmetry, as it is badly
  broken by instanton effects.  At high density, the instantons do not
  play a role, so there is a light $\eta'$, the Goldstone boson for
  broken $U(1)_A$ symmetry, and the $\Sigma$ field carries $U(1)_A$.
  Assign a $U(1)_A$ charge $Q_A=+1$ to $q_L$ and $Q_A=-1$  to $q_R$.
  Then the order parameter \eq{cfl4} carries $Q_A=-4$, and so must the
  $\Sigma$ field.  On the other hand, the QCD Lagrangian is invariant
  of the quark mass $M$ carried $Q_A=2$. Therefore $\Sigma$ cannot
  couple to odd powers of $M$. This assumes that the $\mu=0$
  condensate $\vev{\mybar qq}$ vanishes at high density, an assumption
  on solid ground.

\item As we derived in the previous section, in the EFT below $\mu$,
  the Bedaque-Sch\"afer term $\mu_{BS}$ appears as the time component of an $SU(3)_V$ flavor
  gauge field;  therefore it must also in the chiral Lagrangian.
  Instead of $\partial_0 \Sigma$, we must write $D_0\Sigma =
  \partial_0 \Sigma + i \mu^{(BS)}_L\Sigma-i\Sigma \mu^{(BS)}_R$.

\item The pion decay constant has been calculated and is $O(\mu)$.  Thus 
  $f_\pi \gg \Delta$, unlike in the $\mu=0$ chiral Lagrangian, where
  $f_\pi\sim \Lambda/4\pi$.  Therefore  loop graphs are unimportant to
  leading order in the small quantity $\Delta/\mu$.
\end{enumerate}

Ignoring the $B$ Goldstone boson, as well as the $\eta'$, and
considering only spatially constant Goldstone boson fields, the chiral
Lagrangian takes the form
\beq
\eqn{cflchilag}
\CL &=& f_{\pi}^2 \left[\frac{1}{4}\Tr D_0 \Sigma D_0 \Sigma^\dagger  
+ \frac{a}{2} \Tr
\tilde{M} \left(\Sigma + \Sigma^\dagger\right) + \frac{b}{2}\Tr Q
\Sigma Q \Sigma^\dagger\right]\ \, \nonumber \\ 
& & D_0\Sigma = \partial_0 \Sigma - i\left[
\left(\mu_Q Q + \mu^{(BS)}_L\right)\Sigma -  \Sigma\left(\mu_Q
Q + \mu^{(BS)}_R\right)\right] \,.
\eeq
The decay constant $f_\pi$ has been computed previously
\cite{Son:1999cm}. $Q$ is
the electric charge matrix ${\rm diag}(2/3,-1/3,-1/3)$ while $\mu^{(BS)}_{L,R}$ are
the Bedaque-Sch\"afer terms: $\mu^{(BS)}_L = -\frac{M M^\dagger}{2\mu}\ ,\quad 
\mu^{(BS)}_R = -\frac{ M^\dagger M}{2\mu}$.  I have included a
chemical potential for electric charge,$m_Q$, since in dense matter,
such as the core of stars, there would be the constraint of charge
neutrality.  

The mass term above has been
written in terms of 
 \beq \tilde{M}= M^{-1}\det(M) = \begin{pmatrix} m_d m_s & & \cr & m_u m_s
   & \cr && m_u m_d \cr \end{pmatrix}\ .
\eqn{mtilde}\eeq
One can show that this is the only possible form for the coupling of
$\Sigma$ to $M$ at leading order. It is second order in $M$ as required by the $U(1)_A$
symmetry, it transforms as a $(\mybar 3,3)$, and it has the property
that it vanishes if any two quarks are massless (See problem V.5). 

 The coefficient $a$ has been computed and 
is given by $a=3\frac{\Delta^2}{\pi^2 f_{\pi}^2}$ 
\cite{Son:1999cm}.  The $b$ term accounts 
for electromagnetic corrections to the charged meson masses; it has
not been calculated but is estimated to be of size  $ b\sim \frac{\ \alpha}{4\pi} 
\Delta^2\ .$

The meson masses in terms of the parameters $a$, $b$ are
\beq
m^2_{\pi^-} &=& a (m_u + m_d)m_s + b \cr m^2_{K^-} &=& a (m_u + m_s)m_d + b\cr
m^2_{K^0} &=& a (m_d + m_s)m_u \,.
\eqn{masses}
\eeq

The chiral expansion is in powers of $p/\Delta$, and the
Bedaque-Sch\"afer term appears in the covariant derivative, the theory
breaks down when $\mu^{(BS)} \gtrsim \Delta$, or when any one of the
quark masses satisfies $m^2 > 4\mu\Delta$.  Why is that?  Recall that
for free fermions near the Fermi surface, the energy is given by
$\sqrt{p_F^2 + m^2} = p_F + m^2/2p_F + \ldots$.  A chemical potential
for baryon number ensures that each quark has the same Fermi {\it
  energy} while BCS pairing between states at opposite sides of the
Fermi surface requires that the two quarks have the same Fermi {\it
  momentum}.  Since the pairing gains an energy $\Delta$ per quark,
while maintaining equal $p_F$ between a heavy quark and a light quark
costs $m^2/2p_F$, the pairs will break when $m^2/2p_F \gtrsim
2\Delta$.  Since $\mu=p_F + O(m^2)$, this is the same as saying that
the CFL state breaks down completely at $m^2 \gtrsim 4\mu\Delta$.
Note that at the breakdown point $m^2 = 4\mu\Delta$, the meson masses
are $M^2 \sim \Delta^3/\mu \ll \Delta$, so they are still very light
compared to the cutoff of the theory.  

\subsection{Kaon condensation on top of CFL}
\label{sec:5d}

But is the CFL ground state stable all the way up to strange quark
mass  $m_s^2 = 2\mu\Delta$?  The answer is no. Just as we saw in a
hadronic description of finite density QCD, kaons can alter the ground
state to relieve the stress caused by not having the ideal strangeness.
However, in that case the problem was that we started from a state
with zero strangeness (neutrons) and the system wanted to populate
strange quarks, and so we saw evidence for $K^-$ condensation.    In
the CFL groundstate there are equal numbers of $u$, $d$ and $s$, so as
$m_s$ is turned on, the system will adjust to reduce its strange quark
number, via $K^0$ condensation.

Unlike the mesons of  the chiral Lagrangian we studied in the third
lecture, these mesons have chemical potentials.  Since we choose a
basis where $M=M^\dagger$, the chemical potentials due to $\mu_Q$ and
$\mu^{(BS)}$, where $\mu_Q$ is a real chemical potential added to
ensure charge neutrality, while the Bedaque-Sch\"afer term is a
dynamical term reflecting that mesons want to rearrange the CFL
ground state if
quark masses are unequal.  By expanding the kinetic term to quadratic
order in the mesons, one can read off their individual chemical potentials.
The effective chemical potentials vanish for the $\pi^0$, $\eta$ and $\eta'$,
while for the $\pi^+$, ${\rm K}^+$ and ${\rm K}^0$ mesons they are
\beq
\tilde \mu_{\pi^+} = \mu_Q + \frac{m_d^2 - m_u^2}{2\mu}\ ,\quad
\tilde \mu_{K^+} = \mu_Q + \frac{m_s^2 - m_u^2}{2\mu}\ , \quad
\tilde \mu_{K^0} = \frac{m_s^2 - m_d^2}{2\mu}\ .
\eeq

\begin{figure}[t!]
\begin{center}
\includegraphics[width=.75\textwidth,angle=-90,scale=.7]{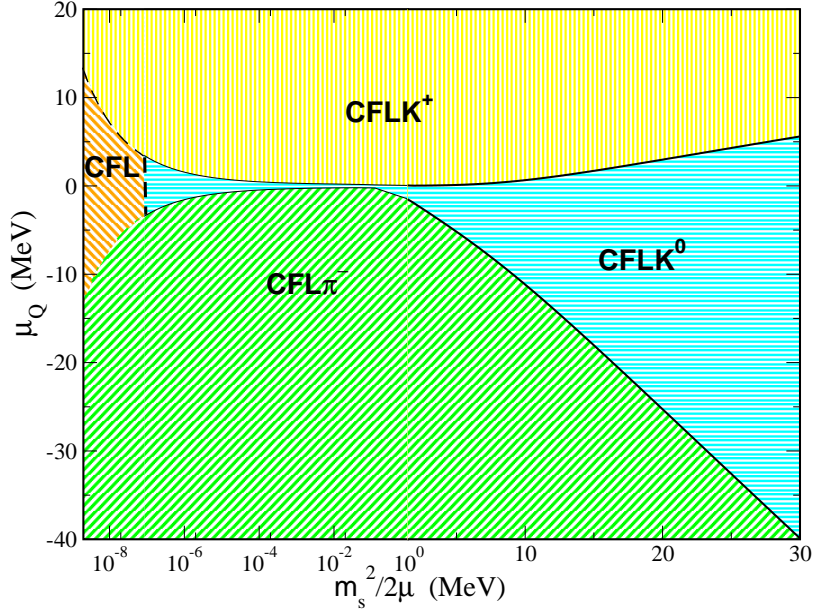}
\end{center}
\caption{\it Meson condensed phases in the neighborhood of the 
symmetric CFL state are shown in the $(m_s^2/2\mu)-\mu_Q$ plane, where $m_s$ is the
strange quark mass (set to $150$~MeV), $\mu$ is the quark number chemical
potential, and $\mu_Q$ is the chemical potential for positive electric
charge. At five times nuclear density $\mu \sim 400$ MeV and $(m_s^2/2\mu)\sim
25$ MeV. Solid and dashed lines indicate first- and second-order transitions
respectively. From ref.~\cite{Kaplan:2001qk}.}
\label{fig:phase}
\end{figure}

Note that as the strange quark is increased, so is the chemical
potential for the $K^+$ and $K^0$.  The fundamental $K^+$ and $K^0$
mesons are made of $\mybar s u$ and $\mybar s d$ quarks;  these CFL
mesons are better thought of as $u$-quark/$s$-hole and
$d$-quark/$s$-hole bound states.  So it makes sense that as $m_s$ is
increased, it becomes more energetically feasible to turn an $s$ quark into a $u$
or $d$ quark, and the most efficient way to do so is to create a kaon.

Recall that Bose-Einstein condensation occurs when a meson's chemical
potential exceeds its mass.  Thus for $\mu_Q=0$, a sufficiently large
quark mass will lead to a second order phase transition in the form
of kaon condensation
\cite{Bedaque:2001je,Kaplan:2001qk,Kaplan:2001hh}. Since
electromagnetic corrections make the $K^+$ heavier than the $K^0$, at
$\mu_Q=0$ one would expect $K^0$ condensation.  However, if $\mu_Q>0$,
positively charged mesons are favored and would get $K^+$
condensation.  Conversely, for sufficiently negative $\mu_Q$ the
system wants negative charge;  $K^-$'s are too costly as they {\it
  add} a strange quark, and so the heavier $\pi^-$ could condense.

The stationary points of the free energy $\Omega$ with respect to
variations of the meson fields are found as solutions to the matrix
equation 
\beq 
\left[\tilde{\mu}\Sigma^\dagger \tilde{\mu} \Sigma - a\,
M\Sigma - b\, Q \Sigma^\dagger Q \Sigma\right] -h.c.=0 \,.
\eqn{sigmin} 
\eeq 
There are distinct solutions for $K^0$, $K^+$ and $\pi^-$
condensation:
\beq
\begin{array}{rlrl}
\Omega_{\pi^\pm} =& -\frac{f_\pi^2}{2} (\tilde \mu_{\pi^\pm}^2
-b)(1-\cos\theta_{\pi^\pm})^2\ ,\quad&\cos\theta_{\pi^\pm}=&\begin{cases}1 &
  M_{\pi^\pm}^2 \ge \tilde \mu_{\pi^\pm}^2\cr & \cr
  \frac{M_{\pi^\pm}^2 -b}{\tilde \mu_{\pi^\pm}^2-b}&  M_{\pi^\pm}^2 \le
  \tilde \mu_{\pi^\pm}^2\end{cases}\cr &&\cr
\Omega_{K^\pm} =& -\frac{f_\pi^2}{2} (\tilde \mu_{K^\pm}^2
-b)(1-\cos\theta_{K^\pm})^2\ ,\quad&\cos\theta_{K^\pm}=&\begin{cases}1 &
  M_{K^\pm}^2 \ge \tilde \mu_{K^\pm}^2\cr & \cr
  \frac{M_{K^\pm}^2 -b}{\tilde \mu_{K^\pm}^2-b}&  M_{K^\pm}^2 \le
  \tilde \mu_{K^\pm}^2\end{cases}\cr&&\cr
\Omega_{K^0} =& -\frac{f_\pi^2}{2} \tilde \mu_{K^0}^2
(1-\cos\theta_{K^0})^2\ ,\quad&\cos\theta_{K^0}=&\begin{cases}1 &
  M_{K^0}^2 \ge \tilde \mu_{K^0}^2\cr & \cr
  \frac{M_{K^0}^2 }{\tilde \mu_{K^0}^2}&  M_{K^0}^2 \le
  \tilde \mu_{K^0}^2\end{cases}
\end{array}
\eqn{scons}
\eeq
where in each case $\Omega$ measures the free energy relative to the
$SU(3)$ symmetric CFL ground state.
Evidently the nontrivial solutions with $\theta\ne 0$ represent phases
with lower free energy than the symmetric CFL phase. Which one has
lower free energy depends on the values for $\mu^{(BS)}$ 
and $\mu_Q$. The phase diagram one finds is shown in Fig.~\ref{fig:phase}.

One might wonder how similar the kaon condensed phase is to the CFL
phase.  The number density of strange quarks participating in BCS pairing in
the CFL phase is $n_s=O(\mu^2 \Delta)$, the volume of a shell of thickness
$\Delta$ at the Fermi surface.  On the other hand, the strange quark
density in the $K^0$ condensate can be computed from the chiral
Lagrangian, and one finds $n_s=-\mu_K f_\pi^2 \sin\theta_{K^0}$.  As $m_s^2
\to \mu\Delta$, $\sin\theta_{K^0}\to 1$ (maximal condensation) and
$n_s = - O(\mu^2\Delta)$, which means that as one turns on $m_s$, by
the time one is at the point where CFL is expected to break down,
the number of strange quarks participating in pairing has been
significantly depleted and the ground state looks quite different than the
$SU(3)$ symmetric CFL ground state.

\begin{figure}[t]
\centerline{\includegraphics[height=2.0in]{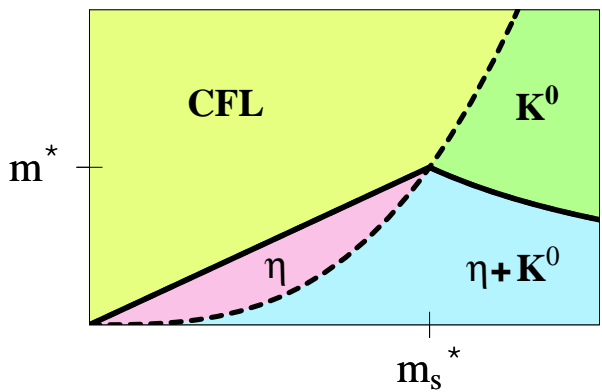} }
\caption{\it The phase diagram as a function of light quark mass $m$ and
  strange quark mass $m_s$ and $m\equiv(m_u+m_d)/2$. The phases marked ``CFL'','' $K^0$'', ``$\eta$'',
  and ``$\eta+K^0$'' are respectively the CFL phase without meson
  condensation, with kaon condensation, with $\eta/\eta'$ condensation, and with
  both $\eta/\eta'$ and $K^0$ condensation.  Phase transitions are represented by 
  a solid line if first order, a dashed line if second order. The
  location of the tetracritical point $(m^*,m_s^*)$ is parametrically
  $m^*=O(\Delta\alpha_s^{3/4})$, $m_s^*=O(\Delta \alpha_s^{1/4})$;
  they are  calculated in ref.~\cite{Kryjevski:2004cw}. }
\label{fig:etacon}\end{figure}

Is kaon condensation the only instability one sees for
$m_s^2/\mu\Delta<1$?  Note that 
$\tilde M_{33}\propto m_u
m_d$ in \eq{mtilde} is extremely small, and therefore small
perturbations could cause  $\Sigma_{33}\ne 1$ in the ground state.
  Indeed, when the $\eta'$ is included in the chiral Lagrangian, one
finds such an instability for condensation of a linear combination of
the $\eta$ and $\eta'$.  Thus at $\mu_Q=0$ one finds the additional
phase structure shown in Fig.~\ref{fig:etacon}.

\hspace{1.25in}

\subsection{Still more phase structure for large $m_s$}
\label{sec:5d}

What happens for higher values of $m_s^2/\mu\Delta$?  There are a host
of possibilities.  As $\mu^{(BS)}$ is turned on, the energy of baryons
(quasi-particles)
carrying anti-strange quarks gets lowered, until eventually on has
massless modes (``gapless superconductivity'') at nonzero quark
masses. This can be seen in the chiral Lagrangian by including the
baryons.  But then new instabilities seem to arise in the baryon
current.  Perhaps these lead to formation of a spatially inhomogeneous
condensate, crystalline superconductivity.  

We see that even at very high densities where QCD is weakly coupled,
dense quark matter exhibits an incredibly rich phase structure, still
far from fully understood.
Effective field theory has played an important role in unraveling
this structure. We can only hope that some day numerical techniques
will be available to explore this phase structure directly from QCD in
regimes where QCD is strongly coupled, and where EFT techniques fail.

\vfill
\noindent
{\bf Acknowledgements}

\medskip

I am grateful to the many people who taught me effective field theory,
most notably my advisor, H. Georgi.  Thanks also to A. Kryjevski and
A. Walker-Loud who
helped proofread these  notes, and to P. Bedaque who invited me to give these lectures.  This work was
 supported by DOE grant DE-FGO3-00ER41132. 

\eject

\bigskip
\noindent
{\bf Problems:}

\bigskip
\hrule
\bigskip
\noindent
{\bf V.1)}
 The Lorentz group is equivalent to a complexified version of
 $SU(2)\times SU(2)$ and irreducible representations can be labeled
 as $(j_1,j_2)$ where the $j$'s are half-integers.  Left-handed
 2-component Weyl fermions transform as
 $(\half,0)$ and right-handed as $(0,\half)$.  Instead of using
 4-component Dirac spinors, a quark can be
 represented by two left-handed Weyl fields $q$   which annihilates
 left-handed quarks,  and $q^c$, which annihilates left-handed
 anti-quarks. This is more convenient than Dirac spinors when baryon
 number is not conserved. Each of these comes in three flavors and
 three colors, with the quantum numbers
\beq
\begin{array}{c|cccc}
&\text{Lorentz} & SU(3)_L & SU(3)_R & SU(3)_c\\ \hline &&& \\
q & (\half,0) & 3 & 1 & 3 \\ &&& \\
q^c & (0,\half) & 1 &\mybar 3 & \mybar 3 
\end{array}
\eeq
 
\begin{enumerate}[a)]
\item Show that there could not be  Lorentz invariant $\vev{q q^c}$ condensate.

\item Show that could be a Lorentz invariant $\vev{qq}$ condensate,
  which would transform  in the reducible $(\mybar 3,1,\mybar 3)\oplus
  (6,1,6)$ representation of $SU(3)_L\times SU(3)_R\times
  SU(3)_c$. (Don't forget that fermion fields anticommute).
\item Show that if the condensate $\vev {q_{\alpha ia} q_{\beta jb}}\propto
  \epsilon_{\alpha\beta}\epsilon_{abx}\epsilon_{ijx}$ (where
  $\alpha,\beta=1,2$ are Lorentz indices,
  $i,j=1,2,3$ are $SU(3)_L$ flavor
  indices,  $a,b=1,2,3$ are color indices, and the index $x$ is summed
  over) corresponds to  a Lorentz singlet condensate in the attractive
  color
  $\mybar 3$ channel. Show that it breaks $SU(3)_L\times SU(3)_c$ down
  to a diagonal $SU(3)$.

\item Consider the analogous $\vev {(q^c)_{\dot\alpha}^{ ia} (q^c)_{\dot\beta}^{ jb}}\propto
  \epsilon_{\dot\alpha\dot\beta}\epsilon^{abx}\epsilon^{ijx}$ (where
  $\dot\alpha,\dot\beta=1,2$ are Lorentz indices,
  $i,j=1,2,3$ are $SU(3)_R$ flavor
  indices,  $a,b=1,2,3$ are color indices) breaks $SU(3)_R\times SU(3)_c$ down
  to a diagonal $SU(3)$.

\item Taken together, show that $SU(3)_L\times SU(3)_R\times SU(3)_c$
  is broken down to a diagonal $SU(3)$, breaking 16 symmetry
  generators.  What happens to the sixteen Goldstone bosons?

\end{enumerate}

\bigskip
\noindent
{\bf V.2)} In \eq{cfl4} I gave a gauge invariant order parameter for
the breaking of $SU(3)\times SU(3)\times U(1)_A$ symmetry.  Find a
gauge invariant parameter for the breaking of baryon number.  What
discrete subgroup of baryon number symmetry is left unbroken?  Can you
see this symmetry in the gauge variant formulation of \eq{cfl}?
(Hint: find all $U(1)_B$ transformations of the quark bilinear which
can be undone by  $SU(3)_c$ transformations).

\bigskip
\noindent
{\bf V.4)} Expand the QCD Lagrangian in terms of $\psi_\pm$ and Derive \eq{psieq}.

\bigskip
\noindent
{\bf V.5)} Start with the QCD Lagrangian expanded in terms of
$\psi_\pm$.  Consider vacuum energy diagrams with a single quark loop,
arbitrary insertions of the mass matrix $M$ and $M^\dagger$, arbitrary
insertions of the quark condensate $\vev{q_{L,i}^a
  q_{L,j}^b}\propto \epsilon^{abx}\epsilon_{ijx}$  and an arbitrary
dressing of gluons.  
\begin{enumerate}[a)]
\item By following chirality around the loop, show that any mass
  dependence of the vacuum energy from such diagrams involves even
  powers of $M$ and/or  $M^\dagger$, and that therefore  there is no
  contribution to the vacuum energy at $O(M)$. (Recall that $M$
  couples $\psi_L^\dagger$  to  $\psi_R$ or $\psi_R^\dagger to
  \psi_L$;  the gluon couples $\psi_L^\dagger$ to $\psi_L$ or
  $\psi_R^\dagger$ to $\psi_R$; and that the condensate couples
  $\psi_L$ to $\psi_L$ or $\psi_R$ to $\psi_R$ (and the conjugate
  couplings).  Draw the analogous diagrams for QCD at $\mu=0$,
  including insertions of the $\mybar q q$ condensate, and show why
  there {\it is} and $O(M)$ contribution in this case, reflected by
  the $\Tr M \Sigma$ term in the chiral Lagrangian, \eq{massterm}.

\item Find a one fermion loop  vacuum energy contribution with two
  insertions of $M$ on the fermion line and insertions of the
  condensate as needed to make the diagram not vanish.  By considering
flavor flowing around the diagram, paying attention to the
$\epsilon$'s in the condensate insertions, show that this vacuum
energy contribution is proportional to $\epsilon_{ijx} \epsilon^{rsx}
M^i_r M^j_s$.  Note that this vanishes if any two quark masses
vanish.  Such diagrams give rise to the $O(M^2)$ terms in the CFL
chiral Lagrangian, \eq{cflchilag}.
\end{enumerate}




\appendix

  \vfill\eject

\section{Dimensional regularization}
\label{sec:7}

\bigskip
\hrule
\bigskip

For your convenience, I am including the standard formulas used in
dimensional regularization.

\subsection{Useful integrals as a function of dimension}
\label{sec:7a}

Consider the following integral in $d$ dimensions with a $Euclidean$ metric:
\beq I_1 \equiv \int \dnk \frac{1}{ (k^2 + a^2)^r} \ .\eeq 
We may evaluate this making in terms of the $\Gamma$ function: 
\beq  \alpha^{-s} \Gamma(s) = \int_0^{\infty} {\rm d}x\ x^{s-1}\ {\rm
e}^{-\alpha x}.\eeq 
Then 
\beq
I_1 &=& \frac{1}{ \Gamma(r)} \int \dnk \int_0^{\infty} {\rm d}x\ x^{r-1}\
{\rm e}^{-x(k^2+a^2)} \cr
&=&\frac{\pi^{d/2}}{ \Gamma(r)} \int_0^{\infty} {\rm d}x\ x^{r-1-d/2}\
{\rm e}^{-x a^2} \cr
&=& \pi^{d/2} a^{d-2r} \frac{\Gamma(r-d/2)}{ \Gamma(r)} \eeq 
Another useful integral is 
\beq I_2 \equiv \int \dnk \frac{k^2}{ (k^2 + a^2)^r} \ .\eeq 
To get this we define
\beq 
I_1(\alpha) &\equiv& \int \dnk \frac{1}{ (\alpha k^2 + a^2)^r} \cr
&=& \alpha^{-d/2} I_1\ ;\eeq 
then by differentiating by $\alpha$ and setting $\alpha=1$ we find
\beq  I_2 = \frac{d \pi^{d/2} a^{d-2r+2} }{ 2(r-1)} \frac{\Gamma (r-1-d/2) }{
\Gamma (r-1)}\ .\eeq 
 
Finally note that 
\beq  
I_3^{\mu \nu} &\equiv& \int \dnk \frac{k^{\mu} k^{\nu} }{ (k^2 + a^2)^r} \cr
&=& \frac{\delta^{\mu \nu} }{ d} I_2\  .\eeq

\subsection{Some properties of the $\Gamma$ function}
\label{sec:7b}

Gamma functions have the property $\Gamma(z+1) = z \Gamma(z)$, with
$\Gamma(1)=1$.  Thus for integers $n \ge 0$,
\beq  \Gamma(n+1) = n!,\qquad n\ge 0\ .\eeq 
Also useful is the value 
\beq  \Gamma\left(\frac{1}{2}\right) = \sqrt{\pi}\ .\eeq 

The Gamma function is singular for non-positive integer arguments. Near
these singularities it can be expanded as
\beq  \Gamma(-n+\epsilon) =\frac{(-1)^n}{ n \!}\left[ \frac{1}{\epsilon}+ \psi(n+1)
+\CO(\epsilon) \right]\ ,\eeq 
where 
\beq 
\psi(n+1) &=& 1 + \frac{1}{2} + \ldots + \frac{1}{ n} - \gamma\ ,\cr
        \gamma &=& 0.5772 \ldots\eeq 
In particular,
\beq 
\Gamma(\epsilon-1) &=& -\frac{1}{\epsilon} +\gamma -1\cr
\Gamma(\epsilon) &=& \frac{1}{\epsilon} - \gamma \eeq

\subsection{Common integrals in $d\to 4$ dimensions}
\label{sec:7c}

It follows  that two of the most useful integrals are given by

\beq  \mu^{2\epsilon} \int \frac{{\rm d}^{4-2\epsilon} q}{
  (2\pi)^{4-2\epsilon}} \frac{1}{ q^2 + m^2} = \frac{m^2}{
  16\pi^2}\left[-\frac{1}{\epsilon} + \gamma - 1 -\ln 4\pi +
  \ln(m^2/\mu^2) + O(\epsilon)\right]\eeq 
\medskip
\beq  \mu^{2\epsilon}\int \frac{{\rm d}^{4-2\epsilon} q}{ (2\pi)^{4-2\epsilon}} \frac{1}{ (q^2 + m^2)^2} 
= \frac{1}{ 16\pi^2}\left[\frac{1}{\epsilon} - \gamma +\ln 4\pi -
  \ln(m^2/\mu^2)+ O(\epsilon)\right]\ .\eeq

\bibliography{eft}
\bibliographystyle{JHEP} 
\end{document}